\documentclass[longauth,twocolumn,traditabstract]{aa}
\usepackage[table,usenames,dvipsnames]{xcolor}
\usepackage{enumerate}
\usepackage{footnpag}
\usepackage{ifthen}
\usepackage{multirow}
\usepackage{upgreek}
\usepackage{graphicx,amsmath}
\usepackage{txfonts}
\usepackage[breaklinks, colorlinks, citecolor=blue, colorlinks=true,linkcolor=blue]{hyperref}
\usepackage{natbib}
\bibpunct{(}{)}{;}{a}{}{,} 
\usepackage{pdfcolmk} 

\def\setsymbol#1#2{\expandafter\def\csname #1\endcsname{#2}}
\def\getsymbol#1{\csname #1\endcsname}

\def\Planck{\textit{Planck}}

\newbox\tablebox    \newdimen\tablewidth
\def\leaderfil{\leaders\hbox to 5pt{\hss.\hss}\hfil}
\def\endPlancktable{\tablewidth=\columnwidth 
    $$\hss\copy\tablebox\hss$$
    \vskip-\lastskip\vskip -2pt}
\def\endPlancktablewide{\tablewidth=\textwidth 
    $$\hss\copy\tablebox\hss$$
    \vskip-\lastskip\vskip -2pt}
\def\tablenote#1 #2\par{\begingroup \parindent=0.8em
    \abovedisplayshortskip=0pt\belowdisplayshortskip=0pt
    \noindent
    $$\hss\vbox{\hsize\tablewidth \hangindent=\parindent \hangafter=1 \noindent
    \hbox to \parindent{$^#1$\hss}\strut#2\strut\par}\hss$$
    \endgroup}
\def\doubleline{\vskip 3pt\hrule \vskip 1.5pt \hrule \vskip 5pt}

\def\L2{\ifmmode L_2\else $L_2$\fi}

\def\DeltaT{\ifmmode \Delta T\else $\Delta T$\fi}
\def\deltat{\ifmmode \Delta t\else $\Delta t$\fi}
\def\fknee{\ifmmode f_{\rm knee}\else $f_{\rm knee}$\fi}
\def\Fmax{\ifmmode F_{\rm max}\else $F_{\rm max}$\fi}
\def\solar{\ifmmode{\rm M}_{\mathord\odot}\else${\rm M}_{\mathord\odot}$\fi}
\def\Msolar{\ifmmode{\rm M}_{\mathord\odot}\else${\rm M}_{\mathord\odot}$\fi}
\def\Lsolar{\ifmmode{\rm L}_{\mathord\odot}\else${\rm L}_{\mathord\odot}$\fi}

\def\inv{\ifmmode^{-1}\else$^{-1}$\fi}
\def\mo{\ifmmode^{-1}\else$^{-1}$\fi}
\def\sup#1{\ifmmode ^{\rm #1}\else $^{\rm #1}$\fi}
\def\expo#1{\ifmmode \times 10^{#1}\else $\times 10^{#1}$\fi}
\def\,{\thinspace}
\def\lsim{\mathrel{\raise .4ex\hbox{\rlap{$<$}\lower 1.2ex\hbox{$\sim$}}}}
\def\gsim{\mathrel{\raise .4ex\hbox{\rlap{$>$}\lower 1.2ex\hbox{$\sim$}}}}

\def\simprop{\mathrel{\raise .4ex\hbox{\rlap{$\propto$}\lower 1.2ex\hbox{$\sim$}}}}
\def\deg{\ifmmode^\circ\else$^\circ$\fi}
\def\pdeg{\ifmmode $\setbox0=\hbox{$^{\circ}$}\rlap{\hskip.11\wd0 .}$^{\circ}
          \else \setbox0=\hbox{$^{\circ}$}\rlap{\hskip.11\wd0 .}$^{\circ}$\fi}
\def\arcs{\ifmmode {^{\scriptstyle\prime\prime}}
          \else $^{\scriptstyle\prime\prime}$\fi}
\def\arcm{\ifmmode {^{\scriptstyle\prime}}
          \else $^{\scriptstyle\prime}$\fi}
\newdimen\sa  \newdimen\sb
\def\parcs{\sa=.07em \sb=.03em
     \ifmmode \hbox{\rlap{.}}^{\scriptstyle\prime\kern -\sb\prime}\hbox{\kern -\sa}
     \else \rlap{.}$^{\scriptstyle\prime\kern -\sb\prime}$\kern -\sa\fi}
\def\parcm{\sa=.08em \sb=.03em
     \ifmmode \hbox{\rlap{.}\kern\sa}^{\scriptstyle\prime}\hbox{\kern-\sb}
     \else \rlap{.}\kern\sa$^{\scriptstyle\prime}$\kern-\sb\fi}
\def\ra[#1 #2 #3.#4]{#1\sup{h}#2\sup{m}#3\sup{s}\llap.#4}
\def\dec[#1 #2 #3.#4]{#1\deg#2\arcm#3\arcs\llap.#4}
\def\deco[#1 #2 #3]{#1\deg#2\arcm#3\arcs}
\def\rra[#1 #2]{#1\sup{h}#2\sup{m}}

\def\dots{\relax\ifmmode \ldots\else $\ldots$\fi}
\def\WHzsr{\ifmmode $W\,Hz\mo\,sr\mo$\else W\,Hz\mo\,sr\mo\fi}
\def\mHz{\ifmmode $\,mHz$\else \,mHz\fi}
\def\GHz{\ifmmode $\,GHz$\else \,GHz\fi}
\def\mKs{\ifmmode $\,mK\,s$^{1/2}\else \,mK\,s$^{1/2}$\fi}
\def\muKs{\ifmmode \,\mu$K\,s$^{1/2}\else \,$\mu$K\,s$^{1/2}$\fi}
\def\muKRJs{\ifmmode \,\mu$K$_{\rm RJ}$\,s$^{1/2}\else \,$\mu$K$_{\rm RJ}$\,s$^{1/2}$\fi}
\def\muKHz{\ifmmode \,\mu$K\,Hz$^{-1/2}\else \,$\mu$K\,Hz$^{-1/2}$\fi}
\def\MJysr{\ifmmode \,$MJy\,sr\mo$\else \,MJy\,sr\mo\fi}
\def\MJysrmK{\ifmmode \,$MJy\,sr\mo$\,mK$_{\rm CMB}\mo\else \,MJy\,sr\mo\,mK$_{\rm CMB}\mo$\fi}
\def\microns{\ifmmode \,\mu$m$\else \,$\mu$m\fi}

\def\muK{\ifmmode \,\mu$K$\else \,$\mu$\hbox{K}\fi}
\def\microK{\ifmmode \,\mu$K$\else \,$\mu$\hbox{K}\fi}
\def\muW{\ifmmode \,\mu$W$\else \,$\mu$\hbox{W}\fi}
\def\kms{\ifmmode $\,km\,s$^{-1}\else \,km\,s$^{-1}$\fi}
\def\kmsMpc{\ifmmode $\,\kms\,Mpc\mo$\else \,\kms\,Mpc\mo\fi}
\setsymbol{LFI:center:frequency:70GHz:units}{70.3\,GHz}
\setsymbol{LFI:center:frequency:44GHz:units}{44.1\,GHz}
\setsymbol{LFI:center:frequency:30GHz:units}{28.5\,GHz}

\setsymbol{LFI:center:frequency:70GHz}{70.3}
\setsymbol{LFI:center:frequency:44GHz}{44.1}
\setsymbol{LFI:center:frequency:30GHz}{28.5}

\setsymbol{LFI:center:frequency:LFI18:Rad:M:units}{71.7\GHz}
\setsymbol{LFI:center:frequency:LFI19:Rad:M:units}{67.5\GHz}
\setsymbol{LFI:center:frequency:LFI20:Rad:M:units}{69.2\GHz}
\setsymbol{LFI:center:frequency:LFI21:Rad:M:units}{70.4\GHz}
\setsymbol{LFI:center:frequency:LFI22:Rad:M:units}{71.5\GHz}
\setsymbol{LFI:center:frequency:LFI23:Rad:M:units}{70.8\GHz}
\setsymbol{LFI:center:frequency:LFI24:Rad:M:units}{44.4\GHz}
\setsymbol{LFI:center:frequency:LFI25:Rad:M:units}{44.0\GHz}
\setsymbol{LFI:center:frequency:LFI26:Rad:M:units}{43.9\GHz}
\setsymbol{LFI:center:frequency:LFI27:Rad:M:units}{28.3\GHz}
\setsymbol{LFI:center:frequency:LFI28:Rad:M:units}{28.8\GHz}
\setsymbol{LFI:center:frequency:LFI18:Rad:S:units}{70.1\GHz}
\setsymbol{LFI:center:frequency:LFI19:Rad:S:units}{69.6\GHz}
\setsymbol{LFI:center:frequency:LFI20:Rad:S:units}{69.5\GHz}
\setsymbol{LFI:center:frequency:LFI21:Rad:S:units}{69.5\GHz}
\setsymbol{LFI:center:frequency:LFI22:Rad:S:units}{72.8\GHz}
\setsymbol{LFI:center:frequency:LFI23:Rad:S:units}{71.3\GHz}
\setsymbol{LFI:center:frequency:LFI24:Rad:S:units}{44.1\GHz}
\setsymbol{LFI:center:frequency:LFI25:Rad:S:units}{44.1\GHz}
\setsymbol{LFI:center:frequency:LFI26:Rad:S:units}{44.1\GHz}
\setsymbol{LFI:center:frequency:LFI27:Rad:S:units}{28.5\GHz}
\setsymbol{LFI:center:frequency:LFI28:Rad:S:units}{28.2\GHz}

\setsymbol{LFI:center:frequency:LFI18:Rad:M}{71.7}
\setsymbol{LFI:center:frequency:LFI19:Rad:M}{67.5}
\setsymbol{LFI:center:frequency:LFI20:Rad:M}{69.2}
\setsymbol{LFI:center:frequency:LFI21:Rad:M}{70.4}
\setsymbol{LFI:center:frequency:LFI22:Rad:M}{71.5}
\setsymbol{LFI:center:frequency:LFI23:Rad:M}{70.8}
\setsymbol{LFI:center:frequency:LFI24:Rad:M}{44.4}
\setsymbol{LFI:center:frequency:LFI25:Rad:M}{44.0}
\setsymbol{LFI:center:frequency:LFI26:Rad:M}{43.9}
\setsymbol{LFI:center:frequency:LFI27:Rad:M}{28.3}
\setsymbol{LFI:center:frequency:LFI28:Rad:M}{28.8}
\setsymbol{LFI:center:frequency:LFI18:Rad:S}{70.1}
\setsymbol{LFI:center:frequency:LFI19:Rad:S}{69.6}
\setsymbol{LFI:center:frequency:LFI20:Rad:S}{69.5}
\setsymbol{LFI:center:frequency:LFI21:Rad:S}{69.5}
\setsymbol{LFI:center:frequency:LFI22:Rad:S}{72.8}
\setsymbol{LFI:center:frequency:LFI23:Rad:S}{71.3}
\setsymbol{LFI:center:frequency:LFI24:Rad:S}{44.1}
\setsymbol{LFI:center:frequency:LFI25:Rad:S}{44.1}
\setsymbol{LFI:center:frequency:LFI26:Rad:S}{44.1}
\setsymbol{LFI:center:frequency:LFI27:Rad:S}{28.5}
\setsymbol{LFI:center:frequency:LFI28:Rad:S}{28.2}

\setsymbol{LFI:white:noise:sensitivity:70GHz:units}{134.7\muKs}
\setsymbol{LFI:white:noise:sensitivity:44GHz:units}{164.7\muKs}
\setsymbol{LFI:white:noise:sensitivity:30GHz:units}{143.4\muKs}

\setsymbol{LFI:white:noise:sensitivity:70GHz}{134.7}
\setsymbol{LFI:white:noise:sensitivity:44GHz}{164.7}
\setsymbol{LFI:white:noise:sensitivity:30GHz}{143.4}

\setsymbol{LFI:white:noise:sensitivity:LFI18:Rad:M:units}{512.0\muKs}
\setsymbol{LFI:white:noise:sensitivity:LFI19:Rad:M:units}{581.4\muKs}
\setsymbol{LFI:white:noise:sensitivity:LFI20:Rad:M:units}{590.8\muKs}
\setsymbol{LFI:white:noise:sensitivity:LFI21:Rad:M:units}{455.2\muKs}
\setsymbol{LFI:white:noise:sensitivity:LFI22:Rad:M:units}{492.0\muKs}
\setsymbol{LFI:white:noise:sensitivity:LFI23:Rad:M:units}{507.7\muKs}
\setsymbol{LFI:white:noise:sensitivity:LFI24:Rad:M:units}{462.2\muKs}
\setsymbol{LFI:white:noise:sensitivity:LFI25:Rad:M:units}{413.6\muKs}
\setsymbol{LFI:white:noise:sensitivity:LFI26:Rad:M:units}{478.6\muKs}
\setsymbol{LFI:white:noise:sensitivity:LFI27:Rad:M:units}{277.7\muKs}
\setsymbol{LFI:white:noise:sensitivity:LFI28:Rad:M:units}{312.3\muKs}
\setsymbol{LFI:white:noise:sensitivity:LFI18:Rad:S:units}{465.7\muKs}
\setsymbol{LFI:white:noise:sensitivity:LFI19:Rad:S:units}{555.6\muKs}
\setsymbol{LFI:white:noise:sensitivity:LFI20:Rad:S:units}{623.2\muKs}
\setsymbol{LFI:white:noise:sensitivity:LFI21:Rad:S:units}{564.1\muKs}
\setsymbol{LFI:white:noise:sensitivity:LFI22:Rad:S:units}{534.4\muKs}
\setsymbol{LFI:white:noise:sensitivity:LFI23:Rad:S:units}{542.4\muKs}
\setsymbol{LFI:white:noise:sensitivity:LFI24:Rad:S:units}{399.2\muKs}
\setsymbol{LFI:white:noise:sensitivity:LFI25:Rad:S:units}{392.6\muKs}
\setsymbol{LFI:white:noise:sensitivity:LFI26:Rad:S:units}{418.6\muKs}
\setsymbol{LFI:white:noise:sensitivity:LFI27:Rad:S:units}{302.9\muKs}
\setsymbol{LFI:white:noise:sensitivity:LFI28:Rad:S:units}{285.3\muKs}

\setsymbol{LFI:white:noise:sensitivity:LFI18:Rad:M}{512.0}
\setsymbol{LFI:white:noise:sensitivity:LFI19:Rad:M}{581.4}
\setsymbol{LFI:white:noise:sensitivity:LFI20:Rad:M}{590.8}
\setsymbol{LFI:white:noise:sensitivity:LFI21:Rad:M}{455.2}
\setsymbol{LFI:white:noise:sensitivity:LFI22:Rad:M}{492.0}
\setsymbol{LFI:white:noise:sensitivity:LFI23:Rad:M}{507.7}
\setsymbol{LFI:white:noise:sensitivity:LFI24:Rad:M}{462.2}
\setsymbol{LFI:white:noise:sensitivity:LFI25:Rad:M}{413.6}
\setsymbol{LFI:white:noise:sensitivity:LFI26:Rad:M}{478.6}
\setsymbol{LFI:white:noise:sensitivity:LFI27:Rad:M}{277.7}
\setsymbol{LFI:white:noise:sensitivity:LFI28:Rad:M}{312.3}
\setsymbol{LFI:white:noise:sensitivity:LFI18:Rad:S}{465.7}
\setsymbol{LFI:white:noise:sensitivity:LFI19:Rad:S}{555.6}
\setsymbol{LFI:white:noise:sensitivity:LFI20:Rad:S}{623.2}
\setsymbol{LFI:white:noise:sensitivity:LFI21:Rad:S}{564.1}
\setsymbol{LFI:white:noise:sensitivity:LFI22:Rad:S}{534.4}
\setsymbol{LFI:white:noise:sensitivity:LFI23:Rad:S}{542.4}
\setsymbol{LFI:white:noise:sensitivity:LFI24:Rad:S}{399.2}
\setsymbol{LFI:white:noise:sensitivity:LFI25:Rad:S}{392.6}
\setsymbol{LFI:white:noise:sensitivity:LFI26:Rad:S}{418.6}
\setsymbol{LFI:white:noise:sensitivity:LFI27:Rad:S}{302.9}
\setsymbol{LFI:white:noise:sensitivity:LFI28:Rad:S}{285.3}

\setsymbol{LFI:knee:frequency:70GHz:units}{29.5\mHz}
\setsymbol{LFI:knee:frequency:44GHz:units}{56.2\mHz}
\setsymbol{LFI:knee:frequency:30GHz:units}{113.7\mHz}

\setsymbol{LFI:knee:frequency:70GHz}{29.5}
\setsymbol{LFI:knee:frequency:44GHz}{56.2}
\setsymbol{LFI:knee:frequency:30GHz}{113.7}

\setsymbol{LFI:knee:frequency:LFI18:Rad:M:units}{16.3\mHz}
\setsymbol{LFI:knee:frequency:LFI19:Rad:M:units}{15.1\mHz}
\setsymbol{LFI:knee:frequency:LFI20:Rad:M:units}{18.7\mHz}
\setsymbol{LFI:knee:frequency:LFI21:Rad:M:units}{37.2\mHz}
\setsymbol{LFI:knee:frequency:LFI22:Rad:M:units}{12.7\mHz}
\setsymbol{LFI:knee:frequency:LFI23:Rad:M:units}{34.6\mHz}
\setsymbol{LFI:knee:frequency:LFI24:Rad:M:units}{46.2\mHz}
\setsymbol{LFI:knee:frequency:LFI25:Rad:M:units}{24.9\mHz}
\setsymbol{LFI:knee:frequency:LFI26:Rad:M:units}{67.6\mHz}
\setsymbol{LFI:knee:frequency:LFI27:Rad:M:units}{187.4\mHz}
\setsymbol{LFI:knee:frequency:LFI28:Rad:M:units}{122.2\mHz}
\setsymbol{LFI:knee:frequency:LFI18:Rad:S:units}{17.7\mHz}
\setsymbol{LFI:knee:frequency:LFI19:Rad:S:units}{22.0\mHz}
\setsymbol{LFI:knee:frequency:LFI20:Rad:S:units}{8.7\mHz}
\setsymbol{LFI:knee:frequency:LFI21:Rad:S:units}{25.9\mHz}
\setsymbol{LFI:knee:frequency:LFI22:Rad:S:units}{15.8\mHz}
\setsymbol{LFI:knee:frequency:LFI23:Rad:S:units}{129.8\mHz}
\setsymbol{LFI:knee:frequency:LFI24:Rad:S:units}{100.9\mHz}
\setsymbol{LFI:knee:frequency:LFI25:Rad:S:units}{38.9\mHz}
\setsymbol{LFI:knee:frequency:LFI26:Rad:S:units}{58.9\mHz}
\setsymbol{LFI:knee:frequency:LFI27:Rad:S:units}{104.4\mHz}
\setsymbol{LFI:knee:frequency:LFI28:Rad:S:units}{40.7\mHz}

\setsymbol{LFI:knee:frequency:LFI18:Rad:M}{16.3}
\setsymbol{LFI:knee:frequency:LFI19:Rad:M}{15.1}
\setsymbol{LFI:knee:frequency:LFI20:Rad:M}{18.7}
\setsymbol{LFI:knee:frequency:LFI21:Rad:M}{37.2}
\setsymbol{LFI:knee:frequency:LFI22:Rad:M}{12.7}
\setsymbol{LFI:knee:frequency:LFI23:Rad:M}{34.6}
\setsymbol{LFI:knee:frequency:LFI24:Rad:M}{46.2}
\setsymbol{LFI:knee:frequency:LFI25:Rad:M}{24.9}
\setsymbol{LFI:knee:frequency:LFI26:Rad:M}{67.6}
\setsymbol{LFI:knee:frequency:LFI27:Rad:M}{187.4}
\setsymbol{LFI:knee:frequency:LFI28:Rad:M}{122.2}
\setsymbol{LFI:knee:frequency:LFI18:Rad:S}{17.7}
\setsymbol{LFI:knee:frequency:LFI19:Rad:S}{22.0}
\setsymbol{LFI:knee:frequency:LFI20:Rad:S}{8.7}
\setsymbol{LFI:knee:frequency:LFI21:Rad:S}{25.9}
\setsymbol{LFI:knee:frequency:LFI22:Rad:S}{15.8}
\setsymbol{LFI:knee:frequency:LFI23:Rad:S}{129.8}
\setsymbol{LFI:knee:frequency:LFI24:Rad:S}{100.9}
\setsymbol{LFI:knee:frequency:LFI25:Rad:S}{38.9}
\setsymbol{LFI:knee:frequency:LFI26:Rad:S}{58.9}
\setsymbol{LFI:knee:frequency:LFI27:Rad:S}{104.4}
\setsymbol{LFI:knee:frequency:LFI28:Rad:S}{40.7}

\setsymbol{LFI:slope:70GHz:units}{$-1.03$\mHz}
\setsymbol{LFI:slope:44GHz:units}{$-0.89$\mHz}
\setsymbol{LFI:slope:30GHz:units}{$-0.87$\mHz}

\setsymbol{LFI:slope:70GHz}{$-1.03$}
\setsymbol{LFI:slope:44GHz}{$-0.89$}
\setsymbol{LFI:slope:30GHz}{$-0.87$}

\setsymbol{LFI:slope:LFI18:Rad:M:units}{$-1.04$\mHz}
\setsymbol{LFI:slope:LFI19:Rad:M:units}{$-1.09$\mHz}
\setsymbol{LFI:slope:LFI20:Rad:M:units}{$-0.69$\mHz}
\setsymbol{LFI:slope:LFI21:Rad:M:units}{$-1.56$\mHz}
\setsymbol{LFI:slope:LFI22:Rad:M:units}{$-1.01$\mHz}
\setsymbol{LFI:slope:LFI23:Rad:M:units}{$-0.96$\mHz}
\setsymbol{LFI:slope:LFI24:Rad:M:units}{$-0.83$\mHz}
\setsymbol{LFI:slope:LFI25:Rad:M:units}{$-0.91$\mHz}
\setsymbol{LFI:slope:LFI26:Rad:M:units}{$-0.95$\mHz}
\setsymbol{LFI:slope:LFI27:Rad:M:units}{$-0.87$\mHz}
\setsymbol{LFI:slope:LFI28:Rad:M:units}{$-0.88$\mHz}
\setsymbol{LFI:slope:LFI18:Rad:S:units}{$-1.15$\mHz}
\setsymbol{LFI:slope:LFI19:Rad:S:units}{$-1.00$\mHz}
\setsymbol{LFI:slope:LFI20:Rad:S:units}{$-0.95$\mHz}
\setsymbol{LFI:slope:LFI21:Rad:S:units}{$-0.92$\mHz}
\setsymbol{LFI:slope:LFI22:Rad:S:units}{$-1.01$\mHz}
\setsymbol{LFI:slope:LFI23:Rad:S:units}{$-0.95$\mHz}
\setsymbol{LFI:slope:LFI24:Rad:S:units}{$-0.73$\mHz}
\setsymbol{LFI:slope:LFI25:Rad:S:units}{$-1.16$\mHz}
\setsymbol{LFI:slope:LFI26:Rad:S:units}{$-0.79$\mHz}
\setsymbol{LFI:slope:LFI27:Rad:S:units}{$-0.82$\mHz}
\setsymbol{LFI:slope:LFI28:Rad:S:units}{$-0.91$\mHz}

\setsymbol{LFI:slope:LFI18:Rad:M}{$-1.04$}
\setsymbol{LFI:slope:LFI19:Rad:M}{$-1.09$}
\setsymbol{LFI:slope:LFI20:Rad:M}{$-0.69$}
\setsymbol{LFI:slope:LFI21:Rad:M}{$-1.56$}
\setsymbol{LFI:slope:LFI22:Rad:M}{$-1.01$}
\setsymbol{LFI:slope:LFI23:Rad:M}{$-0.96$}
\setsymbol{LFI:slope:LFI24:Rad:M}{$-0.83$}
\setsymbol{LFI:slope:LFI25:Rad:M}{$-0.91$}
\setsymbol{LFI:slope:LFI26:Rad:M}{$-0.95$}
\setsymbol{LFI:slope:LFI27:Rad:M}{$-0.87$}
\setsymbol{LFI:slope:LFI28:Rad:M}{$-0.88$}
\setsymbol{LFI:slope:LFI18:Rad:S}{$-1.15$}
\setsymbol{LFI:slope:LFI19:Rad:S}{$-1.00$}
\setsymbol{LFI:slope:LFI20:Rad:S}{$-0.95$}
\setsymbol{LFI:slope:LFI21:Rad:S}{$-0.92$}
\setsymbol{LFI:slope:LFI22:Rad:S}{$-1.01$}
\setsymbol{LFI:slope:LFI23:Rad:S}{$-0.95$}
\setsymbol{LFI:slope:LFI24:Rad:S}{$-0.73$}
\setsymbol{LFI:slope:LFI25:Rad:S}{$-1.16$}
\setsymbol{LFI:slope:LFI26:Rad:S}{$-0.79$}
\setsymbol{LFI:slope:LFI27:Rad:S}{$-0.82$}
\setsymbol{LFI:slope:LFI28:Rad:S}{$-0.91$}

\setsymbol{LFI:FWHM:70GHz:units}{13\parcm01}
\setsymbol{LFI:FWHM:44GHz:units}{27\parcm92}
\setsymbol{LFI:FWHM:30GHz:units}{32\parcm65}

\setsymbol{LFI:FWHM:70GHz}{13.01}
\setsymbol{LFI:FWHM:44GHz}{27.92}
\setsymbol{LFI:FWHM:30GHz}{32.65}

\setsymbol{LFI:FWHM:LFI18:units}{13\parcm39}
\setsymbol{LFI:FWHM:LFI19:units}{13\parcm01}
\setsymbol{LFI:FWHM:LFI20:units}{12\parcm75}
\setsymbol{LFI:FWHM:LFI21:units}{12\parcm74}
\setsymbol{LFI:FWHM:LFI22:units}{12\parcm87}
\setsymbol{LFI:FWHM:LFI23:units}{13\parcm27}
\setsymbol{LFI:FWHM:LFI24:units}{22\parcm98}
\setsymbol{LFI:FWHM:LFI25:units}{30\parcm46}
\setsymbol{LFI:FWHM:LFI26:units}{30\parcm31}
\setsymbol{LFI:FWHM:LFI27:units}{32\parcm65}
\setsymbol{LFI:FWHM:LFI28:units}{32\parcm66}

\setsymbol{LFI:FWHM:LFI18}{13.39}
\setsymbol{LFI:FWHM:LFI19}{13.01}
\setsymbol{LFI:FWHM:LFI20}{12.75}
\setsymbol{LFI:FWHM:LFI21}{12.74}
\setsymbol{LFI:FWHM:LFI22}{12.87}
\setsymbol{LFI:FWHM:LFI23}{13.27}
\setsymbol{LFI:FWHM:LFI24}{22.98}
\setsymbol{LFI:FWHM:LFI25}{30.46}
\setsymbol{LFI:FWHM:LFI26}{30.31}
\setsymbol{LFI:FWHM:LFI27}{32.65}
\setsymbol{LFI:FWHM:LFI28}{32.66}

\setsymbol{LFI:FWHM:uncertainty:LFI18:units}{0.170\arcm}
\setsymbol{LFI:FWHM:uncertainty:LFI19:units}{0.174\arcm}
\setsymbol{LFI:FWHM:uncertainty:LFI20:units}{0.170\arcm}
\setsymbol{LFI:FWHM:uncertainty:LFI21:units}{0.156\arcm}
\setsymbol{LFI:FWHM:uncertainty:LFI22:units}{0.164\arcm}
\setsymbol{LFI:FWHM:uncertainty:LFI23:units}{0.171\arcm}
\setsymbol{LFI:FWHM:uncertainty:LFI24:units}{0.652\arcm}
\setsymbol{LFI:FWHM:uncertainty:LFI25:units}{1.075\arcm}
\setsymbol{LFI:FWHM:uncertainty:LFI26:units}{1.131\arcm}
\setsymbol{LFI:FWHM:uncertainty:LFI27:units}{1.266\arcm}
\setsymbol{LFI:FWHM:uncertainty:LFI28:units}{1.287\arcm}

\setsymbol{LFI:FWHM:uncertainty:LFI18}{0.170}
\setsymbol{LFI:FWHM:uncertainty:LFI19}{0.174}
\setsymbol{LFI:FWHM:uncertainty:LFI20}{0.170}
\setsymbol{LFI:FWHM:uncertainty:LFI21}{0.156}
\setsymbol{LFI:FWHM:uncertainty:LFI22}{0.164}
\setsymbol{LFI:FWHM:uncertainty:LFI23}{0.171}
\setsymbol{LFI:FWHM:uncertainty:LFI24}{0.652}
\setsymbol{LFI:FWHM:uncertainty:LFI25}{1.075}
\setsymbol{LFI:FWHM:uncertainty:LFI26}{1.131}
\setsymbol{LFI:FWHM:uncertainty:LFI27}{1.266}
\setsymbol{LFI:FWHM:uncertainty:LFI28}{1.287}

\setsymbol{HFI:center:frequency:100GHz:units}{100\,GHz}
\setsymbol{HFI:center:frequency:143GHz:units}{143\,GHz}
\setsymbol{HFI:center:frequency:217GHz:units}{217\,GHz}
\setsymbol{HFI:center:frequency:353GHz:units}{353\,GHz}
\setsymbol{HFI:center:frequency:545GHz:units}{545\,GHz}
\setsymbol{HFI:center:frequency:857GHz:units}{857\,GHz}

\setsymbol{HFI:center:frequency:100GHz}{100}
\setsymbol{HFI:center:frequency:143GHz}{143}
\setsymbol{HFI:center:frequency:217GHz}{217}
\setsymbol{HFI:center:frequency:353GHz}{353}
\setsymbol{HFI:center:frequency:545GHz}{545}
\setsymbol{HFI:center:frequency:857GHz}{857}

\setsymbol{HFI:Ndetectors:100GHz}{8}
\setsymbol{HFI:Ndetectors:143GHz}{11}
\setsymbol{HFI:Ndetectors:217GHz}{12}
\setsymbol{HFI:Ndetectors:353GHz}{12}
\setsymbol{HFI:Ndetectors:545GHz}{3}
\setsymbol{HFI:Ndetectors:857GHz}{4}

\setsymbol{HFI:FWHM:Maps:100GHz:units}{9\parcm88}
\setsymbol{HFI:FWHM:Maps:143GHz:units}{7\parcm18}
\setsymbol{HFI:FWHM:Maps:217GHz:units}{4\parcm87}
\setsymbol{HFI:FWHM:Maps:353GHz:units}{4\parcm65}
\setsymbol{HFI:FWHM:Maps:545GHz:units}{4\parcm72}
\setsymbol{HFI:FWHM:Maps:857GHz:units}{4\parcm39}
\setsymbol{HFI:FWHM:Maps:100GHz}{9.88}
\setsymbol{HFI:FWHM:Maps:143GHz}{7.18}
\setsymbol{HFI:FWHM:Maps:217GHz}{4.87}
\setsymbol{HFI:FWHM:Maps:353GHz}{4.65}
\setsymbol{HFI:FWHM:Maps:545GHz}{4.72}
\setsymbol{HFI:FWHM:Maps:857GHz}{4.39}

\setsymbol{HFI:beam:ellipticity:Maps:100GHz}{1.15}
\setsymbol{HFI:beam:ellipticity:Maps:143GHz}{1.01}
\setsymbol{HFI:beam:ellipticity:Maps:217GHz}{1.06}
\setsymbol{HFI:beam:ellipticity:Maps:353GHz}{1.05}
\setsymbol{HFI:beam:ellipticity:Maps:545GHz}{1.14}
\setsymbol{HFI:beam:ellipticity:Maps:857GHz}{1.19}

\setsymbol{HFI:FWHM:Mars:100GHz:units}{9\parcm37}
\setsymbol{HFI:FWHM:Mars:143GHz:units}{7\parcm04}
\setsymbol{HFI:FWHM:Mars:217GHz:units}{4\parcm68}
\setsymbol{HFI:FWHM:Mars:353GHz:units}{4\parcm43}
\setsymbol{HFI:FWHM:Mars:545GHz:units}{3\parcm80}
\setsymbol{HFI:FWHM:Mars:857GHz:units}{3\parcm67}

\setsymbol{HFI:FWHM:Mars:100GHz}{9.37}
\setsymbol{HFI:FWHM:Mars:143GHz}{7.04}
\setsymbol{HFI:FWHM:Mars:217GHz}{4.68}
\setsymbol{HFI:FWHM:Mars:353GHz}{4.43}
\setsymbol{HFI:FWHM:Mars:545GHz}{3.80}
\setsymbol{HFI:FWHM:Mars:857GHz}{3.67}

\setsymbol{HFI:beam:ellipticity:Mars:100GHz}{1.18}
\setsymbol{HFI:beam:ellipticity:Mars:143GHz}{1.03}
\setsymbol{HFI:beam:ellipticity:Mars:217GHz}{1.14}
\setsymbol{HFI:beam:ellipticity:Mars:353GHz}{1.09}
\setsymbol{HFI:beam:ellipticity:Mars:545GHz}{1.25}
\setsymbol{HFI:beam:ellipticity:Mars:857GHz}{1.03}

\setsymbol{HFI:CMB:relative:calibration:100GHz}{$\lsim 1\%$}
\setsymbol{HFI:CMB:relative:calibration:143GHz}{$\lsim 1\%$}
\setsymbol{HFI:CMB:relative:calibration:217GHz}{$\lsim 1\%$}
\setsymbol{HFI:CMB:relative:calibration:353GHz}{$\lsim 1\%$}
\setsymbol{HFI:CMB:relative:calibration:545GHz}{}
\setsymbol{HFI:CMB:relative:calibration:857GHz}{}

\setsymbol{HFI:CMB:absolute:calibration:100GHz}{$\lsim 2\%$}
\setsymbol{HFI:CMB:absolute:calibration:143GHz}{$\lsim 2\%$}
\setsymbol{HFI:CMB:absolute:calibration:217GHz}{$\lsim 2\%$}
\setsymbol{HFI:CMB:absolute:calibration:353GHz}{$\lsim 2\%$}
\setsymbol{HFI:CMB:absolute:calibration:545GHz}{}
\setsymbol{HFI:CMB:absolute:calibration:857GHz}{}

\setsymbol{HFI:FIRAS:gain:calibration:accuracy:statistical:100GHz}{}
\setsymbol{HFI:FIRAS:gain:calibration:accuracy:statistical:143GHz}{}
\setsymbol{HFI:FIRAS:gain:calibration:accuracy:statistical:217GHz}{}
\setsymbol{HFI:FIRAS:gain:calibration:accuracy:statistical:353GHz}{2.5\%}
\setsymbol{HFI:FIRAS:gain:calibration:accuracy:statistical:545GHz}{1\%}
\setsymbol{HFI:FIRAS:gain:calibration:accuracy:statistical:857GHz}{0.5\%}

\setsymbol{HFI:FIRAS:gain:calibration:accuracy:systematic:100GHz}{}
\setsymbol{HFI:FIRAS:gain:calibration:accuracy:systematic:143GHz}{}
\setsymbol{HFI:FIRAS:gain:calibration:accuracy:systematic:217GHz}{}
\setsymbol{HFI:FIRAS:gain:calibration:accuracy:systematic:353GHz}{}
\setsymbol{HFI:FIRAS:gain:calibration:accuracy:systematic:545GHz}{7\%}
\setsymbol{HFI:FIRAS:gain:calibration:accuracy:systematic:857GHz}{7\%}

\setsymbol{HFI:FIRAS:zero:point:accuracy:100GHz:units}{0.8\MJysr}
\setsymbol{HFI:FIRAS:zero:point:accuracy:143GHz:units}{}
\setsymbol{HFI:FIRAS:zero:point:accuracy:217GHz:units}{}
\setsymbol{HFI:FIRAS:zero:point:accuracy:353GHz:units}{1.4\MJysr}
\setsymbol{HFI:FIRAS:zero:point:accuracy:545GHz:units}{2.2\MJysr}
\setsymbol{HFI:FIRAS:zero:point:accuracy:857GHz:units}{1.7\MJysr}

\setsymbol{HFI:FIRAS:zero:point:accuracy:100GHz}{0.8}
\setsymbol{HFI:FIRAS:zero:point:accuracy:143GHz}{}
\setsymbol{HFI:FIRAS:zero:point:accuracy:217GHz}{}
\setsymbol{HFI:FIRAS:zero:point:accuracy:353GHz}{1.4}
\setsymbol{HFI:FIRAS:zero:point:accuracy:545GHz}{2.2}
\setsymbol{HFI:FIRAS:zero:point:accuracy:857GHz}{1.7}

\setsymbol{HFI:unit:conversion:100GHz:units}{0.2415\MJysrmK}
\setsymbol{HFI:unit:conversion:143GHz:units}{0.3694\MJysrmK}
\setsymbol{HFI:unit:conversion:217GHz:units}{0.4811\MJysrmK}
\setsymbol{HFI:unit:conversion:353GHz:units}{0.2883\MJysrmK}
\setsymbol{HFI:unit:conversion:545GHz:units}{0.05826\MJysrmK}
\setsymbol{HFI:unit:conversion:857GHz:units}{0.002238\MJysrmK}

\setsymbol{HFI:unit:conversion:100GHz}{0.2415}
\setsymbol{HFI:unit:conversion:143GHz}{0.3694}
\setsymbol{HFI:unit:conversion:217GHz}{0.4811}
\setsymbol{HFI:unit:conversion:353GHz}{0.2883}
\setsymbol{HFI:unit:conversion:545GHz}{0.05826}
\setsymbol{HFI:unit:conversion:857GHz}{0.002238}

\setsymbol{HFI:colour:correction:alpha=-2:V1.01:100GHz}{0.9893}
\setsymbol{HFI:colour:correction:alpha=-2:V1.01:143GHz}{0.9759}
\setsymbol{HFI:colour:correction:alpha=-2:V1.01:217GHz}{1.0007}
\setsymbol{HFI:colour:correction:alpha=-2:V1.01:353GHz}{1.0028}
\setsymbol{HFI:colour:correction:alpha=-2:V1.01:545GHz}{1.0019}
\setsymbol{HFI:colour:correction:alpha=-2:V1.01:857GHz}{0.9889}

\setsymbol{HFI:colour:correction:alpha=0:V1.01:100GHz}{1.0008}
\setsymbol{HFI:colour:correction:alpha=0:V1.01:143GHz}{1.0148}
\setsymbol{HFI:colour:correction:alpha=0:V1.01:217GHz}{0.9909}
\setsymbol{HFI:colour:correction:alpha=0:V1.01:353GHz}{0.9888}
\setsymbol{HFI:colour:correction:alpha=0:V1.01:545GHz}{0.9878}
\setsymbol{HFI:colour:correction:alpha=0:V1.01:857GHz}{1.0014}

\definecolor{grey}{rgb}{0.5,0.5,0.5} 

\def\ben{\begin{enumerate}}
\def\een{\end{enumerate}}
\def\bi{\begin{itemize}}
\def\ei{\end{itemize}}
\def\be{\begin{equation}}
\def\ee{\end{equation}}
\def\bea{\begin{eqnarray}}
\def\eea{\end{eqnarray}}

\newcommand{\n}{\vec{\hat{n\,}}}

\newcommand{\hi}{\ion{H}{i}}

\begin{document}

\title{\Planck\ 2013 results. XVIII.\\ Gravitational lensing-infrared
  background correlation}
\titlerunning{\Planck\ 2013 results. XVIII. Gravitational
  lensing-infrared background correlation}
\author{\small
Planck Collaboration:
P.~A.~R.~Ade\inst{84}
\and
N.~Aghanim\inst{58}
\and
C.~Armitage-Caplan\inst{89}
\and
M.~Arnaud\inst{71}
\and
M.~Ashdown\inst{68, 6}
\and
F.~Atrio-Barandela\inst{19}
\and
J.~Aumont\inst{58}
\and
C.~Baccigalupi\inst{83}
\and
A.~J.~Banday\inst{92, 10}
\and
R.~B.~Barreiro\inst{65}
% ...
}
\institute{\small
APC, AstroParticule et Cosmologie, Universit\'{e} Paris Diderot, CNRS/IN2P3, CEA/lrfu, Observatoire de Paris, Sorbonne Paris Cit\'{e}, 10, rue Alice Domon et L\'{e}onie Duquet, 75205 Paris Cedex 13, France\\
\and
Aalto University Mets\"{a}hovi Radio Observatory, Mets\"{a}hovintie 114, FIN-02540 Kylm\"{a}l\"{a}, Finland\\
\and
African Institute for Mathematical Sciences, 6-8 Melrose Road, Muizenberg, Cape Town, South Africa\\
\and
Agenzia Spaziale Italiana Science Data Center, c/o ESRIN, via Galileo Galilei, Frascati, Italy\\
\and
Agenzia Spaziale Italiana, Viale Liegi 26, Roma, Italy\\
\and
Astrophysics Group, Cavendish Laboratory, University of Cambridge, J J Thomson Avenue, Cambridge CB3 0HE, U.K.\\
\and
Astrophysics \& Cosmology Research Unit, School of Mathematics, Statistics \& Computer Science, University of KwaZulu-Natal, Westville Campus, Private Bag X54001, Durban 4000, South Africa\\
\and
Atacama Large Millimeter/submillimeter Array, ALMA Santiago Central Offices, Alonso de Cordova 3107, Vitacura, Casilla 763 0355, Santiago, Chile\\
\and
CITA, University of Toronto, 60 St. George St., Toronto, ON M5S 3H8, Canada\\
\and
CNRS, IRAP, 9 Av. colonel Roche, BP 44346, F-31028 Toulouse cedex 4, France\\
\and
California Institute of Technology, Pasadena, California, U.S.A.\\
\and
Centre for Theoretical Cosmology, DAMTP, University of Cambridge, Wilberforce Road, Cambridge CB3 0WA U.K.\\
\and
Centro de Estudios de F\'{i}sica del Cosmos de Arag\'{o}n (CEFCA), Plaza San Juan, 1, planta 2, E-44001, Teruel, Spain\\
\and
Computational Cosmology Center, Lawrence Berkeley National Laboratory, Berkeley, California, U.S.A.\\
\and
Consejo Superior de Investigaciones Cient\'{\i}ficas (CSIC), Madrid, Spain\\
\and
DSM/Irfu/SPP, CEA-Saclay, F-91191 Gif-sur-Yvette Cedex, France\\
\and
DTU Space, National Space Institute, Technical University of Denmark, Elektrovej 327, DK-2800 Kgs. Lyngby, Denmark\\
\and
D\'{e}partement de Physique Th\'{e}orique, Universit\'{e} de Gen\`{e}ve, 24, Quai E. Ansermet,1211 Gen\`{e}ve 4, Switzerland\\
\and
Departamento de F\'{\i}sica Fundamental, Facultad de Ciencias, Universidad de Salamanca, 37008 Salamanca, Spain\\
\and
Departamento de F\'{\i}sica, Universidad de Oviedo, Avda. Calvo Sotelo s/n, Oviedo, Spain\\
\and
Department of Astronomy and Astrophysics, University of Toronto, 50 Saint George Street, Toronto, Ontario, Canada\\
\and
Department of Astrophysics/IMAPP, Radboud University Nijmegen, P.O. Box 9010, 6500 GL Nijmegen, The Netherlands\\
\and
Department of Electrical Engineering and Computer Sciences, University of California, Berkeley, California, U.S.A.\\
\and
Department of Physics \& Astronomy, University of British Columbia, 6224 Agricultural Road, Vancouver, British Columbia, Canada\\
\and
Department of Physics and Astronomy, Dana and David Dornsife College of Letter, Arts and Sciences, University of Southern California, Los Angeles, CA 90089, U.S.A.\\
\and
Department of Physics and Astronomy, University College London, London WC1E 6BT, U.K.\\
\and
Department of Physics, Gustaf H\"{a}llstr\"{o}min katu 2a, University of Helsinki, Helsinki, Finland\\
\and
Department of Physics, Princeton University, Princeton, New Jersey, U.S.A.\\
\and
Department of Physics, University of California, One Shields Avenue, Davis, California, U.S.A.\\
\and
Department of Physics, University of California, Santa Barbara, California, U.S.A.\\
\and
Department of Physics, University of Illinois at Urbana-Champaign, 1110 West Green Street, Urbana, Illinois, U.S.A.\\
\and
Dipartimento di Fisica e Astronomia G. Galilei, Universit\`{a} degli Studi di Padova, via Marzolo 8, 35131 Padova, Italy\\
\and
Dipartimento di Fisica e Scienze della Terra, Universit\`{a} di Ferrara, Via Saragat 1, 44122 Ferrara, Italy\\
\and
Dipartimento di Fisica, Universit\`{a} La Sapienza, P. le A. Moro 2, Roma, Italy\\
\and
Dipartimento di Fisica, Universit\`{a} degli Studi di Milano, Via Celoria, 16, Milano, Italy\\
\and
Dipartimento di Fisica, Universit\`{a} degli Studi di Trieste, via A. Valerio 2, Trieste, Italy\\
\and
Dipartimento di Fisica, Universit\`{a} di Roma Tor Vergata, Via della Ricerca Scientifica, 1, Roma, Italy\\
\and
Discovery Center, Niels Bohr Institute, Blegdamsvej 17, Copenhagen, Denmark\\
\and
Dpto. Astrof\'{i}sica, Universidad de La Laguna (ULL), E-38206 La Laguna, Tenerife, Spain\\
\and
European Southern Observatory, ESO Vitacura, Alonso de Cordova 3107, Vitacura, Casilla 19001, Santiago, Chile\\
\and
European Space Agency, ESAC, Planck Science Office, Camino bajo del Castillo, s/n, Urbanizaci\'{o}n Villafranca del Castillo, Villanueva de la Ca\~{n}ada, Madrid, Spain\\
\and
European Space Agency, ESTEC, Keplerlaan 1, 2201 AZ Noordwijk, The Netherlands\\
\and
Finnish Centre for Astronomy with ESO (FINCA), University of Turku, V\"{a}is\"{a}l\"{a}ntie 20, FIN-21500, Piikki\"{o}, Finland\\
\and
Helsinki Institute of Physics, Gustaf H\"{a}llstr\"{o}min katu 2, University of Helsinki, Helsinki, Finland\\
\and
INAF - Osservatorio Astronomico di Padova, Vicolo dell'Osservatorio 5, Padova, Italy\\
\and
INAF - Osservatorio Astronomico di Roma, via di Frascati 33, Monte Porzio Catone, Italy\\
\and
INAF - Osservatorio Astronomico di Trieste, Via G.B. Tiepolo 11, Trieste, Italy\\
\and
INAF/IASF Bologna, Via Gobetti 101, Bologna, Italy\\
\and
INAF/IASF Milano, Via E. Bassini 15, Milano, Italy\\
\and
INFN, Sezione di Bologna, Via Irnerio 46, I-40126, Bologna, Italy\\
\and
INFN, Sezione di Roma 1, Universit\`{a} di Roma Sapienza, Piazzale Aldo Moro 2, 00185, Roma, Italy\\
\and
IPAG: Institut de Plan\'{e}tologie et d'Astrophysique de Grenoble, Universit\'{e} Joseph Fourier, Grenoble 1 / CNRS-INSU, UMR 5274, Grenoble, F-38041, France\\
\and
IUCAA, Post Bag 4, Ganeshkhind, Pune University Campus, Pune 411 007, India\\
\and
Imperial College London, Astrophysics group, Blackett Laboratory, Prince Consort Road, London, SW7 2AZ, U.K.\\
\and
Infrared Processing and Analysis Center, California Institute of Technology, Pasadena, CA 91125, U.S.A.\\
\and
Institut N\'{e}el, CNRS, Universit\'{e} Joseph Fourier Grenoble I, 25 rue des Martyrs, Grenoble, France\\
\and
Institut Universitaire de France, 103, bd Saint-Michel, 75005, Paris, France\\
\and
Institut d'Astrophysique Spatiale, CNRS (UMR8617) Universit\'{e} Paris-Sud 11, B\^{a}timent 121, Orsay, France\\
\and
Institut d'Astrophysique de Paris, CNRS (UMR7095), 98 bis Boulevard Arago, F-75014, Paris, France\\
\and
Institute for Space Sciences, Bucharest-Magurale, Romania\\
\and
Institute of Astronomy and Astrophysics, Academia Sinica, Taipei, Taiwan\\
\and
Institute of Astronomy, University of Cambridge, Madingley Road, Cambridge CB3 0HA, U.K.\\
\and
Institute of Theoretical Astrophysics, University of Oslo, Blindern, Oslo, Norway\\
\and
Instituto de Astrof\'{\i}sica de Canarias, C/V\'{\i}a L\'{a}ctea s/n, La Laguna, Tenerife, Spain\\
\and
Instituto de F\'{\i}sica de Cantabria (CSIC-Universidad de Cantabria), Avda. de los Castros s/n, Santander, Spain\\
\and
Jet Propulsion Laboratory, California Institute of Technology, 4800 Oak Grove Drive, Pasadena, California, U.S.A.\\
\and
Jodrell Bank Centre for Astrophysics, Alan Turing Building, School of Physics and Astronomy, The University of Manchester, Oxford Road, Manchester, M13 9PL, U.K.\\
\and
Kavli Institute for Cosmology Cambridge, Madingley Road, Cambridge, CB3 0HA, U.K.\\
\and
LAL, Universit\'{e} Paris-Sud, CNRS/IN2P3, Orsay, France\\
\and
LERMA, CNRS, Observatoire de Paris, 61 Avenue de l'Observatoire, Paris, France\\
\and
Laboratoire AIM, IRFU/Service d'Astrophysique - CEA/DSM - CNRS - Universit\'{e} Paris Diderot, B\^{a}t. 709, CEA-Saclay, F-91191 Gif-sur-Yvette Cedex, France\\
\and
Laboratoire Traitement et Communication de l'Information, CNRS (UMR 5141) and T\'{e}l\'{e}com ParisTech, 46 rue Barrault F-75634 Paris Cedex 13, France\\
\and
Laboratoire de Physique Subatomique et de Cosmologie, Universit\'{e} Joseph Fourier Grenoble I, CNRS/IN2P3, Institut National Polytechnique de Grenoble, 53 rue des Martyrs, 38026 Grenoble cedex, France\\
\and
Laboratoire de Physique Th\'{e}orique, Universit\'{e} Paris-Sud 11 \& CNRS, B\^{a}timent 210, 91405 Orsay, France\\
\and
Lawrence Berkeley National Laboratory, Berkeley, California, U.S.A.\\
\and
Max-Planck-Institut f\"{u}r Astrophysik, Karl-Schwarzschild-Str. 1, 85741 Garching, Germany\\
\and
McGill Physics, Ernest Rutherford Physics Building, McGill University, 3600 rue University, Montr\'{e}al, QC, H3A 2T8, Canada\\
\and
MilliLab, VTT Technical Research Centre of Finland, Tietotie 3, Espoo, Finland\\
\and
Niels Bohr Institute, Blegdamsvej 17, Copenhagen, Denmark\\
\and
Observational Cosmology, Mail Stop 367-17, California Institute of Technology, Pasadena, CA, 91125, U.S.A.\\
\and
Optical Science Laboratory, University College London, Gower Street, London, U.K.\\
\and
SB-ITP-LPPC, EPFL, CH-1015, Lausanne, Switzerland\\
\and
SISSA, Astrophysics Sector, via Bonomea 265, 34136, Trieste, Italy\\
\and
School of Physics and Astronomy, Cardiff University, Queens Buildings, The Parade, Cardiff, CF24 3AA, U.K.\\
\and
Space Research Institute (IKI), Russian Academy of Sciences, Profsoyuznaya Str, 84/32, Moscow, 117997, Russia\\
\and
Space Sciences Laboratory, University of California, Berkeley, California, U.S.A.\\
\and
Special Astrophysical Observatory, Russian Academy of Sciences, Nizhnij Arkhyz, Zelenchukskiy region, Karachai-Cherkessian Republic, 369167, Russia\\
\and
Stanford University, Dept of Physics, Varian Physics Bldg, 382 Via Pueblo Mall, Stanford, California, U.S.A.\\
\and
Sub-Department of Astrophysics, University of Oxford, Keble Road, Oxford OX1 3RH, U.K.\\
\and
Theory Division, PH-TH, CERN, CH-1211, Geneva 23, Switzerland\\
\and
UPMC Univ Paris 06, UMR7095, 98 bis Boulevard Arago, F-75014, Paris, France\\
\and
Universit\'{e} de Toulouse, UPS-OMP, IRAP, F-31028 Toulouse cedex 4, France\\
% ...
}

\authorrunning{Planck Collaboration}
\date{}

\abstract
{The multi-frequency capability of the \Planck\ satellite provides
  information both on the integrated history of star formation  (via the cosmic infrared background, or CIB)  
  and on the distribution of dark matter (via the lensing effect on the
  cosmic microwave background, or CMB).
  The conjunction of these two unique probes allows us to measure
  directly the connection between dark and 
  luminous matter in the high redshift ($1 \le z \le 3$)
  Universe. We use a three-point statistic optimized to detect the
  correlation between these two tracers.  Following a thorough
  discussion of possible contaminants and a suite of consistency
  tests, using lens reconstructions at 100, 143 and 217\,GHz and CIB
  measurements at 100--857\,GHz,
  we report the first detection of the correlation between the CIB
  and CMB lensing. The well matched redshift distribution
  of these two signals leads to a detection significance with a peak
  value of $42\,\sigma$ at 545\,GHz and a correlation as high as 80\,\% across these two tracers. Our
  full set of multi-frequency measurements (both CIB auto- and
  CIB-lensing cross-spectra) are consistent with a simple halo-based
  model, with a characteristic mass scale for the halos hosting CIB
  sources of $\log_{10}\left(\mathrm{M}/\Msolar\right) = 10.5 \pm 0.6$. Leveraging
  the frequency dependence of our signal, we isolate the high
  redshift contribution to the CIB, and constrain the star formation
  rate (SFR) density at $z\geq 1$. We measure directly the SFR
  density with around 2$\,\sigma$ significance for three redshift
  bins between $z=1$ and 7, thus opening a new window into the
  study of the formation of stars at early times.} 
{}{}{}{}
\maketitle
\section {Introduction}

This paper, one of a set associated with the 2013 release of data from
the \Planck\footnote{\Planck\ (\url{http://www.esa.int/Planck}) is a project of the European
Space Agency (ESA) with instruments provided by two scientific
consortia funded by ESA member states (in
particular the lead countries France and Italy), with contributions
from NASA (USA) and telescope reflectors provided by a collaboration
between ESA and a scientific consortium led and funded by
Denmark.}\ mission \citep{planck2013-p01}, presents a first detection
of a strong correlation between the infrared background anisotropies
and a lensing-derived projected mass map. The broad frequency coverage of the
\Planck\ satellite provides two important probes of the high redshift
Universe. In the central frequency bands of \Planck\ (70, 100, 143, and 217\,GHz),
cosmic microwave background (CMB) fluctuations dominate over most of the sky.
Gravitational lensing by large-scale structure 
produces small shear and magnification effects in the observed fluctuations,
which can be exploited to reconstruct an integrated measure of the
gravitational potential along the line of sight \cite{Okamoto:2003zw}. 
This ``CMB lensing potential'' is sourced primarily
by dark matter halos located at $1 \lesssim z \lesssim 3$,
halfway between ourselves and the last scattering surface
(see \citealt{Blandford1981,Blanchard1987}, or \citealt{Lewis:2006fu}
for a review).
In the upper frequency bands (353, 545, and 857\,GHz), the dominant 
extragalactic signal is not the CMB, but the cosmic infrared background (CIB),
composed of redshifted thermal radiation from UV-heated dust,
enshrouding young stars. The CIB contains much of the energy
from processes involved in structure formation. According to current
models, the dusty star-forming galaxies (DSFGs), which 
form the CIB have a redshift distribution peaked between $z\sim1$ and $z\sim2$,
and tend to live in $10^{11}$--$10^{13} \Msolar$ dark matter halos
\citep[see, e.g.,][and references therein]{Bethermin:2012a}.

\begin{figure}[!t]
\centering
\includegraphics[width=88mm]{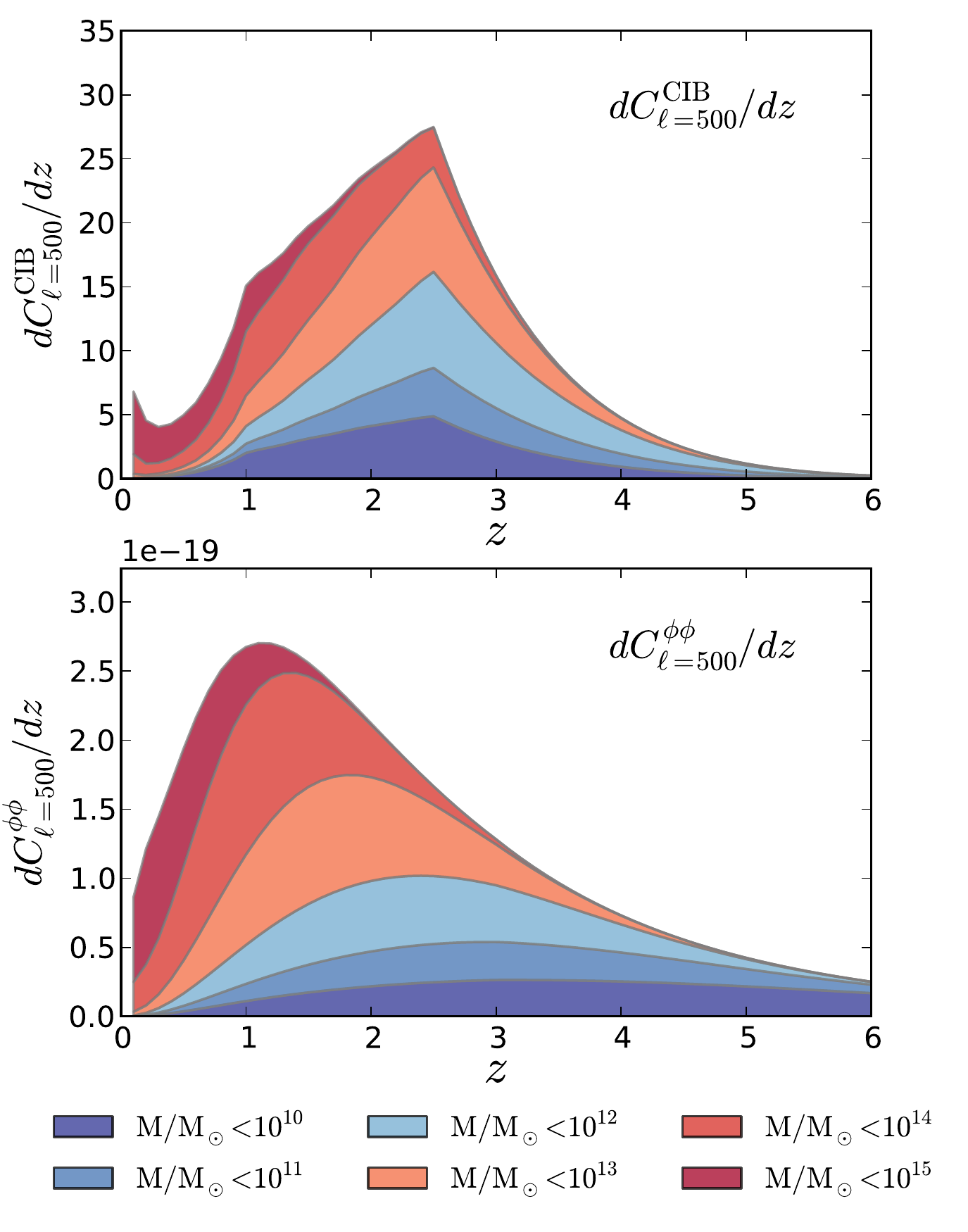}
  \caption[]{Redshift- and mass- integrand for the CIB and CMB lensing
    potential power spectra at $\ell=500$, calculated using the CIB halo
    model of \cite{planck2011-6.6}, evaluated at 217\,GHz. The good
    match between the redshift and halo mass  distributions leads to
    an expected correlation up to 80\,\%. The sharper features in the
    CIB kernel are artefacts from the \cite{Bethermin:2012c} model. We
    note that the low mass, high $z$ behavior of our calculation is
    limited by the accuracy of the mass function we use
    \citep{Tinker:2009mx}. All of our mass integrals use $M_{min} =$
    10$^{5}$~\Msolar.}  
   \label{fig:clcc_clpp_redshifts.pdf}  
\end{figure}

As first pointed out by \cite{Song:2002sg},
the halo mass and redshift dependence 
of the CMB lensing potential and the CIB fluctuations are well matched, and as
such a significant correlation between the two is expected. This point
is illustrated quantitatively in Fig.~\ref{fig:clcc_clpp_redshifts.pdf},
where we plot estimates for the
redshift- and mass- kernels of the two tracers. In this paper we
report on the first detection of this correlation.

Measurements of both CMB lensing and CIB fluctuations are currently
undergoing a period of rapid development.
While the CIB mean was first detected using the FIRAS and DIRBE instruments
aboard {\it COBE\/} \citep{puget1996,fixsen1998,hauser1998},
CIB fluctuations were later detected by the {\it Spitzer Space Telescope\/}
\citep{lagache2007} and by the BLAST balloon experiment
\citep{viero2009} and the {\it Herschel Space Observatory}
\citep{Amblard2011,Viero:2012he}, as well as the new generation
of CMB experiments, including \Planck, which have extended these
measurements to longer wavelengths~\citep{Hall:2009rv,Dunkley:2010ge,planck2011-6.6,Reichardt:2011yv}.
The \Planck\ early results paper: \cite{planck2011-6.6} (henceforth referred
to as PER) presented measurements of the angular power spectra
of CIB anisotropies from arc-minute to degree scales at 217, 353,
545, and 857\,\GHz, establishing \Planck~as a potent probe of the 
clustering of the CIB, both in the linear and non-linear regimes.
A substantial extension of PER is presented in a companion paper
to this work \citep[][henceforth referred to as PIR]{planck2012-PIP}.

The CMB lensing potential, on the other hand, which was first detected
statistically through cross-correlation with galaxy surveys 
(\citealt{Smith:2007rg,Hirata:2008cb}, and more recently \citealt{Bleem:2012gm,Sherwin:2012mr}),
has now been observed directly in CMB maps by the Atacama Cosmology
Telescope and the South Pole Telescope \citep{Das:2011ak,vanEngelen:2012va}.

\Planck's frequency coverage, sensitivity and survey area, 
allow high signal-to-noise measurements of both the CIB and the CMB
lensing potential. 
Accompanying the release of this paper, \cite{planck2013-p12} reports the first
measurement and characterisation of the CMB lensing potential with the \Planck\ data, which
has several times more statistical power than previous measurements, over a large fraction (approximately $70\%$ of the sky).
We will use this measurement of the lensing potential in cross-correlation with measurements of the CIB in the \Planck HFI bands
to make the first detection of the lensing-infrared background correlation.
In addition to our measurement, we discuss the implications for models
of the CIB fluctuations. The outline of this paper is as follows.  In 
Sect.~\ref{sec:data} we describe the data we will use, followed by
a description of our pipeline for correlating the CIB and lensing signals
in Sect.~\ref{sec:cross_corr}. Our main result is
presented in Sect.~\ref{sec:results}, with a description of our error budget,
consistency tests and an array of systematic tests in
Sect.~\ref{sec:systematics}.
We discuss the implications of the measured correlation for CIB modelling in
Sect.~\ref{sec:interpretation}. 

%

\section{Data sets}
\label{sec:data}

\subsection{\Planck~maps}
\label{sec:planckmaps}

\Planck\ \citep{tauber2010a, planck2011-1.1} is the third generation
space mission to measure the anisotropy of the CMB.
It observes the sky with high sensitivity in nine frequency bands
covering 30--857\,GHz at an angular resolution
from 31\arcm\ to 5\arcm. The Low Frequency Instrument (LFI;
\citealt{Mandolesi2010, Bersanelli2010, planck2011-1.4}) covers the
30, 44, and 70\,GHz bands with radiometers that incorporate amplifiers cooled to 20\,\hbox{K}.  The
High Frequency Instrument (HFI; \citealt{Lamarre2010, planck2011-1.5})
covers the 100, 143, 217, 353, 545, and 857\,GHz bands with bolometers
cooled to 0.1\,\hbox{K}.  Polarization is measured in all but the
highest two bands \citep{Leahy2010, Rosset2010}.  A combination of
radiative cooling and three mechanical coolers produces the
temperatures needed for the detectors and optics
\citep{planck2011-1.3}.  Two data processing centres (DPCs) check and
calibrate the data and make maps of the sky \citep{planck2011-1.7,planck2011-1.6}.  \Planck's sensitivity, angular resolution, and
frequency coverage make it a powerful instrument for Galactic and
extragalactic astrophysics as well as cosmology.  Early astrophysics
results are given in Planck Collaboration VIII--XXVI 2011, based on
data taken between 13~August 2009 and 7~June 2010.  Intermediate
astrophysics results are now being presented in a series of papers
based on data taken between 13~August 2009 and 27~November 2010. This
paper uses data corresponding to the second \Planck~data  release,
with data acquired in the period up to 27~November 2010 and undergoing improved processing.   

We use the \Planck\ HFI temperature maps at all six frequencies, i.e.,
100, 143, 217, 353, 545, and 857\,GHz. The maps at each frequency
were created using almost three full-sky surveys.  Here we give
an overview of the HFI map-making
process with additional details given in \citet{planck2011-1.7,planck2013-p03}. The data are
organized as time-ordered information, hereafter TOI. The attitude of
the satellite as a function of time is provided by two star trackers
on the spacecraft. The pointing for each bolometer is computed by
combining the attitude with the location of the bolometer in the focal
plane, as determined by planet observations. The raw bolometer TOI for
each channel is first processed to produce cleaned timelines and to set flags that mark bad data
(for example data immediately following a cosmic ray strike on the detector).
This TOI processing includes: (1) signal demodulation and filtering;
(2) deglitching, which flags the strong part of any glitch and subtracts
the tails;
(3) conversion from instrumental units (volts) to physical units (watts of
absorbed power, after a correction for the weak non-linearity of the response);
(4) de-correlation of thermal stage fluctuations;
(5) removal of the systematic effects induced by 4\,K cooler mechanical
vibrations; and
(6) deconvolution of the bolometer time response.
Focal plane reconstruction and beam shape estimation is made using observations
of Mars. The simplest description of the beams, an elliptical Gaussian,
leads to full-width at half-maximum (FWHM) values of 9.65, 7.25, 4.99,
4.82, 4.68 and 4.33 \arcmin as given in Table 
  4 of \citet{planck2013-p03}. As explained in this paper, the
inter-calibration accuracy between channels is better than the absolute calibration. This leads us to
adopt conservative absolute calibration uncertainties of 0.64,
  0.53, 0.69, 2.53, 10., 10.\,\% at 100, 143 217, 353, 545 and 857
  \,GHz respectively. We convert between emissivities given in \MJysr
  (with the photometric convention $\nu I_\nu={\rm constant}$) and
  temperatures in \muK, using the measured bandpass filters (see PER
  and PIR for details).   

For the sake of consistency testing (presented in particular in
Sect.~\ref{sec:systematics}), we will sometimes use temperature maps where
only a fraction of the TOI is used to generate the sky map. In
particular, throughout this paper we use the terminology ``half-ring''
(HR) maps to refer to maps made using the first and second half of the stable
pointing period, ``survey'' for individual full-sky survey maps (note
that the third survey is incomplete for the particular data release used
in the intermediate papers), and ``detset'' for maps made using
two independent sets of detectors per frequency
\citep[for details see][]{planck2011-1.7}.

We create three masks to exclude regions with bright Galactic and
extragalactic foreground emission.
The first mask accounts for diffuse Galactic emission as observed in
the \Planck\ data.  To allow us to test for the effects of residual
Galactic emission on our results
we create three different versions of this mask,
each with a different masked area, such that
20, 40 or 60\,\% of the sky is unmasked.
Each version of this mask is created directly from the \Planck\ 353\,GHz map,
from which we remove the CMB using the 143\,GHz channel as a CMB template
before smoothing by a Gaussian with FWHM of $5^\circ$.
The map is then thresholded such that the mask has the
required sky fraction.  Although the Galactic emission is stronger at
857\,GHz, we expect the 353\,GHz mask to better trace dust emission at
the lower frequencies we use. The mask therefore accounts for Galactic
dust and Galactic CO emission as explained in \citet{planck2013-p06}. We will
not worry about synchrotron emission, which is important at low
frequencies, since its contribution at 100\,GHz and at high Galactic
latitudes is small, and, as with the dust component, will be
uncorrelated with the lensing potential. The second mask covers bright
point sources. This mask is created using algorithms tailored to detect
point sources in the \Planck\ data and is optimized for each
frequency, as detailed in \cite{planck2011-1.10} and \cite{planck2011-1.10sup}. 
The third mask is designed to remove extended high-latitude Galactic
dust emission (``cirrus''), as traced by external \hi\ data, as we
will describe in Sect.~\ref{sec:hidata}. While the first two masks are
described in \cite{planck2013-p06}, the latter is specific to our
cross-correlation analysis, as it provides a method to  reduce the
large-scale noise in our measurement, and the 3-point nature of our 
estimate ensures that it will not introduce a bias (although we test for
this in Sect.~\ref{sec:systematics}). Ultimately, when we combine the
three masks we obtain an effective sky fraction of 16, 30 and 43\,\%
for the 20, 40 and 60\,\% Galactic masks, respectively.

\begin{figure}[!t]
  \centering
  \includegraphics[width=88mm,clip=true,trim=0.4cm 1.7cm 0.2cm 1cm]{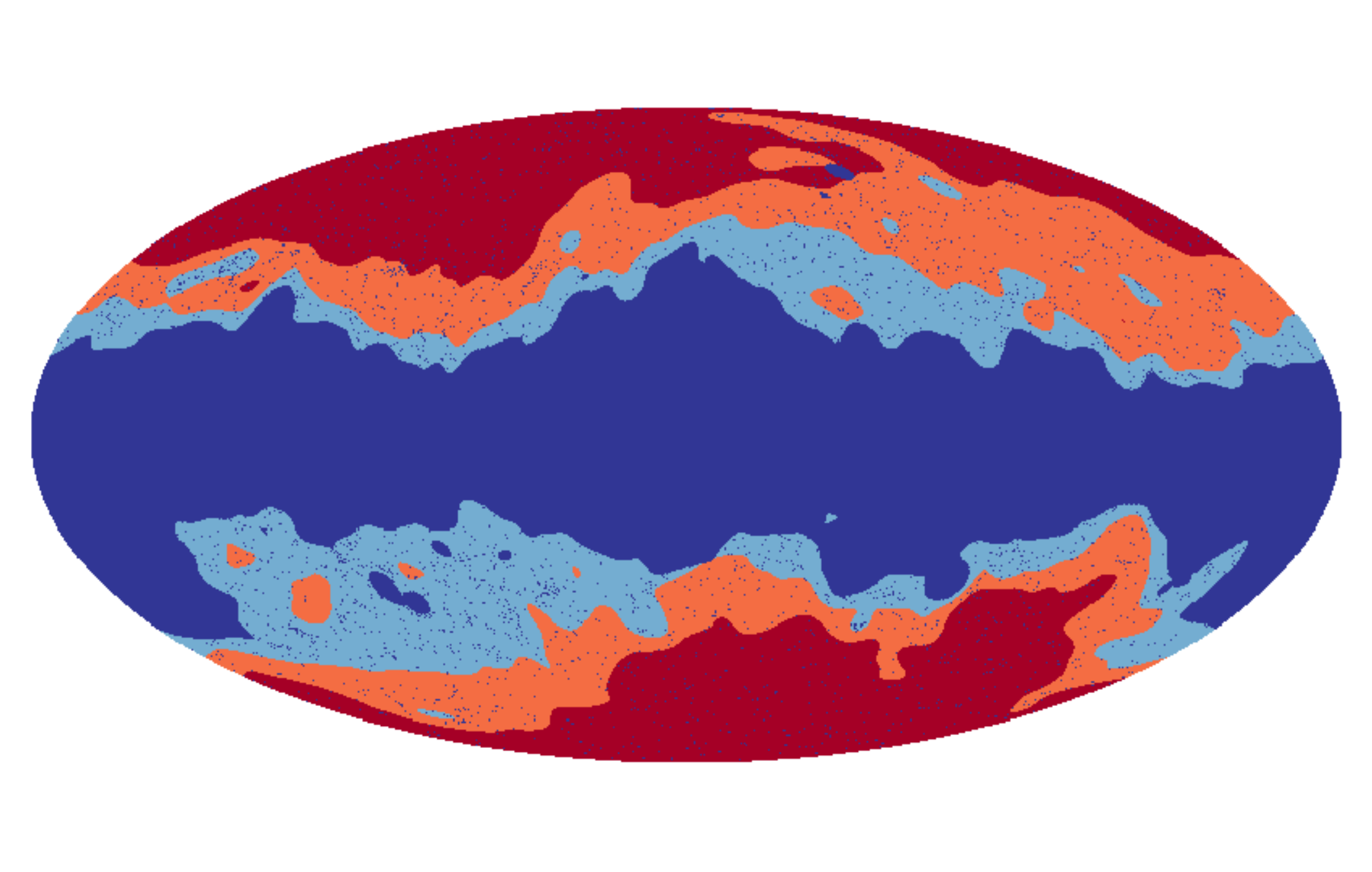}
  \caption[]{Combined Galactic, point-source and \hi\ mask with sky fractions 16, 30 and 43\,\%.}
   \label{fig:galactic_masks}  
\end{figure}

\subsection{External data sets}

\subsubsection{\hi\ maps}
\label{sec:hidata}

We use measurements of 21-cm emission from Galactic neutral hydrogen (\hi)
as a cirrus monitor. Outside of our Galactic and point source masks we use the
\hi\ data to construct a template of the dust emission in regions where the
\hi\ column density is low (less than $N_{\mathrm{HI}}\le 2\times
10^{20}~\mathrm{cm}^{-2}$), and we mask regions where it is high, since in
these regions the \hi\ and dust emission are not well correlated
\citep[][PER]{boulanger1996,Boulanger1988}.  The masking procedure
that we use is described in detail in \cite{planck2011-7.12}.
It consists of subtracting the \hi\ dust template from the \Planck\
temperature map at 857\,GHz and calculating the skewness of the residuals
in $5\,{\rm deg}^2$ regions. If the skewness is larger than a given value then
the region is masked. This is an improvement over the usual
cut-off in \hi\ column density.
We use the latest release from the Leiden/Argentina/Bonn (LAB) survey \citep{Kalberla:2005ts}, which consists of the
Leiden/Dwingeloo Survey (LDS) \citep{hartmannburton1997} north of $-30^\circ$
declination, combined with the Instituto Argentino de Radioastronomia Survey
\citep{arnal2000,bajaja2005} south of $-25^\circ$ declination.
The angular resolution of the combined map is approximately $0.6^\circ$ FWHM.
The LAB Survey is the most sensitive Milky Way \hi\ survey to date, with
the greatest coverage both spatially and kinematically.
We make use of projections of the LAB maps onto $N_{\rm side}=512$
{\tt HEALPix} maps performed by \citet{landslosar2007} and made available
through the LAMBDA
website\footnote{\url{http://lambda.gsfc.nasa.gov/product/foreground/}}.
The local standard of rest velocity coverage spans the interval
$-450\,{\rm km}\,{\rm s}^{-1}$ to $+400\,{\rm km}\,{\rm s}^{-1}$, at a
resolution of $1.3\,{\rm km}\,{\rm s}^{-1}$, with an rms
brightness-temperature noise of 0.07--0.09\,K, and with additional
errors due to defects in the correction for stray radiation that are less than
20--40\,mK for most of the data.

\subsubsection{IRIS/\textit{IRAS\/} maps}
\label{sec:iris}

As a consistency test we will use an additional tracer of the CIB that
derives from re-processed {\it IRAS\/}
maps at 60 and 100$\,\mu$m. This new generation of {\it IRAS\/} maps,
known as IRIS, benefits from improved zodiacal light subtraction, a
calibration and zero level compatible with DIRBE,
and an improved de-striping procedure
\citep{mamd2005}. {\it IRAS\/} made two full-sky maps (HCON-1 and HCON-2),
as well as a final map that covers 75\,\% of the sky (HCON-3).
The three maps had identical processing that included deglitching, checking
of the zero-level stability, visual examination for glitches
and artifacts, and zodiacal light removal. The three HCONs were then
co-added, taking into account the inhomogeneous sky coverage maps, to
generate the average map (HCON-0). Note that the \citet{finkbeiner1999}
maps are also constructed from the {\it IRAS\/} 100\,$\mu$m data,
and as such we will not investigate their 
cross-correlation properties since the IRIS map contains the same
information. For simplicity we will assume that the effective IRIS beam is uniform across the sky
and described by a Gaussian with FHWM of 4.3\arcm.

\section{Cross-correlation formalism and implementation}
\label{sec:cross_corr}

We now describe our statistical formalism and its
implementation, with additional technical details given in the
appendices. Our analysis consists of cross-correlating a full-sky
reconstruction of the CMB lensing potential with a temperature map.

\subsection{Reconstructing the CMB lensing potential}
\label{sec:lens_recon}

The CMB is lensed by the gravitational potential of all matter along the
photon trajectory from the last scattering surface to us. The lensed CMB
is a remapping of the unlensed CMB with the lensed temperature equal to
$\tilde{\Theta}(\n) = \Theta(\n + \nabla \phi)$,
where $\Theta(\n)$ is the unlensed CMB
temperature and $\phi$ is the lensing potential. 
We use the methodology described in \citet{planck2013-p12} to obtain
estimates $\hat{\phi}_{LM}$ of the lensing potential in harmonic space,
using the standard \citet{Okamoto:2003zw} quadratic estimator.

Complete details on the lens reconstruction procedure, which we use are
given in \citet{planck2013-p12}, although we review it briefly in point form here.
Our estimates of $\hat{\phi}$ are obtained by the following set of steps:
\begin{enumerate}
\item Inverse variance filter the CMB map.
\item Use the filtered CMB map as the input to a \textit{quadratic lensing estimator}, which is designed to extract the off-diagonal contributions to the CMB covariance matrix induced by lensing.
\item Subtract a ``mean-field bias'', which corrects for known non-lensing contributions to the covariance matrix, including instrumental noise inhomogeneity, beam asymmetry, and the Galaxy+point source mask.
\end{enumerate}
The output from this pipeline is an estimate of the lensing
potential in harmonic space $\hat{\phi}_{LM}$ and an associated noise
power spectrum $N_L^{\phi\phi}$, which we use to weight our cross-correlation estimates.
We also produce a set of simulated lens reconstruction, which we use to establish our statistical error bars.

Our nominal lens reconstructions use the 143\,GHz channel, however there
is almost equivalent power to measure lensing using the 217\,GHz channel.
Combining both channels would reduce the noise power spectrum of our lens
reconstruction by approximately $25\,\%$, compared with using either
individually (the improvement is significantly less than $50\,\%$
because a significant portion of the lens reconstruction noise is due
to the finite number of CMB modes, which we are able to observe, and is
correlated between the two channels).  We choose to focus on 143\,GHz
here because it is significantly less
susceptible to CIB contamination.  We will use lens reconstructions based
on the 100 and 217\,GHz data for consistency tests.

\subsection{Decreasing the foreground noise}
\label{sec:fg_cleaning}

An important source of noise (but, as we will explain below, not bias) in our
cross-correlation measurement is Galactic foreground emission.
Dust emission is the dominant Galactic component at
HFI frequencies above 217\,GHz
(see Sect.~\ref{sec:stat_error} for a quantitative discussion).
In order to reduce the Galactic dust emission we create a dust template
and subtract it from the temperature maps described in
Sect.~\ref{sec:planckmaps}. At 100 and 143\,GHz the CMB signal is significantly brighter than
the dust emission outside the Galactic mask. We therefore
do not create and subtract a dust template at these frequencies. Note
that while we could use other frequency maps to trace the CMB and
remove it, to quantify the non-negligible amount of CIB that would be removed 
this way is not easy given the uncertainties in the cross-frequency
CIB correlation structure.

We rely on the well documented (but complex) correlation between
Galactic \hi\ and dust \citep[e.g.,][PER]{Boulanger1988,boulanger1996,lagache1998} to reduce the
contamination by subtracting the \hi-correlated dust component.
As was performed in PER, we split the \hi\ map into two velocity
components: a low-velocity local component (LC) typical of  high-latitude \hi\
emission, and a component of intermediate-velocity
clouds (IVC). We found that the inclusion of a high-velocity component
makes a negligible difference to the dust-cleaned map.
Unlike the dedicated high-resolution \hi\ observations used in PER and PIR
that only partially cover the sky, here we use the full-sky,
low resolution LAB survey introduced in Sect.~\ref{sec:hidata} as our
\hi\ tracer.  Although the resolution of this survey is lower than the
\Planck\ resolution, it allows us to perform dust cleaning on large scales,
where our cross-correlation measurement has high signal-to-noise ratio.
The emissivity of the dust varies across the sky, and so the correlation
between the dust and \hi~emission is expected to vary. To account for this
we divide the sky into regions where we assume that the dust-\hi~correlation
is constant. For the sake of convenience, we use regions of
approximate size 13 (52)\,${\rm deg}^2$ defined by the {\tt Healpix} pixels at
resolution $N_{\rm side}=16$ (8) that are outside the Galactic mask. We test
that our conclusions do not depend on this resolution. 

The details of our procedure is as follows.
We subtract the 143\,GHz \Planck\ temperature map from each of the
217--857\,GHz temperature maps to remove the CMB signal (this CMB subtraction
is only done for the purposes of creating the dust template).  We upgrade each
of the $N_{\rm side}=512$ LAB maps compiled in \citet{landslosar2007}
to the \Planck\ map resolution of $N_{\rm side}=2048$.
Within each region we then simultaneously fit for the amplitude of each
\hi\ velocity component in the CMB-subtracted maps, and use the
two coefficients per region to assemble a full-sky (minus the mask)
dust template for each of the 217--857\,GHz channels. We smooth each
template with a Gaussian of FWHM $10^\prime$ to remove the discontinuity at the
patch boundaries, and then subtract the template from the original
(CMB-unsubtracted) \Planck\ maps.

\begin{figure*}[!t]
  \centering
  \includegraphics[width=180mm]{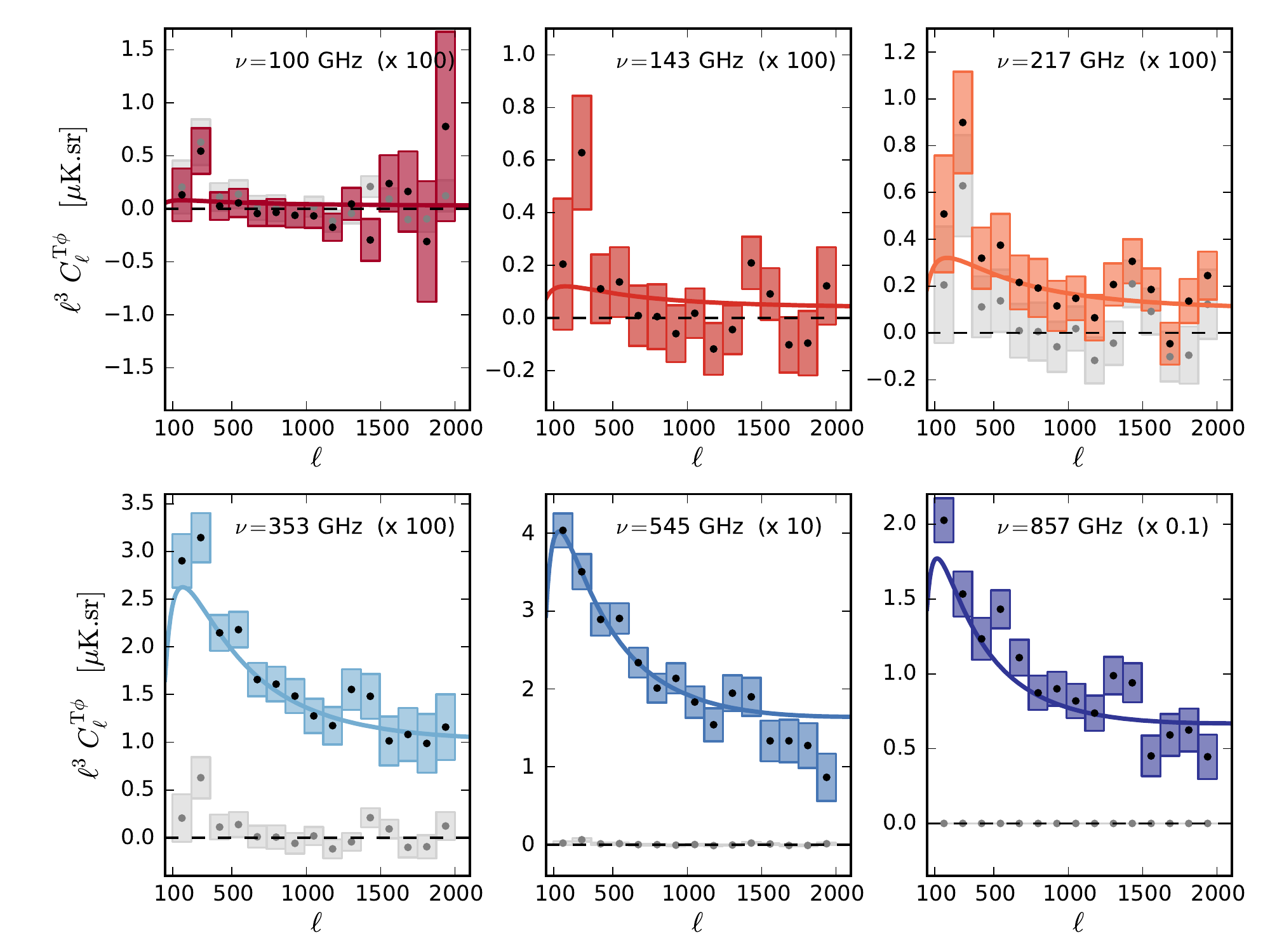}
  \caption[]{Angular cross-spectra between the reconstructed lensing
    map and the temperature map at the six HFI frequencies. The error bars
    correspond to the scatter within each band. The solid line is the
    expected result based on the PER model and is \emph{not\/} a fit to these
    data (see Fig.~\ref{fig:autos_fit} for an adjusted model),
    although it is already a satisfying model. In each panel we also
    show the correlation between the lens reconstruction at 143\,GHz and the
    143\,GHz temperature map in grey. This is a simple
    illustration of the frequency scaling of our measured signal
    and also the strength of
    our signal as compared to possible intra-frequency systematic errors.}
   \label{fig:nominal_results}  
\end{figure*}

We note that the accuracy of this procedure would be difficult to
evaluate for all possible uses of the map, i.e., whether it might constitute a robust
component separation method remains to be demonstrated. However, in
the case of our cross-correlation analysis the dust-removal requirements
are less severe, since the dust emission only contributes to our measurement
as noise. We will describe later in Sect.~\ref{sec:inst_and_syste} the effect
on the cross-spectrum of removing this emission, and will place limits on
the residual Galactic contamination in Sect.~\ref{sec:cib_bispectrum}.

\subsection {Measuring cross-correlations}

To estimate the cross-correlation between the lensing potential and a
tracer $t$, we
calculate
\be
\hat{C}^{t\phi}_\ell = \frac{1}{2\ell+1} \sum_{m} \hat{t}_{\ell m}
\hat{\phi}_{\ell m}^*.
\ee
As the CIB has an approximately $\ell^{-1}$ dependence
and the lensing potential has an $\ell^{-2}$ dependence,
we multiply the cross-spectra by $\ell^3$,
and then bin it in 15 linearly spaced bins between $100 < \ell < 2000$.
As we will discuss in Sect.~\ref{sec:systematics}, modes with $\ell < 100$
are not considered, due to possible lens mean-field systematic effects, and modes
with $\ell > 2000$ are removed due to possible extragalactic
foreground contamination. We have tested that our results are robust
to an increase or decrease in the number of $\ell$-bins.

We expect the error bars to be correlated across bins to some extent,
due to pseudo-$C_\ell$ mixing induced by the mask, and between frequencies,
because the lens reconstruction noise is common. 
In addition, any foregrounds that are present in multiple
channels will introduce correlated noise. The foreground mask will also induce
a coupling between different modes of the unmasked map. This extra coupling
can be calculated explicitly using the mixing matrix formalism introduced in
\citet{Hivon:2001jp}.  Using this formalism and our best-fit models we
have evaluated the correlation between different bins of the
cross-correlation signal for our nominal binning scheme. We find that the
mask-induced correlation is less than 2\,\% across all bins at all frequencies.
We will thus neglect it in our analysis. For this reason, and based on the
results we obtain from simulations, we do not
attempt to ``deconvolve'' the mask from the
cross-spectrum~\citep[see e.g.,][]{Hivon:2001jp}
and instead correct for the power lost through masking the maps by a
single sky fraction, $f_{\rm sky}$, ignoring the mode coupling.

As will be discussed later in Sect.~\ref{sec:modeling},
when we fit models to the cross-spectrum we will assume that the noise
correlation between bins can be neglected and that the band-powers are flat.

\subsection{Simulating cross-correlations}
\label{sec:simulations}

In order to validate our measurement pipeline and to confirm that our
estimate of the cross-spectrum is unbiased we create simulated maps
of the lensed CMB and CIB that have the expected statistical
properties. 

Using the \Planck~only favored $\Lambda$CDM cosmology as described in
\citet{planck2013-p11} we generate a theoretical prediction of the
lensing potential spectrum using {\tt CAMB} \citep{Lewis:1999bs}, from which we
generate 300 maps of $\phi$ that are used to lens 300 CMB realizations
using the approach described in~\cite{planck2013-p12}. We then use the PER best-fit CIB model to generate CIB auto- and CIB-$\phi$
cross-spectra, from which we create CIB realizations that are correctly
correlated with $\phi$ in each HFI band.
The PER model that we use describes the CIB clustering at HFI frequencies
using a halo approach, and simultaneously reproduces known number count
and luminosity function measurements.  At each frequency we add a
lensed CMB realization to each of the CIB realizations
and then smooth the maps using a symmetric beam
with the same FWHM as the beam described in Sect.~\ref{sec:planckmaps}.
Once this set of realizations has been generated we apply the
reconstruction procedure described above to produce an estimate of the lensing
potential map, and then calculate the cross-power spectrum using our
measurement pipeline.

These simulations will miss some complexities inherent in the
\Planck\ mission. They do not take into account inhomogeneous and
correlated noise, and we do not simulate asymmetric beam effects.
In addition, we do not simulate any foreground components, and we instead take
a different approach to determine their contribution.
While simplistic, we believe that our simulations are good enough for
the purposes of this particular measurement.  In Sect.~\ref{sec:systematics}
we will discuss possible limitations, as well as how we test for systematic
effects that are not included in the simulations.

We use the simulated maps to check that our pipeline correctly recovers
the cross-spectrum that was used to generate the simulations.
For the $\ell$-bins used in our analysis, we find that the recovered spectrum is
unbiased (to within the precision achievable with 300 simulations),
and with a noise level consistent with expectations. The noise in the
recovered spectrum is discussed in Sect.~\ref{sec:stat_error}.

\section {A strong signal using \Planck\ HFI data}
\label{sec:results}

\begin{figure*}[!t]
 \centering
 \includegraphics[width=180mm]{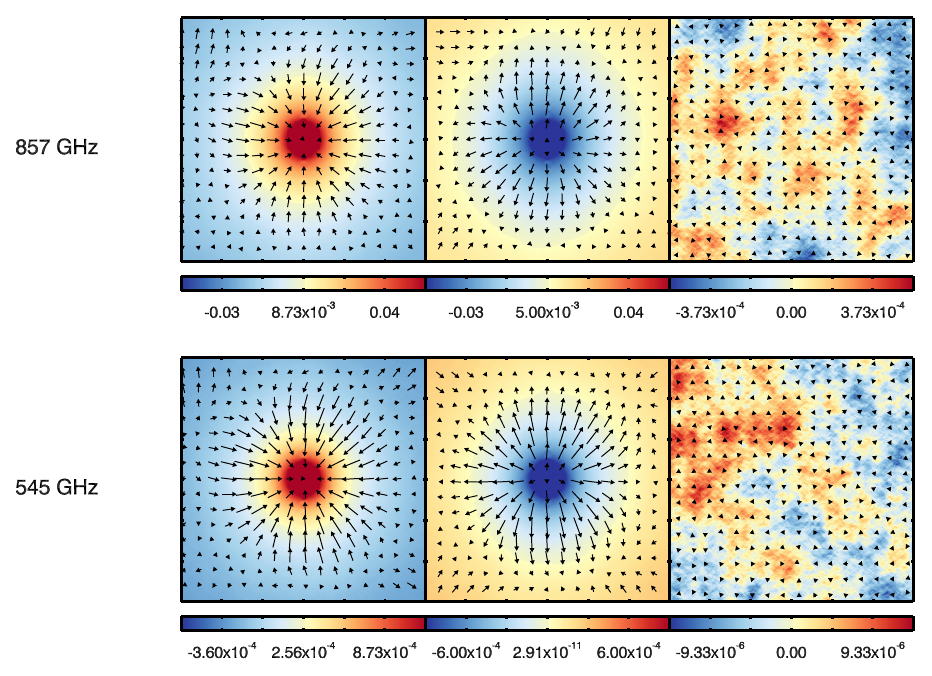}
 \caption[]{Temperature maps of size $1\,{\rm deg}^2$ at 545 and 857\,GHz
    stacked on the 20,000 brightest peaks (left column), troughs
    (centre column) and random map locations (right column). The
    stacked (averaged) temperature maps is in K. The arrows
    indicate the lensing deflection angle deduced from the
    gradient of the band-pass filtered lensing potential map stacked on the
    same peaks. The longest arrow corresponds to a deflection of
    6.3\arcs, which is only a fraction of the total deflection
    angle because of our filtering. This stacking allows us to visualize in
    real space the lensing of the CMB by the galaxies that generate the CIB.
    A small and expected offset ($\simeq$1\arcmin) was corrected by
    hand when displaying the deflection field.} 
  \label{fig:stacking}
\end{figure*}

We now describe the result of applying our pipeline to
our nominal data set, i.e., the lens reconstruction at 143\,GHz and
the foreground reduced \Planck\ HFI temperature maps with a 40\,\% Galactic
mask, which when combined with the point source mask and \hi\ mask leaves
30.4\,\% of the sky unmasked. The results are presented in Fig.~\ref{fig:nominal_results}.
The error bars correspond to the naive scatter measured within each
bin. The thin black line corresponds to the expected CIB-lensing
correlation \emph{predicted\/} using the PER CIB model (the HOD
parameters of the PER 217\,GHz best-fit model were used at 100 and
143\,GHz since no CIB clustering measurement at these frequencies is
available). As can be seen from these plots, the noise is strongly
correlated across frequencies, especially at the lowest frequencies
where the CMB dominates the error budget. A detailed analysis of the
uncertainties and potential systematic errors attached to this result
is presented in Sect.~\ref{sec:systematics}.  

As clearly visible in Fig.~\ref{fig:nominal_results}, a strong
signal is detected. To set a reference point and naively quantify its
statistical significance when taken at face value, we define a
detection significance as follows. We count the number of standard deviations
as the quadrature sum of the significance in the different multipole bins:   
\be
\label{eqn:detsig}
s_{\nu} = \sqrt{\sum_{i=1}^{15}{
 \left(\frac{C_i^{\rm{T}\phi}}{\Delta C_i^{\rm{T}\phi}}\right)^2}}.
\ee
For our nominal parameters this gives us
$3.6\,\sigma$, $4.3\,\sigma$, $8.3\,\sigma$, $31\,\sigma$, $42\,\sigma$,
and $32\,\sigma$, at, respectively, 100, 143, 217, 343, 545 and 857\,GHz.
Note that these numbers include an additional 20\,\% contribution to the
statistical error to account for mode correlations
(which we discuss in Sect.~\ref{sec:stat_error}), but do not include
systematic errors or our point source correction. As a comparison, in each
panel we plot the correlation between the lens
reconstruction at 143\,GHz and the 143\,GHz map in grey. This shows the
frequency scaling of our measured signal and also the strength
of the signal, as compared to possible intra-frequency systematic
effects. This will be studied in depth in Sect.~\ref{sec:systematics}.

This first pass on our raw data demonstrates a strong detection that
is in good agreement with the expected CIB-lensing signal. To get a
better intuition for this detection, we show in
Fig.~\ref{fig:stacking} the real-space correlation between the
observed temperature and the lens deflection angles.
This figure allows us to visualize the correlation between the CIB and the CMB
lensing deflection angles for the first time.
These images were generated using the following stacking technique.  We first
mask the 545 and 857\,GHz temperature maps with our combined mask that includes
the 20\,\% Galaxy mask, and identify 20,000 local maxima and minima in these
maps.  We also select 20,000 random locations outside the masked region to
use in a null test.
We then band pass filter the lens map between $\ell = 400$--600 to remove
scales larger than our stacked map as well as small-scale noise.
We stack a $1\,{\rm deg}^2$ region around each point in both the
filtered temperature map and lensing potential map, to generate stacked
CIB and stacked lensing potential images.
We take the gradient of the stacked lensing potential to calculate the
deflection angles, which we display in Fig.~\ref{fig:stacking} as arrows.
The result of the stacking over the maxima, minima and random points is
displayed from
left to right in Fig.~\ref{fig:stacking}. The strong correlation seen
already in the cross-power spectrum is clearly visible in both
the 545 and 857\,GHz extrema, while the stacking on random
locations leads to a lensing signal consistent with noise. From
simulations, we expect a small off-set ($\simeq 1\arcmin'$) in the
deflection field. This offset was corrected for in this plot. We have verified in simulations that this
is due to noise in the stacked lensing potential map that shifts the peak.
As expected, we see that the temperature maxima of the CIB, which contain
a larger than average number of galaxies, deflect light inward, i.e.,
they correspond to gravitational potential
wells, while temperature minima trace regions with fewer galaxies and
deflect light outward, i.e., they correspond to gravitational potential hills.

\section{Statistical and systematic error budget}
\label{sec:systematics}

The first pass of our pipeline suggests a strong correlation of the
CIB with the CMB lensing potential.
We now turn to investigate the strength and the origin of this signal.
We will first discuss the different contributions to the statistical error
budget in Sect.~\ref{sec:stat_error}, and then possible systematic effects in 
Sect.~\ref{sec:inst_and_syste}. Although the most straightforward
interpretation of the signal is that it arises from dusty star-forming
galaxies tracing the large-scale mass distribution, in
Sect.~\ref{sec:astro_contam} we consider other potential astrophysical
origins for the observed  correlation.

\subsection{Statistical error budget}
\label{sec:stat_error}

In this section we discuss any noise contribution that does not lead to a
bias in our measurement. The prescription adopted throughout this paper is
to obtain the error estimates from the naive Gaussian analytical error bars
calculated using the measured auto-spectra of the CIB and lensing potential.
We find that these errors are approximately equal to 1.2 times the naive
scatter within an $\ell$-bin, and we will sometimes use this prescription where
appropriate for convenience (as will be stated in the text).
This is justified in Appendix~\ref{app:stat_errors} where we consider six
different methods of quantifying the statistical errors using both simulations
and data. The Gaussian analytical errors, $\Delta \hat{C}_\ell^{{\rm T}\phi}$,
are calculated using the naive prescription
\be
\label{eqn:anal_error}
\begin{split}
f_{\rm sky} \, (2\ell+1) \, \Delta\ell \,
 \left(\Delta \hat{C}_\ell^{{\rm T}\phi}\right)^2 &= \hat{C}_\ell^{{\rm TT}}
 \hat{C}_\ell^{\phi \phi} + \left(C_\ell^{{\rm T} \phi} \right)^2,
\end{split}
\ee
where as before $f_{\rm sky}$ is the fraction of the sky that is unmasked,
$\Delta\ell$ = 126 for our 15 linear bins between $\ell=100$ and
$\ell=2000$, $\hat{C}_\ell^{{\rm TT}}$ and $\hat{C}_\ell^{\phi \phi}$ are the spectra
measured using the data, and $C_\ell^{{\rm T}\phi}$ is the model cross
spectrum.  This last term provides a negligible contribution
due to the large noise bias on $\hat{C}_\ell^{\phi \phi}$, as we now
describe.

\begin{figure}[!t]
  \centering
  \includegraphics[width=88mm]{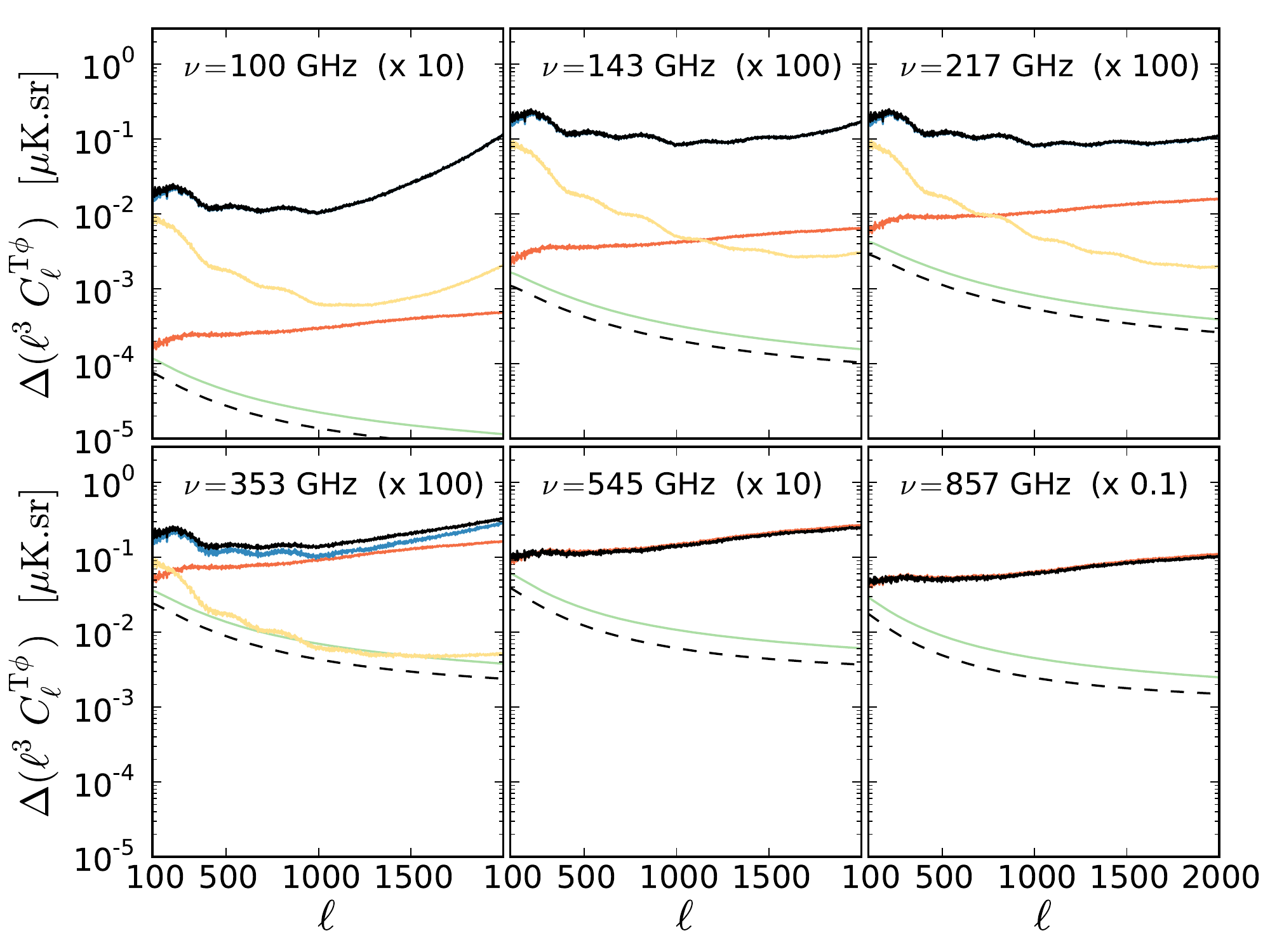}
  \caption[]{Naive analytical estimates of the contribution to the
    $C_\ell^{\mathrm{T}\phi}$ variance as a function of multipole and
    frequency as given in Eq.~\ref{eqn:error_breakdown}. We assume the
    same bin sizes as in Fig.~\ref{fig:nominal_results}. The different
    lines are the contribution to the analytical error from the signal
    only: $C_\ell^{\phi\phi} C_\ell^{\rm CIB} + \left(C_\ell^{{\rm CIB} \, \phi}\right)^2$ (green),
    noise only: $\hat{C}_\ell^{\phi\phi,{\rm N}}\hat{C}_\ell^{{\rm CIB,N}}$
    (blue), and the mixed signal and noise terms:
    $C_\ell^{\phi\phi}\hat{C}_\ell^{\rm CIB, N}$ (yellow) and
    $\hat{C}_\ell^{\phi\phi,{\rm N}} C_\ell^{\rm CIB}$ (orange).
    The total contribution is the solid black line, and the theory spectrum,
    $\left(C_\ell^{CIB \, \phi}\right)^2$, is the dashed line.}
  \label{fig:error_by_l}
\end{figure}

The statistical error has two sources: instrumental and
astrophysical. The measured auto-spectra in Eq.~\ref{eqn:anal_error}
contain a signal and noise contribution:
$\hat{C}_\ell^{XX}=C_\ell^{XX}+C_\ell^{XX, N}$.
It is informative to split the right hand side of
Eq.~\ref{eqn:anal_error} into four pieces:
\be
\label{eqn:error_breakdown}
\begin{split}
rhs &= \left[ C_\ell^{\phi \phi} C_\ell^{{\rm CIB}} +
  \left(C_\ell^{{\rm CIB} \phi}\right)^2 \right] +
       C_\ell^{\phi \phi,{\rm N}} C_\ell^{{\rm CIB},{\rm N}} \\
    &+ C_\ell^{\phi \phi} C_\ell^{{\rm CIB},{\rm N}} +
       C_\ell^{\phi \phi,{\rm N}} C_\ell^{\rm CIB}.
\end{split}
\ee
Here the first term is a signal-only piece, the second is a noise-only piece,
and the remaining two terms are mixed signal and noise pieces.
To discuss the relative importance of these terms, we will use for the
signal terms the model spectra, and for the noise terms we
subtract the model spectra from the measured spectra: 
$\hat{C}_\ell^{XX, N}=\hat{C}_\ell^{XX}-C_\ell^{XX}$.
With this definition, the noise will include the instrumental contribution, as
well as other astrophysical signals including the CMB, which we do not remove
from our data for reasons previously mentioned. We show the different terms in
Fig.~\ref{fig:error_by_l}. Up to 353\,GHz the measured temperature spectrum,
$\hat{C}_\ell^{{\rm TT}}$, is dominated by the CMB at low $\ell$ and the
instrumental noise at high $\ell$. At higher frequencies Galactic emission
dominates at low $\ell$ and the CIB at high $\ell$. 
For all frequencies up to 353\,GHz the dominant contribution to the errors
in our signal comes from the noise-only term (in blue), which is proportional
to the temperature noise spectrum. At 353\,GHz and above the mixed
signal-noise term $C_\ell^{\phi\phi,{\rm N}} C_\ell^{\rm CIB}$ (orange)
becomes important and is the largest contribution at 545 and 857\,GHz
at high $\ell$.

\subsection{Instrumental and observational systematic effects}
\label{sec:inst_and_syste}

A number of systematic errors affect the \Planck\ HFI analysis and we briefly
discuss some of them here. A more complete discussion can be found in
\cite{planck2011-1.7}. We will illustrate how the very nature of our
measurement, a 3-point function, makes it particularly robust to many
systematic effects, and we will check for their signatures using null
tests. For example, there is no noise bias in the 3-point measurement,
and many effects that can lead to biases in the auto-spectrum of $\phi$ do not affect us.     

\subsubsection{Potential sources of systematic error}

We begin by describing our knowledge of known systematic effects, before
discussing others that could bias our result. To account for an error
in the calibration of the temperature maps, we simply add in
quadrature a calibration uncertainty to our error bars. In Sect.~\ref{sec:blind_tests} we use
null tests to check that these errors are consistent with the data.
In addition we use the null tests to search for evidence that the calibration
has changed between surveys, for example due to gain drifts.
We account for beam errors in a conservative manner by using a constant
error at each frequency equal to the maximum error in the beam multipoles,
$B_\ell$, at any $\ell$ (see PIR for details).
The $B_\ell$  uncertainty is
larger at high-$\ell$ with, for example, values at $\ell=1500$ of
79.5\,\%,  8.2\,\%, 0.53\,\%, 0.95\,\%, 0.31\,\%, and 0.70\,\%,
at 100--857\,GHz, respectively. The calibration
error is therefore larger than the beam error at all $\ell$ between 
217 and 857\,GHz but smaller at high $\ell$ in the 100 and 143\,GHz
channels. We add the beam error in quadrature in an $\ell$- and
frequency-dependent manner. 
As discussed in \citet{planck2013-p12}, uncertainties in the beam transfer
(as well as the fiducial CMB power spectrum $C_{\ell}^TT$) propagate
directly to a normalization uncertainty in the lens reconstruction.
Based on the beam eigenmodes of \cite{planck2013-p03c},
it is estimated in \cite{planck2013-p12} that beam uncertainty
leads to an effective normalization uncertainty of approximately $0.2\%$
and 143 and 217\,GHz, and $0.8\%$ at 100\,GHz.
To be conservative, on top of the calibration and
beam error we will add in quadrature a 2\,\%  uncertainty on the overall lens normalization.

CMB lens reconstruction recovers modes of the lensing potential
through their anisotropic distorting effect on small-scale hot and
cold spots in the CMB. The quadratic estimator, which we use
to reconstruct the lensing potential is optimized to measure the lensing induced
statistical anisotropy in CMB maps.  However, other sources of statistical
anisotropy, such as the sky mask, inhomogeneous noise, and beam asymmetries,
produce signals, which can potentially overlap with lensing. 
These introduce a ``mean-field'' bias, which we estimate using
Monte Carlo simulations and subtract from our lensing estimates.
Innaccuracies in the simulation procedure will lead to errors in this correction, particularly if the correction is large.
The mean-field introduced by the
application of a Galaxy and point-source mask, for example,
which can be several orders above magnitude larger than the 
lensing signal at $\ell < 100$.
This is discussed further in Appendix B of \cite{planck2013-p12}.
The mask mean-field is a particular concern for us because it
has the same phases as the harmonic transform of the mask.
If our masked CIB maps have a non-zero monopole, for example,
it will correlate strongly with any error in the mask mean-field
correction. 
For this reason we do not use any data below $\ell = 100$ in our analysis. 

To summarize, we do not expect these known systematic effects to be
present at a significant level.  Nevertheless,
we still perform a set of consistency tests that would be
sensitive to them or other unknown effects.

\subsubsection{Null tests}
\label{sec:blind_tests}

The \Planck\ scanning strategy, its multiple frequency bands and its numerous
detectors per frequency, offer many opportunities to test the
consistency of our signal (see Sect.~\ref{sec:planckmaps}).
We focus on such tests in this 
section. Our aim is to reveal any systematic effects that could lead to 
a spurious correlation.
For all of the tests presented,  we will quote a $\chi^2$ value
as well as the number of degrees of freedom ($N_{\rm dof}$) as
a measure of the consistency with the expected null result.
Throughout this section,
black error bars in plots will correspond to the measured scatter within an
$\ell$ bin multiplied by 1.2, as was justified in
Sect.~\ref{sec:stat_error} and Appendix~\ref{app:stat_errors},
and will also include a CIB calibration error and a beam error,
while the coloured boxes correspond to the
analytical errors of the corresponding signal (i.e., not the difference
corresponding to the null test). Plotting these two error bars
illustrates how important any deviation could be to our
signal. Throughout this section, we will illustrate our findings with
the 545\,GHz correlation, since it is our prime band for this measurement,
but our conclusions hold at other frequencies.

\begin{figure}[!t]
  \centering
  \includegraphics[width=88mm, clip=true, trim=0.5cm 0 0 0]{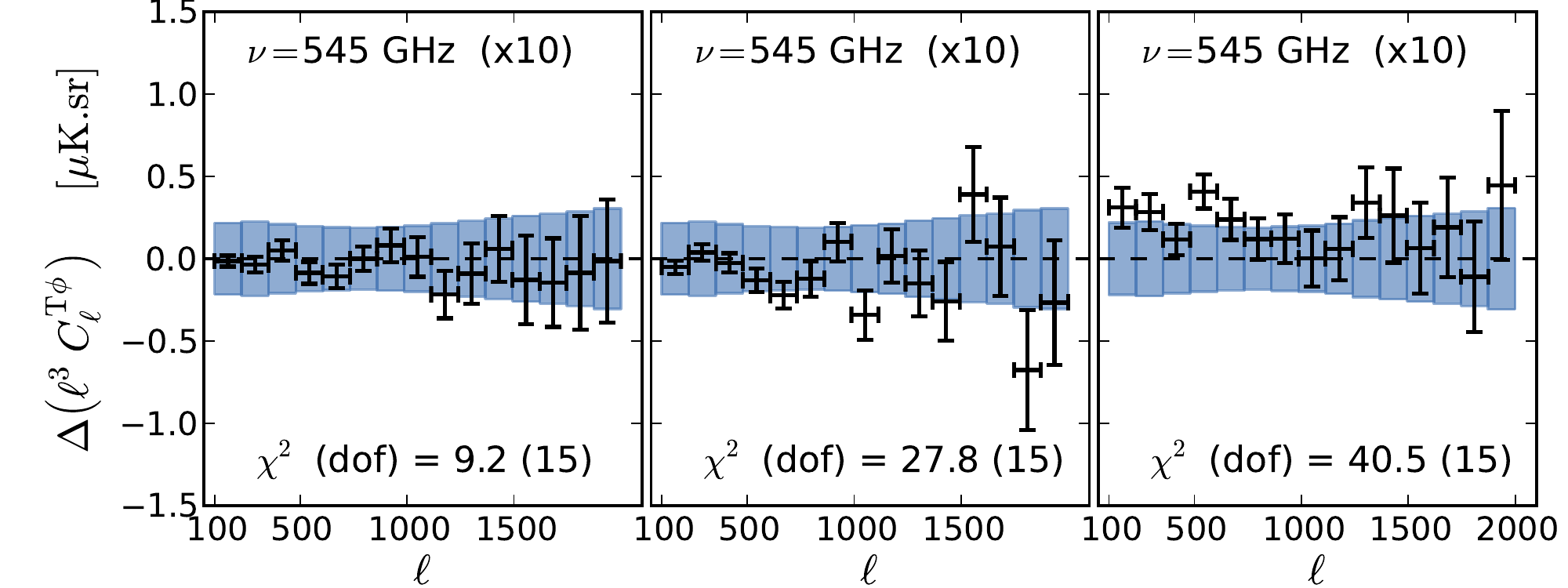}
  \caption[]{Null tests at 545\,GHz. {\it Left:}
    difference spectra obtained by nulling the signal in the HR temperature
    map before correlating it with our nominal $\phi$ reconstruction.
    {\it Middle:} temperature signal nulled using
    different detectors at 545\,GHz.  {\it Right:}
    temperature signal nulled using the first and second survey maps. The
    black error bars correspond to the scatter measured within an
    $\ell$-bin, while the coloured bands correspond to the analytical
    estimate. Except for the survey null test (see text for details),
    these tests are passed satisfactorily except, as illustrated 
    by the quoted $\chi^2$ and $N_{\rm dof}$, thus strengthening
    confidence in our signal.}
  \label{fig:null_test_545}
\end{figure}

The first test we conduct is to take the temperature difference between the two
half-ring (HR) maps to cancel any signal contribution, and therefore
investigate the consistency of our measurements
with our statistical errors on all time scales. We null the temperature maps and
correlate with our nominal lensing map. The results are shown in the left panel
of Fig.~\ref{fig:null_test_545}. We see a significant deviation from
null only when considering survey differences. This particular failure
can probably be explained by apparent gain drifts due to nonlinearity
in the analog-digital conversion
\citep{planck2013-p03,planck2013-p03f}, not yet corrected at this
frequency. Note however that the predicted variation is about 1\%
while the deviation from null would call for a variation of
1.5-2\%. But in any case, its amplitude is too small to significantly
affect our measurement.

We see a significant deviation from
null only when considering survey differences. This particular failure
can probably be explained by apparent gain drifts due to nonlinearity
in the analog-digital conversion
\citep{planck2013-p03,planck2013-p03f}, not yet corrected at these
frequency. But in any case, its amplitude is too small to
significantly affect our measurment.

The second test uses multiple detectors at a given
frequency that occupy different parts of the focal plane. These detector sets
are used to construct the ``detset'' maps that were described in
Sect.~\ref{sec:planckmaps}. The two ``detset'' maps are subtracted and then
correlated with our nominal lens reconstruction.
This test is particularly sensitive to long term noise properties or gain
variations, as we do not expect these to be correlated from detector to
detector. Since this detector division breaks the focal plane
symmetry, it is also a good check for beam asymmetry
effects. Here again, we do not find any significant deviation, as
illustrated in the middle panel of Fig.~\ref{fig:null_test_545}.

The third test we conduct makes use of the redundant sky coverage,
using multiple surveys to cancel the signal. As above, we null
the temperature signal and correlate with the nominal lens reconstruction.
This test is particularly sensitive to
any long term, i.e., month timescale drifts that could affect our
measurement. It is also a good test for any beam asymmetry
effects, as individual pixels are observed with a different set of
orientations in each survey. Since only the first two surveys are
complete for this particular data release, we only use the two full
survey maps to avoid complications with the partially completed third
survey. Here again, we do not find any significant deviation, as 
illustrated in the right panel of Fig.~\ref{fig:null_test_545}.

To conclude, this first set of stringent consistency tests have shown
that there is no obvious contamination of our measurements due to instrumental
effects. In addition, the reasonable $\chi^2/N_{\rm dof}$ obtained gives
us confidence in our statistical noise evaluation. Although we measure
the noise directly from the data, this success was not guaranteed.

\subsection{Astrophysical contamination}
\label{sec:astro_contam}

We now turn to possible astrophysical biases to our
measurement. We will discuss successively known
astrophysical contaminants that can either come
from Galactic or extragalactic origin. Once again, besides our knowledge
of these signals, we will rely heavily on consistency tests made
possible by having multiple full sky frequency maps.

\begin{figure}[!t]
  \centering
  \includegraphics[width=88mm, clip=true, trim=0.5cm 0 0 0]{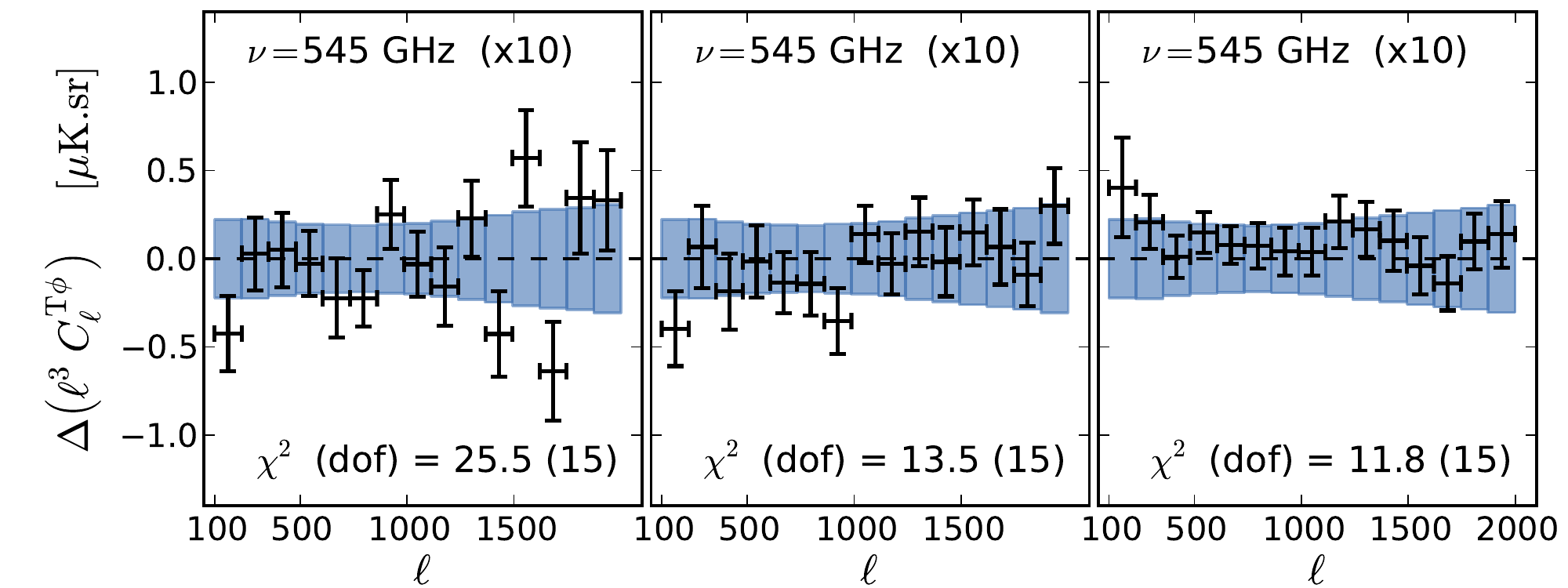}
  \caption[]{{\it Left:} difference between the 
    cross-spectra measured using the 20\,\% Galactic mask (20\,\% is the
    unmasked sky fraction) from that measured with our default 40\,\% Galactic
    mask.  {\it Middle:} spectra obtained when differencing the
    60\,\% and 40\,\% Galaxy mask measurements. For both left and middle
    panels and all Galactic masks, the same point source and \hi\ masks
    are used, which removes an additional fraction of the sky.  {\it Right:}
    difference between the cross-spectra calculated with the
    \hi\ cleaned temperature maps from those with no \hi\ cleaning. This
    cross-spectrum is thus the correlation between the \hi\ template and
    the $\phi$ reconstruction. The error bars are calculated in the same way as
    in Fig.~\ref{fig:null_test_545}.  Again, the null tests are passed with
    an acceptable $\chi^2$.}
  \label{fig:null_test_545b}
\end{figure}
\begin{figure}[!t]
  \centering
  \includegraphics[width=88mm, clip=true, trim=0.5cm 0 0 0]{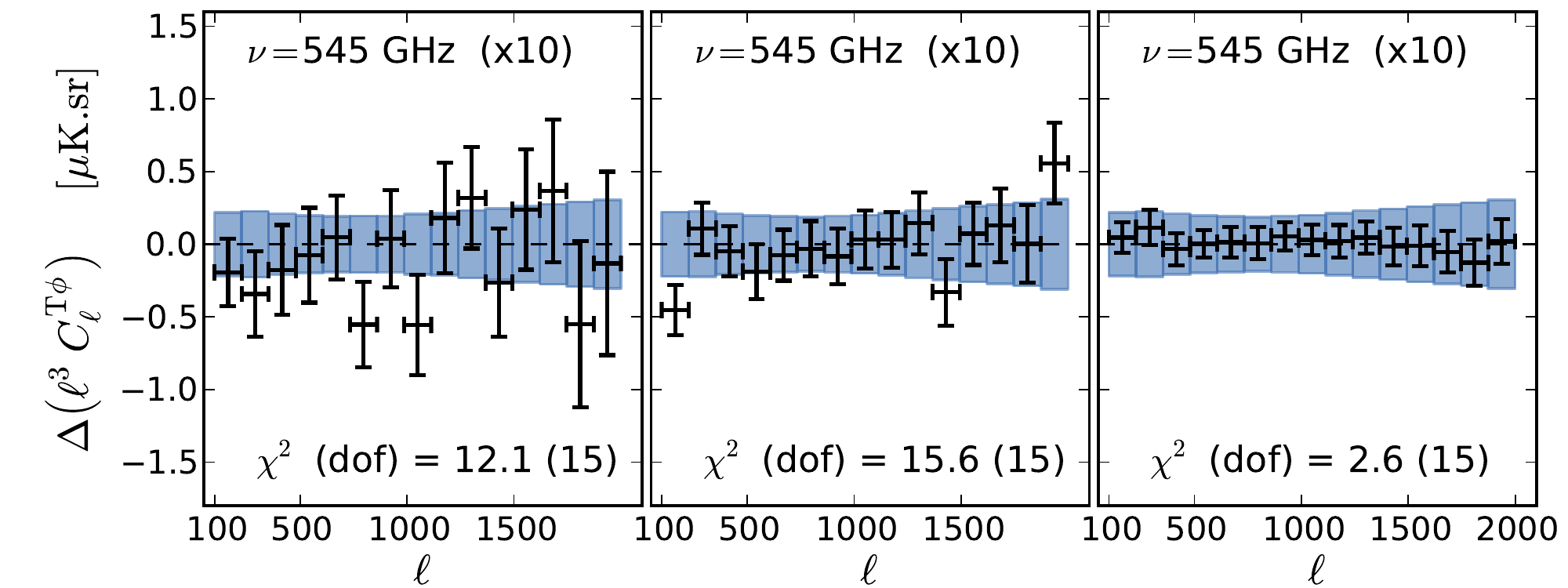}
  \caption[]{{\it Left:} difference between cross-spectra
    calculated using the lens reconstruction
    at 100\,GHz with the nominal 143\,GHz reconstruction. We see
    an overall shift, which leads to a high reduced $\chi^2$. This
    shift can be explained by the expected overall normalization
    uncertainties of the 100\,GHz and 143\,GHz reconstructions. While
    this uncertainty is not included in the $\chi^2$ or the solid bars, it
    is included later in our analysis in Sect.~\ref{sec:modeling}.
    {\it Middle:} same as the left panel, but the 217\,GHz
    reconstruction is used instead of the 100\,GHz reconstruction.
    {\it Right:} difference between cross-spectra when we
    consider the 143\,GHz lens reconstruction calculated with a
    less restrictive Galaxy mask (that excludes only 20\,\% of the sky) and
    the nominal reconstruction mask that excludes 40\,\% of the sky.
    }  \label{fig:null_test_545c}
\end{figure}

\subsubsection{Galactic foregrounds}
\label{sec:gal_fg}

Galactic foregrounds have two possible effects on our measurement.
The first is the introduction of an extra source of noise.
The second is that contamination of the lensing reconstruction by any Galactic
signal, e.g., synchrotron, free-free or dust, which could then
correlate with foreground emission present in the temperature maps,
remains a source of bias that has to be investigated.
We will show that this bias is small.  To do so, we
take three approaches. We first investigate various
Galactic masks, then perform the lensing reconstruction at various
frequencies, and finally investigate the effect of a dust-cleaning
procedure.  

First, we consider two additional masks, either more aggressive or more
conservative than our fiducial one. Both were introduced in
Sect.~\ref{sec:planckmaps}. The first one leaves approximately
20\,\% of the sky unmasked, while the second one leaves approximately 60\,\% of
the sky. Given the strong dependence of Galactic foregrounds on
Galactic latitude, any Galactic contamination should vary strongly
when we switch between masks. Comparing the measurements using these
masks with our fiducial 40\,\% mask in the left and centre panels of
Fig.~\ref{fig:null_test_545b},
we do not see any substantial deviation from our fiducial measurements.
This excludes strong Galactic contamination of our results.   

Second, we perform the lens reconstruction at
100 and 217\,GHz, different from the fiducial frequency of 143\,GHz, and
compare their correlation with the temperature maps. Given the strong
dependence of the Galactic emission with frequency, $T\propto
\nu^{-3}$ for synchrotron and $T\propto \nu^2 $ for dust in this
frequency range, any contamination of our signal would have a
strong frequency dependence.  The comparison with the 100\,GHz (217\,GHz)
reconstruction is presented in the left (centre)
panel of Fig.~\ref{fig:null_test_545c}. The right panel shows the
difference of the cross-spectra calculated using the 143\,GHz reconstruction
with a more aggressive Galaxy mask (20\,\% instead of 40\%), to reduce
possible Galactic contaminants in the reconstruction, and the nominal
reconstruction. The three differences are consistent with null as
demonstrated by the quoted $\chi^2$ and $N_{\rm dof}$.  

Third, we investigate more specifically how cirrus, the dominant Galactic
contaminant for our higher
frequency channels, affects our measurements. We rely on the dust
cleaning procedure detailed in Sect.~\ref{sec:fg_cleaning} that aims to
remove the \hi-correlated dust component. This procedure leads to
a decrease in the variance measured outside the mask of 22, 65, 73 and
73\,\% in the 217, 353, 545 and 857\,GHz maps, respectively. This
frequency dependence is expected given the dust scaling.
However, in Fig.~\ref{fig:null_test_545b}, where we show the differences
between the cleaned and non-cleaned cross-spectra, we observe that the
large scale \hi\ cleaning, even though it makes a substantial impact on the
power within our map, only makes a small change at low-$\ell$ in the
cross-spectrum, as well as reducing the noise at all multipoles.
If we quantify the effect  of our ``local'' \hi\ cleaning on the detection significance level computed using only
statistical errors, we find that the significance is increased by
4, 4, 28, and 36\,\% at 217, 353, 545 and 857\,GHz,
respectively. Also, not surprisingly, we observe that for
frequencies up to 353\,GHz where the statistical errors are dominated 
by the CMB, the \hi\ cleaning has almost no effect on the cross-spectra.
From the three studies in this section we conclude that
there are no obvious signs of Galactic foreground contamination in our
cross-correlation.

\subsubsection{Point source contamination}
\label{sec:pointsources}

We now discuss another well-known potential source of
contamination, namely the contribution of unresolved point sources
visible either through their radio or dust emission. Our concern is
that a correlation between a spurious lens reconstruction caused by
unresolved point sources
can correlate with sources in the temperature map.
Although in Sect.~\ref{sec:gal_fg} our null test using lens reconstructions
at different frequencies suggests that unresolved point sources are not an
obvious contaminant, we will now perform a more detailed test designed
specifically to search for point source contamination.
Following \cite{smith2007,SJOLensingPointSources}, we will construct
a \textit{point source estimator} designed to be more sensitive than the
lensing estimator to point source contamination. 
Our focus here will be on possible contamination from the 
point source shot-noise bispectrum.
In Sect.~\ref{sec:cib_bispectrum} we will discuss contamination
from a scale dependent bispectrum.

Our (unnormalized) quadratic estimator, which is designed to detect point source
contributions is given by
\be
(\bar{\Theta}^{143}(\hat{n}))^2_{LM}
\equiv \sum_{LM} Y_{LM}^{\ast}(\hat{n}) \, (\bar{\Theta}^{143}(\hat{n}))^2,
\label{eqn:qe_slm}
\ee
where $\bar{\Theta}$ is the inverse-variance filtered sky map.
This estimator is simply the square of the inverse-variance filtered sky map, which is
a more sensitive probe of point sources than the standard lensing estimator.

\begin{figure*}[!t]
  \centering
  \includegraphics[width=180mm]{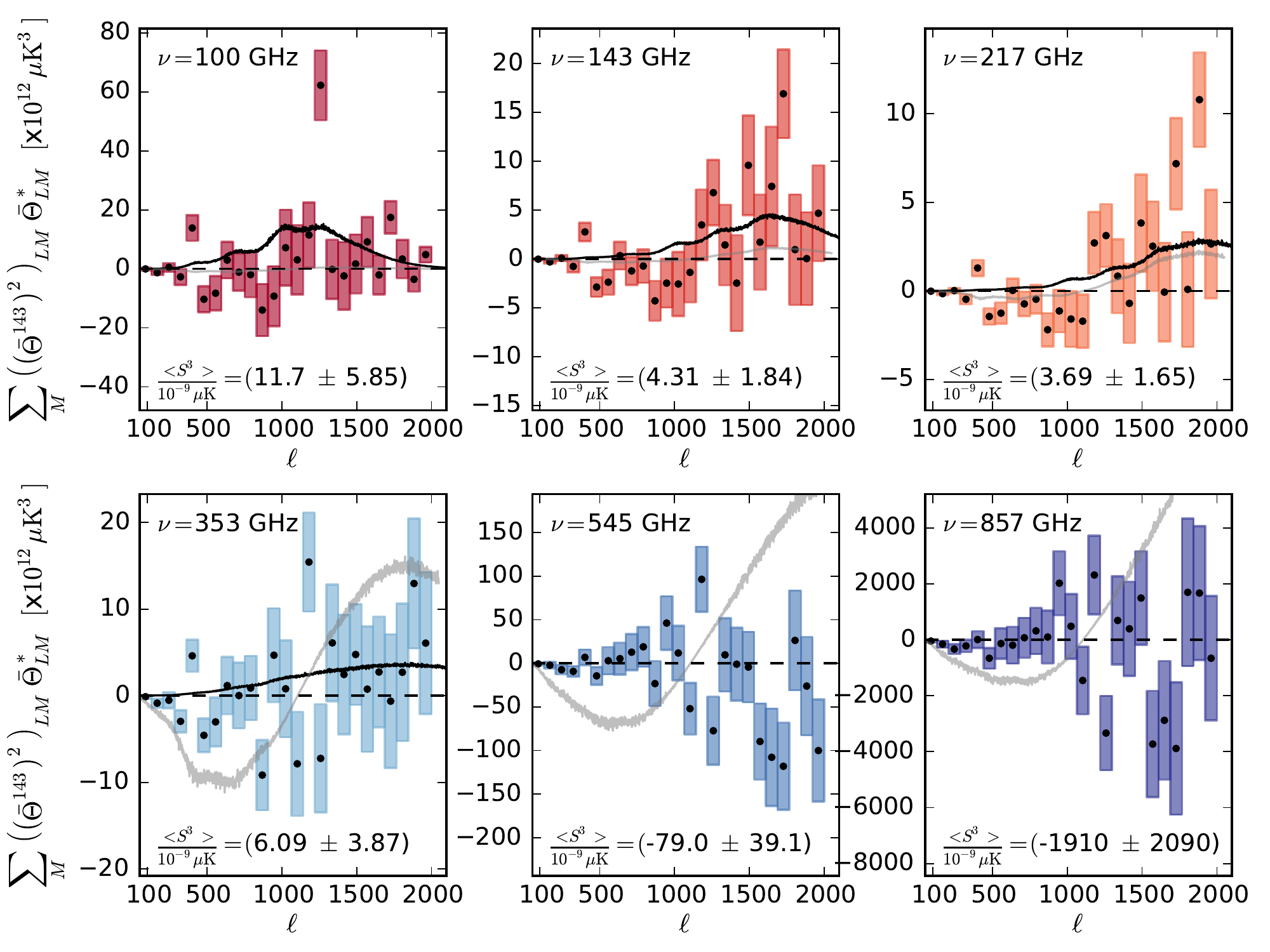}
  \caption[]{Results from the point source contamination estimator of Eq.~\eqref{eqn:qe_slm}.
    The best-fit cross-spectra associated with shot noise are plotted in black.
    We do not show the best-fit at 545 and
    857\,GHz since the signal-to-noise ratio is low. The grey line is
    a prediction for the bias from the CMB lensing - infrared correlation, and has been subtracted from the
    spectra (plotted as black points). We see that with the
    subtraction of the bias from CMB lensing, the measured
    bispectrum-related spectrum is generally consistent either with zero, or
    with the shape expected for shot noise.} 
  \label{fig:bspec_spec}
\end{figure*}

In Fig.~\ref{fig:bspec_spec} we plot the cross-spectrum of $(\bar{\Theta}^{143}(\hat{n}))^2_{LM}$
measured at 143\,GHz and $\bar{\Theta}_{LM}^{\nu}$ for the full set of HFI channels.
This cross-spectrum is probing the same point source contributions that
could bias our estimates of $C_{\ell}^{T\phi}$, however with a greater signal-to-noise ratio.

There is one complication here, which is that just as lens reconstruction may be biased by
point source contributions, the point source estimator is correspondingly biased by lensing.
The bias to the plotted cross-spectra is given by
\begin{multline}
\langle (\bar{\Theta}^{143}(\hat{n}))^2_{LM} \bar{\Theta}_{LM}^{\nu *} \rangle_{\phi} =
\frac{C_L^{\nu \phi}}{C_L^{\nu \nu}} \sum_{\ell_1 m_1}\sum_{\ell_2 m_2}
\frac{G_{L \ell_1 \ell_2}^{-M m_1 m_2}}{C^{\rm tot}_{\ell_1}C^{\rm tot}_{\ell_2}} \\
\times \left( (-1)^{m_1} I_{\ell_2 L \ell_1}^{m_2 -M -m_1} \tilde{C}_{\ell_1} +
        (-1)^{m_2} I_{\ell_1 L \ell_2}^{m_1 -M -m_2} \tilde{C}_{\ell_2} \right)\ ,
\end{multline}
where $G_{\ell_1 \ell_2 \ell_3}^{m_1 m_2 m_3} = \int d\hat{n} \, Y_{\ell_1 m_1}(\hat{n}) Y_{\ell_2 m_2}(\hat{n}) Y_{\ell_3 m_3}(\hat{n})$,
$\tilde{C}_{\ell}$ is the unlensed CMB spectrum,
$C_\ell^{\rm tot}$ is the spectrum of $\Theta^{143}(\hat{n})$,
$I_{\ell_1 \ell_2 \ell_3}^{m_1 m_2 m_3}$ is as defined in~\citet{Okamoto:2003zw},
and we have used the fact that for our inverse-variance filtering,
$\bar{\Theta}_{\ell m} \approx \Theta_{\ell m} / C^{\rm tot}_{\ell}$.
We have calculated this contribution using our measured $C_L^{T\phi}$ and subtracted it from the data points of Fig.~\ref{fig:bspec_spec}. 

We can consider the effect of shot noise on this cross-spectrum.
With the shot-noise bispectrum defined by
\be
\langle \Theta_{\ell_1 m_1} \Theta_{\ell_2 m_2} \Theta_{\ell_3 m_3}^{\nu} \rangle_{S^3} =
G_{\ell_1 \ell_2 \ell_3}^{m_1 m_2 m_3} \left< S^3 \right>
\ee
the bias to the plotted cross-spectrum is given as
\be
\langle (\bar{\Theta}^{143}(\hat{n}))^2_{LM} \bar{\Theta}_{LM}^{\nu} \rangle_{S^3} =
\left<S^3\right> \frac{(-1)^M}{C_L^{\nu \nu}} \sum_{\ell_1 m_1} \sum_{\ell_2 m_2}
\frac{G_{L \ell_1 \ell_2}^{-M m_1 m_2} G_{L \ell_1 \ell_2}^{M m_1 m_2}}{C^{\rm tot}_{\ell_1}C^{\rm tot}_{\ell_2}} .
\ee
This bias is plotted for best-fit values of $\langle S^3 \rangle$ as the black lines in Fig.~\ref{fig:bspec_spec}.
To minimize systematic effects that might bias the $\left< S^3 \right>$
estimator, we have estimated $S^3$ from the spectra of Fig.~\ref{fig:bspec_spec}
between multipoles between $\ell=500$
and 2000.
The fitted $\left<S^3\right>$ amplitudes are given in Table~\ref{tab:ptsrc}.

\begin{table}[!b]            
\begingroup
\newdimen\tblskip \tblskip=5pt
\caption{Point source estimator.  The measured quantity $\left<S^3\right>$, as
  defined in Eq.~\ref{eqn:ps_estimator} is given as a function of frequency.}
\label{tab:ptsrc}          
\nointerlineskip
\vskip -3mm
\footnotesize
\setbox\tablebox=\vbox{
   \newdimen\digitwidth 
   \setbox0=\hbox{\rm 0} 
   \digitwidth=\wd0 
   \catcode`*=\active 
   \def*{\kern\digitwidth}
   \newdimen\signwidth 
   \setbox0=\hbox{+} 
   \signwidth=\wd0 
   \catcode`!=\active 
   \def!{\kern\signwidth}
\halign{\hbox to 0.75in{#\leaderfil}\tabskip 2.2em& \hfil#\hfil& \hfil#\hfil\tabskip=0pt\cr
\noalign{\doubleline}
\noalign{\vskip -3pt}
\omit\hfil Frequency\hfil& \omit\hfil$\left<S^3\right>$\hfil&
 (No.\ of $\sigma $)\hfil\cr  
\omit\hfil [\GHz]\hfil& $[\times 10^9 \mu K^3]$& \cr
\noalign{\vskip 3pt\hrule\vskip 3pt}
100& $!11.7 \pm *5.8$& !(2.0)\cr
143& $!*4.3 \pm *1.8$& !(2.3)\cr
217& $!*3.7 \pm *1.6$& !(2.2)\cr
353& $!*6.1 \pm *3.9$& !(1.6)\cr
545& $*-79  \pm *39$& ($-$2.0)\cr
857& ($-1.9 \pm 2.1)\times 10^3$& ($-$0.9)\cr
\noalign{\vskip 3pt\hrule\vskip 3pt}}}
\endPlancktable                   
\endgroup
\end{table}       

These amplitudes match our expectations, for example
see~\citet{planck2011-6.1}. We observe a decrease in the amplitude of the point
source contribution going from 100 to 217\,GHz, which corresponds to a dominant
contribution from radio point sources. We do not see any evidence of a dusty
galaxy contribution to the shot-noise bias. These conclusions have
been verified using less restrictive point source masks that cover
fewer sources. 

With estimates of $S^3$ in hand, we estimate a corresponding bias
to $C_\ell^{T\phi}$, given by 
\be
\left< \hat{\phi}_{LM} \bar{\Theta}_{LM}^{\nu *} \right>_{S^3} =
(-1)^M \frac{\left<S^3\right>}{C_L^{\nu\nu}}
\sum_{\ell_1 m_1} \sum_{\ell_2 m_2}
\frac{W_{\ell_1 \ell_2 L}^{m_1 m_2 M}}{C_{\ell_1}^{\rm tot}C_{\ell_2}^{\rm tot}}
G_{\ell_1 \ell_2 L}^{m_1 m_2 -M}
\label{eqn:ps_estimator}
\ee
where
$W_{\ell_1 \ell_2 L}^{m_1 m_2 M} =
(-1)^{m_2}
( I_{\ell_2 L \ell_1}^{-m_2 M m_1} \tilde{C}_{\ell_1} + I_{\ell_1 L \ell_2}^{m_1 M -m_2} \tilde{C}_{\ell_2})
/2 {\cal R}_L^{\phi \phi}$
with ${\cal R}_L^{\phi \phi}$ defined in~\citet{planck2013-p12}.
We show this contribution later in Fig.~\ref{fig:astro_contamination} as the dotted
line. While non-zero, we see that the point source shot noise contribution is
always negligible in the $\ell$ range we consider, except at lower
frequencies where the radio point sources are important (but still not
strong enough to lead to any clear signal in the cross-spectra).

\subsubsection{SZ contamination}
\label{sec:sz_cont}

\begin{figure}[!t]
  \centering
  \includegraphics[width=88mm,clip=true,trim=0cm 0 0 0]{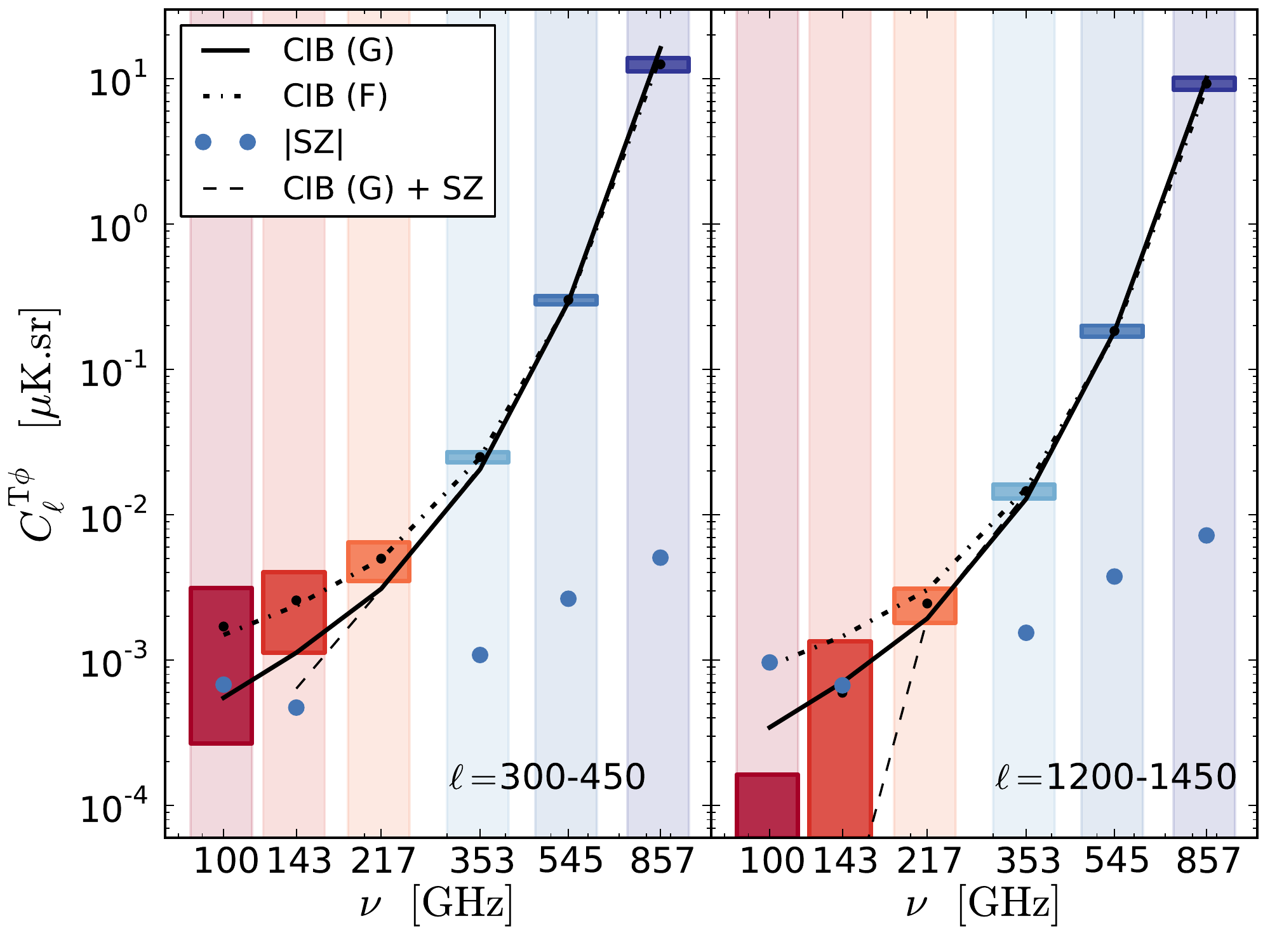}
  \caption[]{Frequency spectrum of our cross-spectra averaged within
    $\ell$-bins (black points with associated error bars). The
    light shaded regions correspond to the HFI frequency bands. The solid
    black curve corresponds to the best-fit CIB assuming a
    \cite{gispert2000} spectrum, while the dot-dashed line assumes a
    \cite{fixsen1998} spectrum. The dashed black line corresponds to the
    best-fit model when allowing for an SZ component in addition to the
    \cite{gispert2000} CIB shape. The blue dots correspond to the
    associated absolute value of the best-fit SZ component. We conclude from
    this plot that the SZ effect is not an important contaminant.}
  \label{fig:sz_spectra}
\end{figure}

A fraction of CMB photons travelling from the surface of recombination are
scattered by hot electrons in galaxy clusters. In the most massive
clusters approximately 1\,\% of CMB photons passing through them get 
scattered. On average, their energy will be increased, which leads
to a measurable spectral distortion. This is the so-called thermal
Sunyaev-Zeldovich (SZ) effect~\citep{1970Ap&SS...7....3S}. At the
location of a galaxy cluster the CMB appears colder at frequencies
below about $220$\,GHz and hotter at higher frequencies, with a
temperature change proportional to the cluster optical depth to
Compton scattering and to the electron temperature. Since hot electrons
in clusters also trace the large scale matter potential that is traced by
CMB lensing, we expect an SZ-induced contamination
in our measurement at some level. We will show below
that the level of contamination is negligible. In these calculations
we ignore the small relativistic corrections to the
thermal SZ spectrum~\citep[e.g.,][]{2000ApJ...536...31N}. We also
ignore the kinetic SZ signal coming from the bulk motion of hot
electrons in clusters, since it is subdominant to the thermal
signal~\citep{1980MNRAS.190..413S,Reichardt:2011yv,Hand:2012ui}.  

The frequency dependence of the SZ signal in our map depends on the
detector bandpasses and is
\be
\label{eqn:sz_spectrum}
f(\nu) = \frac{\int d\nu \; h(\nu) \; g(\nu)}{\int d\nu \; h(\nu)},
\ee
where $h(\nu)$ is the detector bandpass and $g(\nu)$ is the SZ
frequency dependence, which in the non-relativistic limit is $g(\nu)
= x \; (e^x+1)/(e^x-1) - 4$, with $x = h\nu/k_B T_{\rm CMB}$.
The effect of the bandpass only makes a large difference at
217\,GHz near the null of the SZ signal. The thermal SZ affects our
measurement in two ways. First, since the SZ emission in our maps is not a
Gaussian random field~\citep[e.g.,][]{Wilson:2012sr} it introduces a
spurious signal into our lens reconstruction
that will correlate with the SZ signal in our CIB map. As shown in
\citet{SJOLensingPointSources}, this is well approximated by a
Poisson noise term and is thus already addressed by our treatment of point
sources in Sect.~\ref{sec:pointsources}. The spurious lensing signal
will also correlate with other components in our map such as the CIB.
However, we ignore these terms since they will be smaller than
those that correlate directly with the SZ
emission. Additionally, a contribution comes from SZ emission in our CIB map
that correlates with the lensing potential itself. The latter is the
dominant term and we discuss it in this section.

To measure a contribution from the SZ-lensing correlation
we attempt to separate the SZ and CIB emission based on their
differing spectral shapes. We consider all frequencies from 100
to 857\,GHz, but we will illustrate this procedure by
considering only two $\ell$ bands: $\ell = 300$--450; and
$1200$--$1450$. The first is well inside the linear regime, while the
second receives a more important non-linear contribution.
However, we have checked that if we consider different $\ell$-bins we obtain
similar conclusions.  We model the signal within each
$\ell$ band as $s_{\ell}(\nu) = a_{1,\ell} c(\nu) + a_{2,\ell} f(\nu)$,
where $c(\nu)$ and $f(\nu)$ are, respectively, the CIB frequency dependence
(as proposed in \citealt{fixsen1998} or \citealt{gispert2000})
and the SZ frequency dependence
obtained from Eq.~\ref{eqn:sz_spectrum}. For each $\ell$ band,  we will
solve for $a_{1,\ell}$ and $a_{2,\ell}$ minimizing the associated
$\chi^2$ while forcing both amplitudes to be positive. As an
approximation to the error in each multipole band we calculate the
scatter of the signal within the band and multiply it by 1.2, as
discussed in Sect.~\ref{sec:stat_error}.

In Fig.~\ref{fig:sz_spectra} we show the measured frequency spectrum within
each $\ell$ band, along with the best-fit SZ-lensing and CIB-lensing spectra.
For the CIB-only fit with the \cite{gispert2000} frequency dependence we find a
relatively poor fit in the lowest $\ell$-bin, $\chi^2$ (dof) = 15.5
(5), but an improved fit in the higher $\ell$-bin, $\chi^2$ (dof) =
4.15 (5). Including the SZ component gives $\Delta \chi^2$ = 0.52 and 1.34 in the
low and high $\ell$ bins for one extra degree of freedom.
When we use the \cite{fixsen1998} frequency dependence we find an improved fit,
with $\chi^2$ (dof) = 2.25 (5) and 5.49 (5) in the low and high-$\ell$-bins,
respectively. Overall, the improvement in the $\chi^2$/dof when including the SZ
component does not justify inclusion of the SZ component in the model,
with the poor fit driven by the lowest frequency bands where the CIB
scaling is rather unconstrained. In fact, our measurements might constitute
the first constraints to date on this scaling. From these results we
conclude that including the SZ-lensing correlation in our data does not
improve the fit in the $\ell$ range of interest to us and thus we do
not consider it necessary to correct for.

As an extra validation of this result, we now verify its consistency with
current models of the CIB and SZ emission.  For this purpose, we use the
calculation of the correlation from~\citet{SJOLensingPointSources}, based
on~\citet{Babich:2008uw}, which models the SZ emission as a
statistically isotropic signal modulated by a biased density contrast,
where the bias depends on the cluster mass and redshift.
To obtain an estimate of the contribution to the cross-spectrum at
217--857\,GHz we assume that the
measured cross-spectrum at 143\,GHz is entirely due to thermal SZ
emission (note that we do this to find what we believe to be an upper
limit on the SZ contribution at 217--857\,GHz; for the reasons stated
above we do not expect the 143\,GHz correlation to be due to SZ).
Since the SZ signal at 143\,GHz gives a decrement in the CMB,
and the CIB emission gives an enhanced signal, it is possible that
this approach could still underestimate the SZ signal. We find that in
order to fit the cross-spectrum at 143\,GHz using only the SZ-lensing
correlation requires an amplitude of ($2.4 \pm 1.6$) times our
calculated SZ-lensing cross-spectrum. In
Fig.~\ref{fig:astro_contamination} the dashed line shows the magnitude
of this SZ signal scaled to each frequency using Eq.~\ref{eqn:sz_spectrum}.
The small contribution it makes at 217--857\,GHz further suggests
that we can neglect this component. At 217\,GHz the signal is negative, while
at higher frequencies it is positive.

\subsubsection{ISW contamination}

The Integrated Sachs-Wolfe (ISW) effect describes the redshifting (blueshifting) of
photons travelling through gravitational potential wells (hills) that decay as 
the photons travel through them~\citep{Sachs:1967er}. 
The induced modulation of the CMB mean by
the gravitational potential generates CMB fluctuations that correlate with
the lensing potential, which also traces out the gravitational
potential perturbations~\citep{Seljak:1998nu,Goldberg:1999xm,Lewis:2011fk}. Note
that because the mean of the CIB is relatively much smaller than its 
fluctuation, the ISW induced CIB fluctuations make a negligible change to total CIB anisotropy.
 The CMB ISW induced signal has the same frequency dependence as the CMB and so is only
a significant contaminant for us at low frequencies. We evaluate this
signal using a theoretical calculation performed in {\tt
  CAMB}~\citep{Lewis:1999bs}. The results are shown as the solid line
in Fig.~\ref{fig:astro_contamination}. It is a negligible contribution
at all frequencies, except in the lowest $\ell$-bin of the lowest
frequencies, where the measured cross-spectrum is consistent with zero.

\subsubsection{CIB bispectrum}
\label{sec:cib_bispectrum}

\begin{figure*}[!t]
  \centering
  \includegraphics[width=180mm]{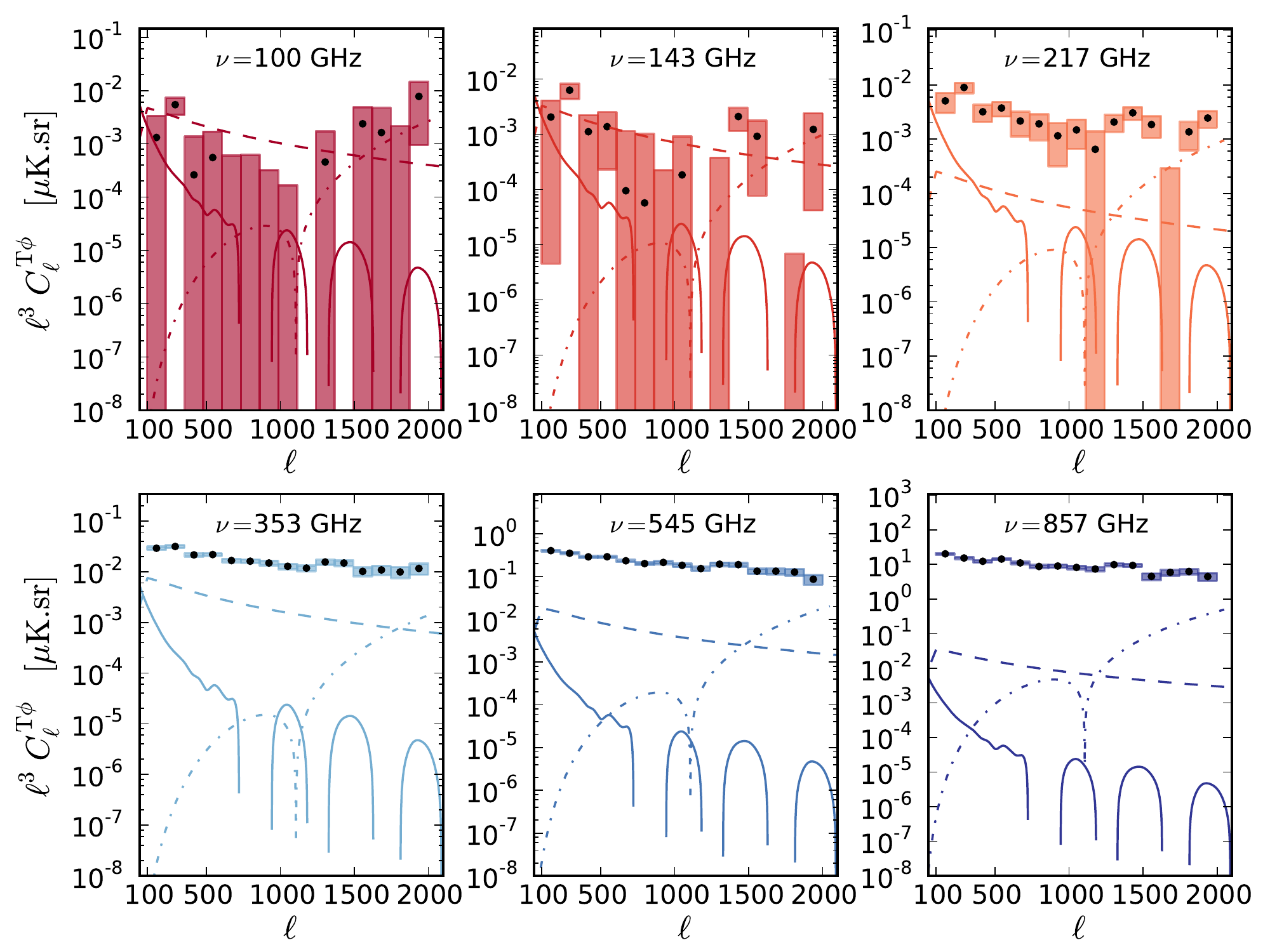}
  \caption[]{Foreground components at each frequency. The data points and
    error bars show our results. The dashed line is an estimated upper limit
    on the magnitude
    of the SZ contamination derived in Sect.~\ref{sec:sz_cont}. We show
    the absolute value of this contribution,  which is negative at
    frequencies less than 217\,GHz. The  dot-dashed line is the
    extragalactic point source contribution, with an
    amplitude measured from our data as derived in
    Sect.~\ref{sec:pointsources}. Again we show the absolute value,
    with the signal being negative below $\ell \sim 1200$. The oscillating
    solid line corresponds to the calculated ISW
    contamination.} 
  \label{fig:astro_contamination}
\end{figure*}

\begin{figure}[!t]
  \centering
  \includegraphics[width=88mm]{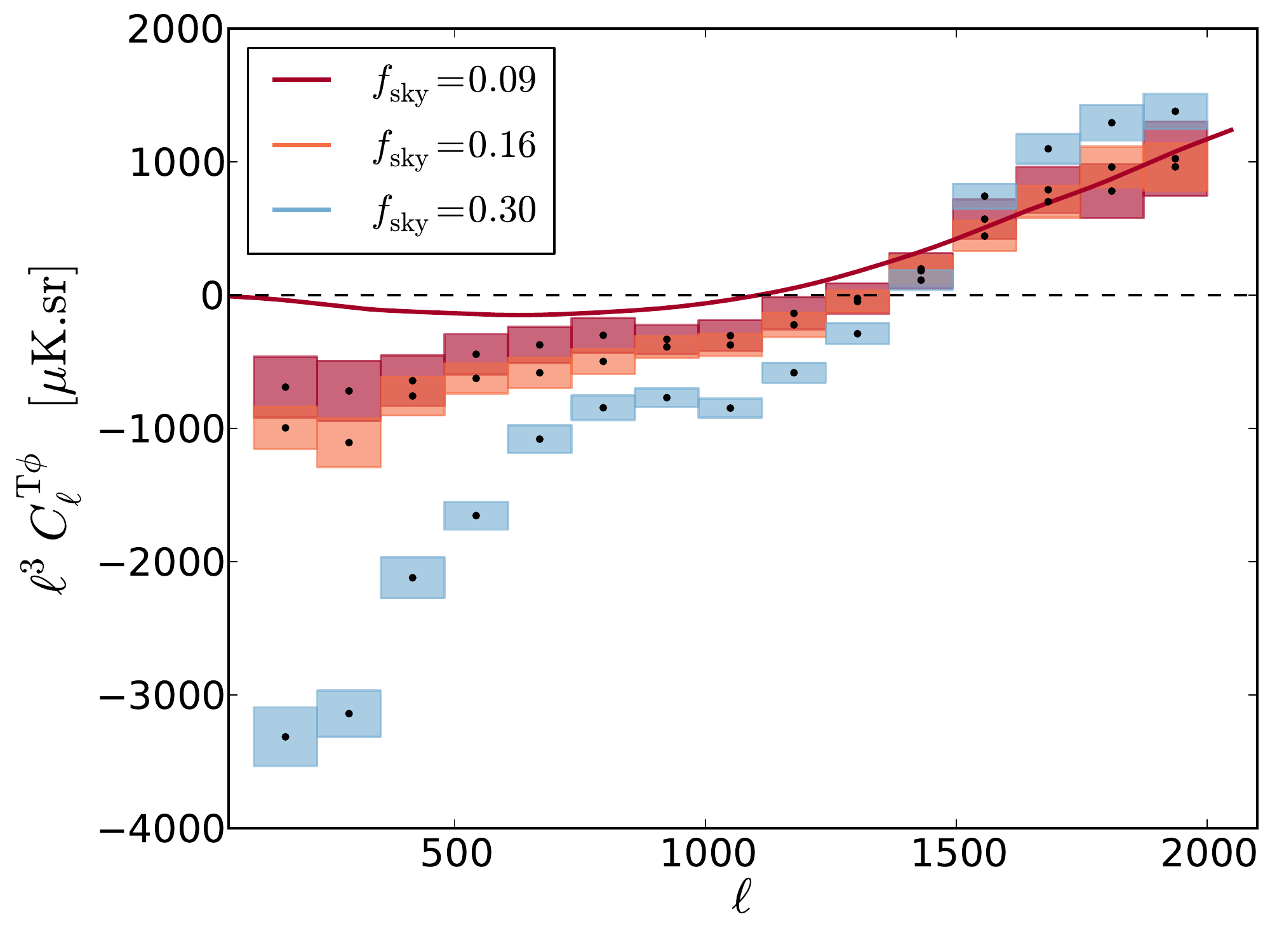}
  \caption[]{Cross-spectrum of the 545\,GHz lens reconstruction
    correlated with the 545\,GHz temperature map with different Galactic masks.
    The legend gives the visible sky fractions.
    The solid line represents the analytic unclustered shot-noise contribution
    fit to the $f_{\rm sky}=0.09$ points above $\ell=1300$.}
  \label{fig:cib_bspec}
\end{figure}

Having calculated the bias from the point source shot noise in
Sect.~\ref{sec:pointsources}, we now discuss a more complicated form of the
unresolved point source 3-point function that could be
present in our data, namely the clustering contribution.  While
unknown (although the first detection was recently reported in
\citet{Crawford:2013uka}), the CIB bispectrum is potentially a direct
contaminant to our measurement. Because of the non-linear clustering of DSFGs (PER), it
has to exist. But because of the very large redshift kernel that
characterizes the CIB, this non-Gaussian effect will be washed out,
reducing its importance.  Nevertheless, we ought to study it carefully.

If important, this effect would show up as a departure of the data from the
best-fit curve in Fig.~\ref{fig:bspec_spec}, since the best-fit model
that we used assumes only a Poissonian shot-noise contribution. We do not
see any significant deviation in Fig.~\ref{fig:bspec_spec}. Still, in
order to isolate this effect we create cross-spectra with increased
sensitivity to the clustered point source signal.  We do this
by calculating the lens reconstruction at 100\,GHz and 545\,GHz, where,
respectively, the radio and dusty point source contribution is stronger. The
545\,GHz map has a much larger Galactic dust signal than our nominal
143\,GHz map. However, unlike in our fiducial estimates, here we do not
attempt to project out dust contamination from the map used to perform
our lens reconstruction 
as this would also remove some of the CIB signal in the bispectrum.
As was found in Sect.~\ref{sec:gal_fg} the cross-correlation between
the 100\,GHz reconstruction and the 100\,GHz temperature map does not
show any large difference with the cross-spectrum obtained using the
143\,GHz signal. We are thus not sensitive enough to detect
a bias from the clustering of radio sources using this method.
However, we do detect a strong cross-correlation between the raw
545\,GHz lens reconstruction and the 545\,GHz temperature map. This
cross-spectrum is shown in Fig.~\ref{fig:cib_bspec} for three
different Galaxy masks. The line shows the point source shot-noise
template derived in Sect.~\ref{sec:pointsources},
fit to the cross-spectrum with the
10\,\% Galaxy mask at $\ell$ above 1300. If the signal were entirely due to
extragalactic point sources, then the signal would be independent of
masking, and we do see a convergence of the signal at high $\ell$ as
the size of the Galactic mask
is increased. At low $\ell$, however, there is a large Galactic
contribution and the convergence with the reduced mask size is less
clear. We thus conclude that a strong contribution from Galactic dust is
present in this measurement at all $\ell$.

We do not attempt to calculate accurately the shape of the clustering
contribution to the CIB bispectrum here, since it is beyond the scope
of this work, even though a simple prescription for it has recently been proposed in
\cite{Lacasa:2011ej}. To separate the Galactic from non-Galactic
contributions in our bispectrum measurement is difficult,
even if a strong Galactic signal is clearly present, given the strong dependence of the signal
on variations of the Galactic mask in Fig.~\ref{fig:cib_bspec}.  However,
the combination of dust cleaning that we perform in our nominal pipeline,
coupled with the fact that our nominal pipeline uses the 143\,GHz map for
lens reconstruction, means that we do not observe any dependence with
masking in our measurement, as seen in Fig.~\ref{fig:null_test_545b}.
Because of this, the CIB bispectrum is unlikely to be a large contribution
to our measurement. Furthermore, even if we were to assume that all of the signal seen in
Fig.~\ref{fig:cib_bspec} was extragalactic in nature, using the \cite{gispert2000} frequency
scaling for the CIB (also appropriate for the Galactic dust in fact,
\citealt{planck2011-7.12}), the roughly $-1700\,\mu$K.sr observed at
$\ell=400$ for the 40\,\% Galactic mask would only lead to a
$-0.02\,\mu$K.sr signal in Fig.~\ref{fig:astro_contamination}, which is an order
of magnitude smaller than our measured signal. To conclude, although our
analysis does not lead to a clean measurement of the CIB bispectrum,
we can safely assume that it is not a contaminant to our measurement.  

\begin{table*}[!t] 
\begingroup
\newdimen\tblskip \tblskip=5pt
\caption{Cross-spectrum detection band-powers. All values are in units
  of $\mu$K.sr.  The extragalactic foreground contribution,
  $C_\ell^{\rm fore}$ has been removed from $C_\ell$. Both statistical
  and systematic errors are given (see Sect.~\ref{sec:systematics} for
  details).}
\label{tab:bandpowers}
\nointerlineskip
\vskip -3mm
\footnotesize
\setbox\tablebox=\vbox{
   \newdimen\digitwidth 
   \setbox0=\hbox{\rm 0} 
   \digitwidth=\wd0 
   \catcode`*=\active 
   \def*{\kern\digitwidth}
   \newdimen\signwidth 
   \setbox0=\hbox{+} 
   \signwidth=\wd0 
   \catcode`!=\active 
   \def!{\kern\signwidth}
\halign{\hbox to 0.4in{#}\tabskip 0.2em& \hfil#&
 \hfil#\hfil& \hfil#\hfil& \hfil#\hfil&
 \hfil#\hfil& \hfil#\hfil& \hfil#\hfil&
 \hfil#\hfil& \hfil#\hfil& \hfil#\hfil&
 \hfil#\hfil& \hfil#\hfil& \hfil#\hfil&
 \hfil#\hfil& \hfil#\hfil& \hfil#\hfil\tabskip 0pt\cr
\noalign{\doubleline}
\omit& $\ell_{\rm mean}$****& *163& *290& *417& !*543& !*670& !*797& !*923&
 !*1050& !*1177& !*1303& !*1430& !*1557& !*1683& !*1810& *1937\cr
\noalign{\vskip 3pt\hrule\vskip 3pt}
\underline{100\,GHz}& $\ell^3 C_\ell\times1000*$& *1.32& *5.44& *0.26& **0.54& *$-$0.46& *$-$0.37& *$-$0.64& *$-$0.69& *$-$1.72& **0.59& *$-$2.63& **2.96& **2.61& *$-$1.56& *10.01\cr
\omit& $\Delta\left(\ell^3 C^{\rm stat}_\ell\right)\times1000*$& *2.48& *2.15& *1.30& **1.33& **1.17& **1.27& **1.17& **1.12& **1.29& **1.51& **1.97& **2.66& **3.78& **5.69& **8.93\cr
\omit& $\Delta\left(\ell^3 C^{\rm sys}_\ell \right)\times10000$& *0.28& *1.14& *0.05& **0.12& **0.09& **0.07& **0.13& **0.14& **0.37& **0.10& **0.62& **0.50& **0.34& **0.65& **1.63\cr
\omit& $\ell^3 C^{\rm fore}_\ell\times10000$& *-0.00& $-$0.01& $-$0.03& *$-$0.08& *$-$0.16& *$-$0.25& *$-$0.28& *$-$0.14& **0.33& **1.31& **3.05& **5.79& **9.78& *15.29& *22.48\cr
\noalign{\vskip 3pt\hrule\vskip 3pt}
\underline{143\,GHz}& $\ell^3 C_\ell\times1000*$& *2.05& *6.28& *1.11& **1.37& **0.09& **0.05& *$-$0.60& **0.18& *$-$1.16& *$-$0.39& **2.21& **1.13& *$-$0.65& *$-$0.39& **2.05\cr
\omit& $\Delta\left(\ell^3 C^{\rm stat}_\ell\right)\times1000*$& *2.49& *2.16& *1.30& **1.32& **1.15& **1.23& **1.07& **0.94& **0.98& **0.93& **0.99& **0.99& **1.05& **1.22& **1.48\cr
\omit& $\Delta\left(\ell^3 C^{\rm sys}_\ell \right)\times10000$& *0.42& *1.30& *0.23& **0.28& **0.02& **0.01& **0.12& **0.04& **0.24& **0.09& **0.43& **0.19& **0.21& **0.20& **0.25\cr
\omit& $\ell^3 C^{\rm fore}_\ell\times10000$& *-0.00& *-0.00& $-$0.01& *$-$0.03& *$-$0.06& *$-$0.09& *$-$0.10& *$-$0.05& **0.12& **0.48& **1.13& **2.14& **3.61& **5.65& **8.31\cr
\noalign{\vskip 3pt\hrule\vskip 3pt}
\underline{217\,GHz}& $\ell^3 C_\ell\times1000*$& *5.08& *8.99& *3.19& **3.75& **2.15& **1.91& **1.15& **1.47& **0.66& **2.11& **3.15& **2.04& *$-$0.15& **1.84& **3.16\cr
\omit& $\Delta\left(\ell^3 C^{\rm stat}_\ell\right)\times1000*$& *2.49& *2.17& *1.31& **1.34& **1.16& **1.24& **1.07& **0.93& **0.97& **0.90& **0.95& **0.90& **0.89& **0.94& **1.02\cr
\omit& $\Delta\left(\ell^3 C^{\rm sys}_\ell \right)\times10000$& *1.07& *1.90& *0.68& **0.79& **0.46& **0.41& **0.24& **0.31& **0.14& **0.44& **0.65& **0.39& **0.10& **0.29& **0.52\cr
\omit& $\ell^3 C^{\rm fore}_\ell\times10000$& *-0.00& *-0.00& $-$0.01& *$-$0.03& *$-$0.05& *$-$0.08& *$-$0.09& *$-$0.04& **0.10& **0.41& **0.96& **1.83& **3.09& **4.83& **7.10\cr
\noalign{\vskip 3pt\hrule\vskip 3pt}
\underline{353\,GHz}& $\ell^3 C_\ell\times1000*$&*$\cdots$&*$\cdots$& 21.47& *21.79& *16.56& *16.08& *14.83& *12.76& *11.76& *15.60& *14.98& *10.44& *11.33& *10.67& *12.76\cr
\omit& $\Delta\left(\ell^3 C^{\rm stat}_\ell\right)\times1000*$&*$\cdots$&*$\cdots$& *1.88& **1.87& **1.75& **1.81& **1.79& **1.81& **1.96& **2.13& **2.34& **2.55& **2.78& **3.07& **3.44\cr
\omit& $\Delta\left(\ell^3 C^{\rm sys}_\ell \right)\times1000*$&*$\cdots$&*$\cdots$& *0.69& **0.70& **0.53& **0.52& **0.48& **0.41& **0.38& **0.50& **0.48& **0.33& **0.35& **0.32& **0.37\cr
\omit& $\ell^3 C^{\rm fore}_\ell\times10000$&*$\cdots$&*$\cdots$& $-$0.01& *$-$0.04& *$-$0.09& *$-$0.13& *$-$0.15& *$-$0.07& **0.17& **0.68& **1.59& **3.02& **5.10& **7.97& *11.72\cr
\noalign{\vskip 3pt\hrule\vskip 3pt}
\underline{545\,GHz}& $\ell^3 C_\ell\times100**$&*$\cdots$&*$\cdots$& 28.92& *29.05& *23.38& *20.12& *21.37& *18.32& *15.38& *19.36& *18.78& *12.94& *12.67& *11.70& **7.13\cr
\omit& $\Delta\left(\ell^3 C^{\rm stat}_\ell\right)\times100**$&*$\cdots$&*$\cdots$& *2.09& **1.97& **1.91& **1.87& **1.94& **2.00& **2.12& **2.31& **2.44& **2.60& **2.73& **2.87& **3.04\cr
\omit& $\Delta\left(\ell^3 C^{\rm sys}_\ell \right)\times100**$&*$\cdots$&*$\cdots$& *2.95& **2.96& **2.38& **2.05& **2.18& **1.87& **1.57& **1.98& **1.94& **1.36& **1.36& **1.30& **0.88\cr
\omit& $\ell^3 C^{\rm fore}_\ell\times1****$&*$\cdots$&*$\cdots$&*$\cdots$&*$\cdots$&*$\cdots$&*$\cdots$&*$\cdots$&*$\cdots$&*$\cdots$&*$\cdots$&*$\cdots$&*$\cdots$&*$\cdots$&*$\cdots$&*$\cdots$\cr
\noalign{\vskip 3pt\hrule\vskip 3pt}
\underline{857\,GHz}& $\ell^3 C_\ell\times1****$&*$\cdots$&*$\cdots$& 12.34& *14.32& *11.08& **8.73& **9.00& **8.19& **7.37& **9.85& **9.35& **4.42& **5.75& **5.99& **4.09\cr
\omit& $\Delta\left(\ell^3 C^{\rm stat}_\ell\right)\times1****$&*$\cdots$&*$\cdots$& *1.40& **1.27& **1.21& **1.14& **1.14& **1.14& **1.18& **1.25& **1.30& **1.36& **1.39& **1.43& **1.48\cr
\omit& $\Delta\left(\ell^3 C^{\rm sys}_\ell \right)\times1****$&*$\cdots$&*$\cdots$& *1.26& **1.46& **1.13& **0.89& **0.92& **0.83& **0.75& **1.01& **0.96& **0.46& **0.60& **0.64& **0.45\cr
\omit& $\ell^3 C^{\rm fore}_\ell\times1****$&*$\cdots$&*$\cdots$&*$\cdots$&*$\cdots$&*$\cdots$&*$\cdots$&*$\cdots$&*$\cdots$&*$\cdots$&*$\cdots$&*$\cdots$&*$\cdots$&*$\cdots$&*$\cdots$&*$\cdots$\cr
\noalign{\vskip 3pt\hrule\vskip 3pt}}}
\endPlancktablewide 
\endgroup
\end{table*}

\subsection{Final statistical and systematic error budget}

Throughout the suite of tests for instrumental and observational
systematic errors presented in Sect.~\ref{sec:inst_and_syste}, as well as the
suite of tests for possible astrophysical contaminants presented in
Sect.~\ref{sec:astro_contam}, we have established the robustness of our
measurement. The fact that our consistency tests do not lead
to any significant deviation gives us confidence in our error
budget. As described in Sect.~\ref{sec:inst_and_syste} we add  to them
an overall calibration uncertainty, beam uncertainty, and lens  
normalization uncertainty, consistent with the \Planck\ data processing
paper \citep{planck2011-1.7}. We gather the measured band-powers in
Table~\ref{tab:bandpowers},
along with our statistical and systematic errors. These band-powers have been
corrected for the point source component measured in
Sect.~\ref{sec:pointsources}, whose amplitude is also given in
Table~\ref{tab:bandpowers}. 

Once all systematic effects are factored in, we claim a detection
significance of
3.6 (3.5), 4.3 (4.2), 8.3 (7.9), 31 (24), 42 (19), and 32 (16) $\sigma$
statistical (statistical and systematic)
at 100, 143, 217, 353, 545 and 857\,GHz, respectively.

\section {Interpretation and discussion}
\label{sec:interpretation}

The correlation we have investigated leads to a very strong signal at most
HFI frequencies. After a thorough examination of possible
instrumental and astrophysical origins, we interpret it as originating from
the spatial correlation between the sources of the CIB and the matter
responsible for the gravitational deflection of CMB photons. In this 
section, we build on this result and interpret the
measurement using both angular and frequency information. 

Before doing so, we highlight the spectral information contained in the
signal. It was shown in PER that both the frequency spectrum of the CIB
mean and fluctuation rms are well approximated by the two modified
blackbody spectra proposed by \cite{fixsen1998} and \cite{gispert2000},
with a slight preference for the latter. We expect our measurements to
follow the same spectral energy distribution (SED). Following the procedure
outlined in Sect.~\ref{sec:sz_cont}, we plot in
Fig.~\ref{fig:sz_spectra} the best-fit CIB component with either a
\cite{fixsen1998} -- dot-dashed black line -- or \cite{gispert2000} -- solid
black line -- SED. We do so for two $\ell$-bins. We can see that, indeed,
for a given $\ell$-bin, our measurements qualitatively follow the
expected CIB spectrum. Unlike PER, we do not find a preference for the
\cite{gispert2000} shape in our low $\ell$-bin and only a slight one
in the high $\ell$ bin. It is worth emphasizing that by carrying out 
a cross-correlation measurement we can obtain constraints at the lowest 
frequency, which is usually heavily contaminated by the CMB (see PIR for 
discussion). This is particularly interesting, because these measurements
are simultaneously the most sensitive to high-$z$ star formation processes and
the most discrepant with either of the SEDs, i.e., they are both
systematically low by about 0.5$\,\sigma$. The models presented in this
section will allow us to use both the spectral dependence and the relative
amplitudes of the $\ell$-bins that was lost in Fig.~\ref{fig:sz_spectra}.
We now describe the general methodology we will use, before describing our
models in detail.

\subsection{Model comparison methodology}
\label{sec:modeling}

For the purpose of model fitting, we will utilize both the CIB-lensing
cross-spectra measured in this paper and the CIB auto-spectra obtained
from PIR.  We use the CIB-lensing cross-spectra for two purposes: to improve
constraints on the model parameters; and
to provide a consistency test of models fit to the CIB auto-
and frequency cross-spectra alone. As will be seen in PIR, the
cross-spectra of the CIB at different frequencies provide powerful
constraints on the CIB emissivity.

We use the log-likelihood,
\be
\ln{\mathcal{L}(p)} = -\frac{1}{2} \sum_{\ell} \sum_{ij}
 \left[ \hat{C}^i_{\ell} - \tilde{C}^i_{\ell}(p) \right]
 \left(N^{ij}_{\ell}\right)^{-1}
 \left[ \hat{C}^j_{\ell} - \tilde{C}^j_{\ell}(p) \right]^{\dagger},
\label{eqn:log_likelihood}
\ee
where $\hat{C}$ and $\tilde{C}$ are the data and theory spectra with
parameters $p$, the $i$ and $j$ indices denote the type of spectra
(e.g., $100$\,\GHz\,$\times$\,$\phi$ or $100$\,\GHz\,$\times\,100$\,\GHz),
and $N$ is the covariance matrix that includes both statistical and
systematic errors. We make the approximation that the covariance matrix is
diagonal, i.e., we treat the errors for different bins of each auto- and
cross-spectrum as being uncorrelated. The small ($\simeq 2\,\%$) mask-induced
mode-coupling between neighbouring bins supports this
approximation. However, calibration and beam errors (which are
correlated between the auto- and cross-spectra at a given frequency),
as well as the lens normalization
error (which is also correlated across spectra) are not accounted for
in this approximation. In addition, the lens reconstruction has some
sensitivity to all modes of the temperature maps, and so different
$\phi$ modes are correlated to some degree. We also neglect the fact
that the contribution to the error from the CIB signal itself
(the orange line in Fig.~\ref{fig:error_by_l})
is also substantially correlated from frequency to frequency.  However,
our evaluation using simulations suggests that these
effects are too small to significantly affect our procedure. We thus
resort to simply adding the beam, calibration and normalization
uncertainties in quadrature to the statistical errors. The posterior
probability distributions of model parameters are determined using now
standard Markov Chain Monte Carlo techniques \citep[e.g.,][]{Knox:2001fz,Lewis:2002ah}. 

\subsection{Two modelling approaches}
\label{sec:modeling_app}

\begin{figure}[!t]
  \centering
  \includegraphics[width=88mm]{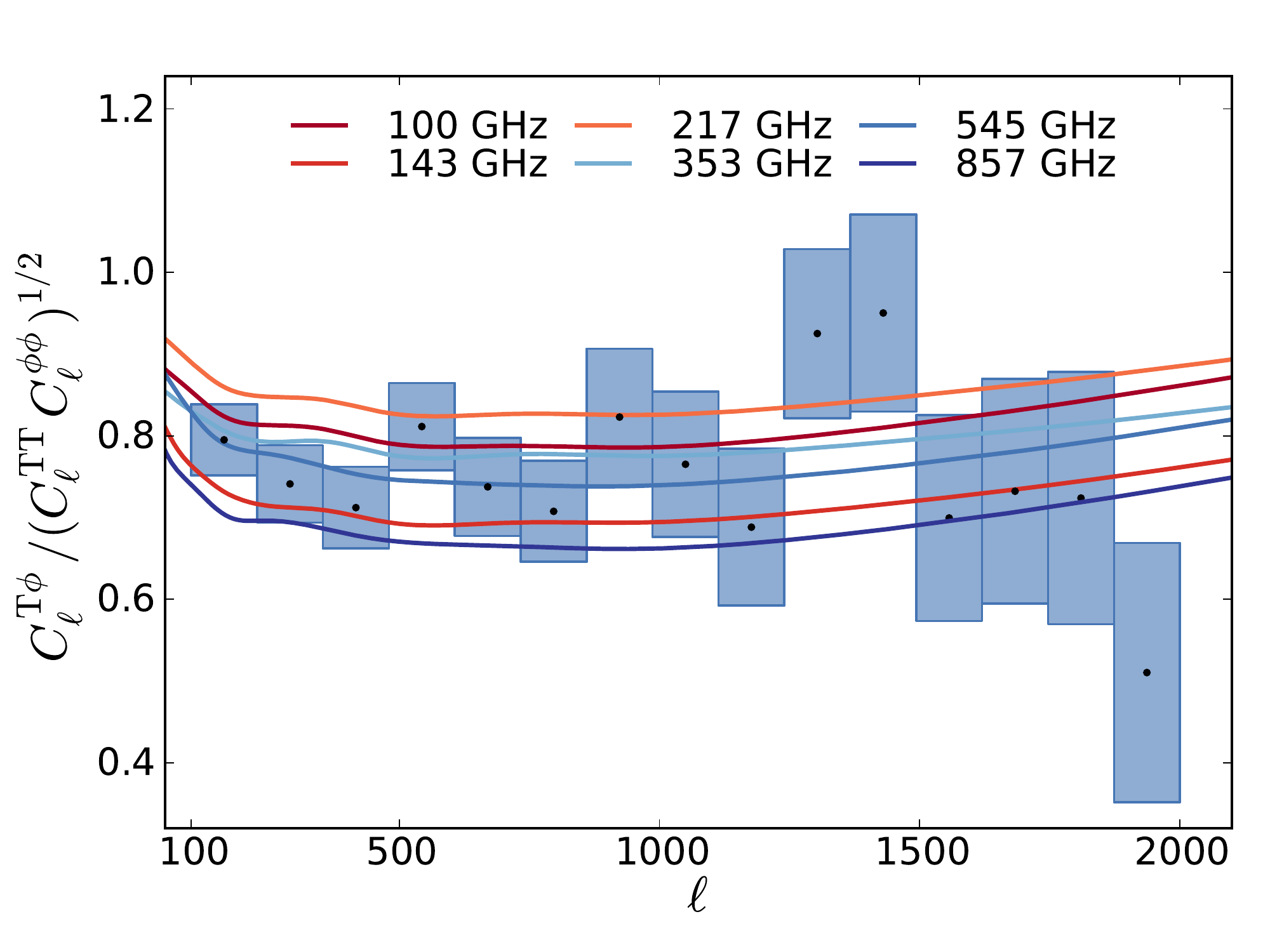}
  \caption[]{Cross-correlation coefficients calculated from the model
    $\phi$ spectrum and best-fit halo model at each frequency. 
    The CIB is a spectacular tracer of CMB lensing, and
    vice-versa. The data points represent the measured
    cross-correlation divided by the best-fit auto power spectra models at 545~GHz. }
  \label{fig:rcib_phi}
\end{figure}

The strength of the correlation signal should come as no surprise, given our
current knowledge of CMB lensing and the CIB. The PER model predicts
a high correlation between the CIB and the lensing potential. As
clearly illustrated in Fig.~\ref{fig:clcc_clpp_redshifts.pdf}, the
broad overlap of the redshift distributions of the CIB with the lensing kernel
peaking at $z \approx 2$--3 leads to a correlation of 60--80\,\%.
This is comparable to our measurements
at all of the HFI frequencies, as illustrated in Fig.~\ref{fig:rcib_phi},

In models of the cross-correlation, the underlying properties we can probe
come from a combination of the lensing potential and the characteristics of the DSFGs, in
particular their emissivity and clustering properties. Mostly driven
by linear physics, the former is well understood theoretically,
as confirmed by recent observations \citep{Smith:2008an,Hirata:2008cb,Das:2011ak,vanEngelen:2012va}. Assuming
the currently favored $\Lambda$CDM cosmology, we can consider it to be known to better than 
10\,\% in the multipole range of interest to us, an uncertainty
dominated by the uncertainty in the normalization of the primordial power
spectrum. Given that this is much smaller than the theoretical
uncertainties related to DSFGs, we will fix the cosmology to the
currently favoured \Planck~alone flat $\Lambda$CDM model in
\cite{planck2013-p11}, and will focus our analysis on the modelling of the DSFGs. 

At a given redshift we model the fluctuations in the mean CIB emission,
$\bar{j}$, as being proportional to the fluctuations in the number of
galaxies, $n_{\rm g}$ \citep{Haiman:1999hh},
\be
\delta j \propto \bar{j}\ \frac{\delta n_{\rm g}}{n_{\rm g}}\ .
\ee
With this hypothesis, the goal of the CIB modelling becomes twofold:
first, to better understand the statistical properties of the dusty
galaxy number density, $\delta n_{\rm g}$; and second, to reconstruct the mean
emissivity of the CIB as a function of redshift. 

In this paper we will use two different models of the CIB
emission. The first model (described in Sect.~\ref{sec:model_linbias} and inspired by \citealt{Hall:2009rv}) uses a
single bias parameter at all frequencies with the mean CIB emissivity
modelled as a two parameter Gaussian. This model is not designed to be
physically realistic, and furthermore we will marginalize over an arbitrary
amplitude in this case.  Nevertheless, we present this simple model
to show that our measurements are
quite consistent with broad expectations of the CIB. The second model,
described in Sect.~\ref{sec:model_independent}, is a natural extension
of the Halo Occupation Density (HOD) approach used in PER
\citep[see also][and references therein]{penin2011b}. But unlike the results
obtained in PER we now use a single HOD to describe the spectra at all
frequencies. This is possible by allowing for deviations from the
\cite{bethermin2011} model (hereafter B11) that was used to fix the emissivity.  

\subsubsection{Linear bias model}
\label{sec:model_linbias}

\begin{figure}[!t]
  \centering
  \includegraphics[width=88mm,clip=true,trim=0cm 0 0 0]{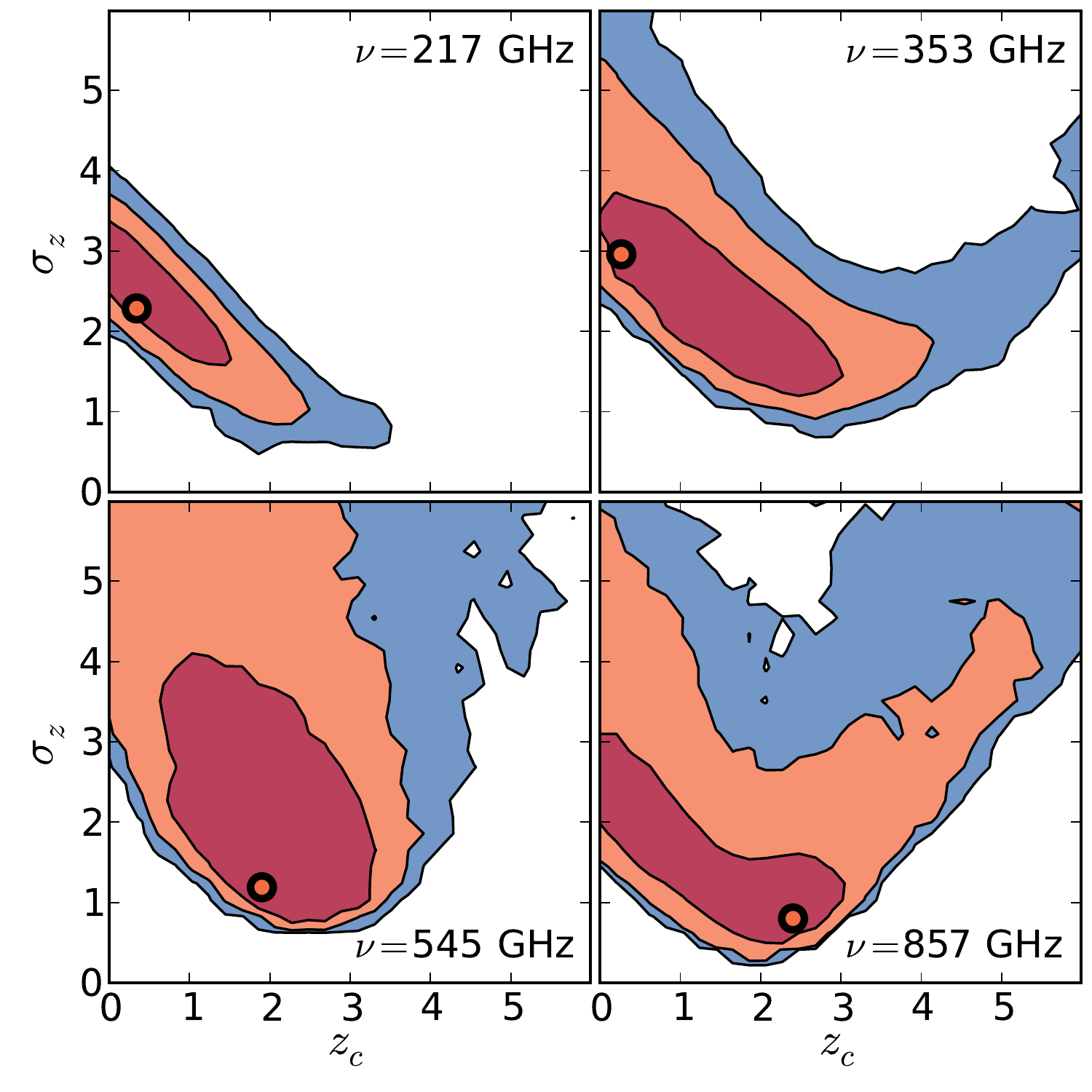}
  \caption[]{Marginalized 2-D distributions of $z_{\rm c}$ and
    $\sigma_z$ for the linear bias model,
    fit to all frequencies simultaneously.
    The orange dots indicate the parameter values at the minimum $\chi^2$.}
 \label{fig:hall_bias}
\end{figure}

As a first pass at interpreting our measurement, we will consider a
redshift independent linear bias model with a simple parametric
SED. This model was found to provide a reasonable fit to the
auto-spectra in the linear regime in PER.
Throughout this paper we use the Limber approximation,
and in this section, since we are
using a linear model, we write the relevant angular spectra as
\be
C_{\ell}^{X Y} = \int_0^{\chi_{\ast}} d\chi \; W^X(\chi) W^Y(\chi)
 \; P_{\delta \delta}(k=\ell/\chi, \chi),
\label{eqn:hall_spec}
\ee
where $X$ and $Y$ are either the CIB at frequency $\nu$ or the lensing
potential $\phi$, the integral is over $\chi$, the comoving distance along
the line of sight, $\chi_{\ast}$ is the comoving distance to the last
scattering surface, $P_{\delta \delta}(k,\chi)$
is the matter power spectrum at distance $\chi$, and the $W^X$
functions are the redshift weights for each of the signals $X$:
\be
\begin{split}
  W^{\nu}(\chi)  &= b \frac{a \bar{j}_{\nu}(\chi) }{\chi}; \\
  W^{\phi}(\chi) &= - \frac{3}{\ell^2} \Omega_m H_0^2 \frac{\chi}{a}
   \left(\frac{\chi_{\ast}-\chi}{\chi_{\ast}\chi}\right).
\end{split}
\ee
Here $b$ is the DSFG bias that we assume to be redshift
independent, $a$ is the scale factor, $\bar{j}_{\nu}(\chi)$ is the
mean CIB emissivity at frequency $\nu$, as defined in PER, $\Omega_m$
is the matter density today in critical density units and $H_0$ is the
Hubble parameter today. We use the \citet{Hall:2009rv} model for the
CIB kernel, which treats the  CIB emissivity as a Gaussian in redshift:
\be
\label{Eq:cib_def}
\bar{j}_{\nu}(\chi) \propto a \; \chi^2
 \exp{\left[-(z-z_{\rm c})^2 / 2 \sigma_z^2 \right]} \; f_{\nu (1+z)}
\ee
where we use a modified blackbody frequency dependence
\be
f_{\nu(1+z)} \propto \nu^{\,\beta} B_{\nu}(T_d).
\ee
We fix the dust temperature to $T_{\rm d} = 34$\,K, the spectral
index to $\beta=2$ \citep{hall2010},
and assume a constant bias $b$. We include a single normalization
parameter for $j$, which we marginalize over. Since the normalization
and bias parameters are degenerate in Eq.~\ref{eqn:hall_spec},
if we were to only use the measured auto- and cross-spectra
this approach would be equivalent to marginalizing over a frequency
independent bias parameter. However, we will further constrain our model
using the FIRAS data, which breaks this degeneracy.
We constrain the $z_{\rm c}$ and $\sigma_z$ parameters at each frequency,
giving us a total of 13 free parameters.

For 217--857\,GHz, we use the FIRAS measurements of the CIB mean intensity from
\cite{lagache2000} as an additional constraint to our model.
The mean intensity is simply
\be
I_{\nu} = \int_0^{\chi_{\ast}} d\chi \; a \bar{j}_{\nu}(\chi) \ .
\label{eqn:firas_model}
\ee
Using this equation and the measured FIRAS mean and uncertainty
we calculate a $\chi^2$ value
and add this to the $\chi^2$ in Eq.~\ref{eqn:log_likelihood}.
Since there are no FIRAS constraints at 100 and 143\,GHz,
as well as no CIB auto-spectra measurements, and
noisier cross-spectra measurements at these frequencies,
our constraints for the 100 and 143\,GHz redshift parameters are weaker than
for the other parameters. 

\begin{figure}[!t]
  \centering
  \includegraphics[width=88mm, clip=true, trim=0.3cm 0 0 0]{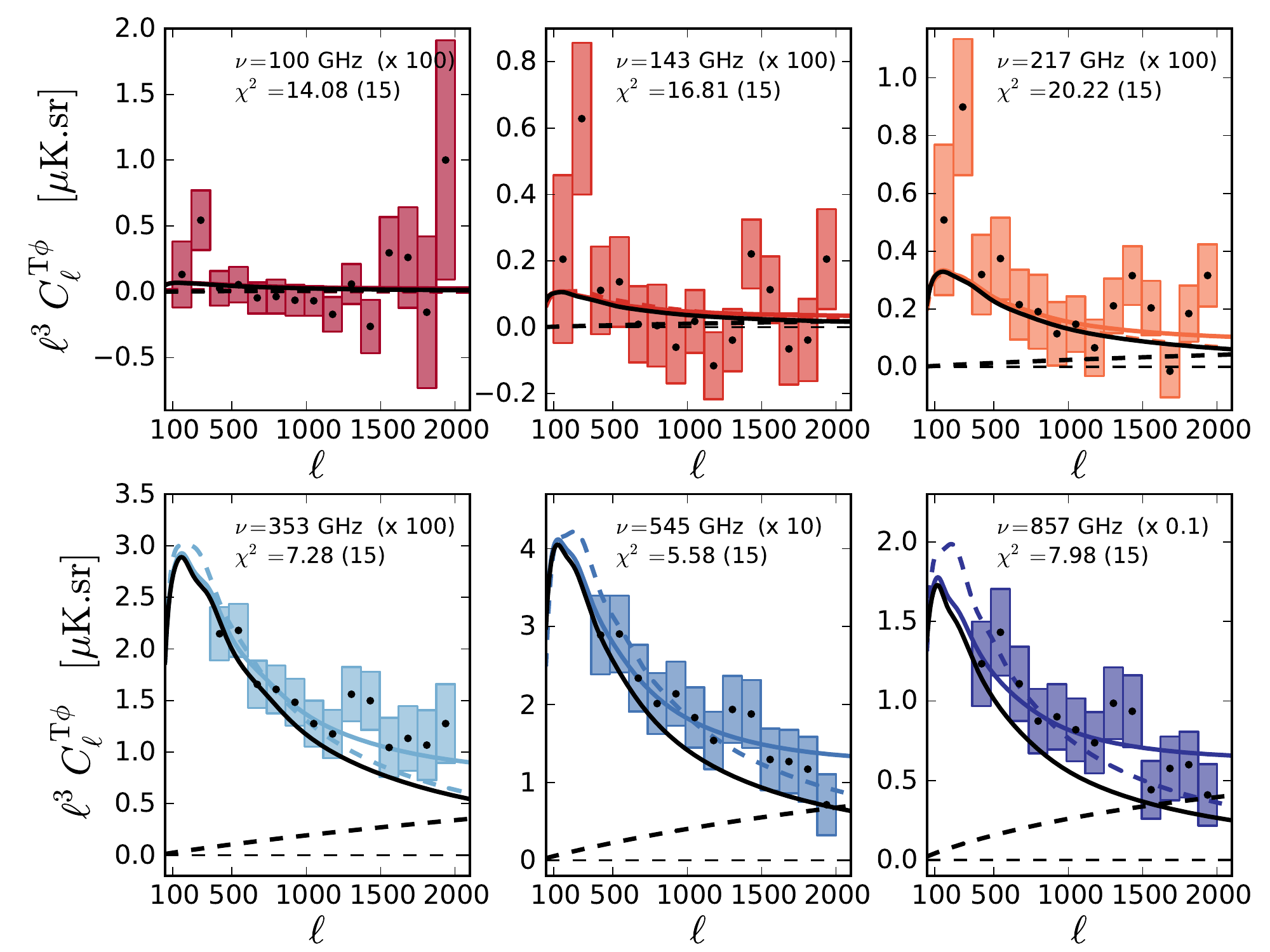}
  \caption[]{Measured cross-spectra with the best-fit $j$ reconstruction model
    fit to both the CIB auto- and CIB-lensing cross-spectra (solid coloured),
    and the best-fit linear bias model (dashed coloured).
    The $\chi^2$ values quoted in each panel are the contribution to the
    global $\chi^2$ from the data in the panel for the halo model, and
    loosely indicate the goodness of fit (see text for details).
    The one and two-halo contributions are shown as the dashed and solid
    black lines, respectively. A light dashed black horizontal line is indicating the zero level.}
  \label{fig:cross_fit}
\end{figure}
\begin{figure}[!t]
  \centering
  \includegraphics[width=88mm]{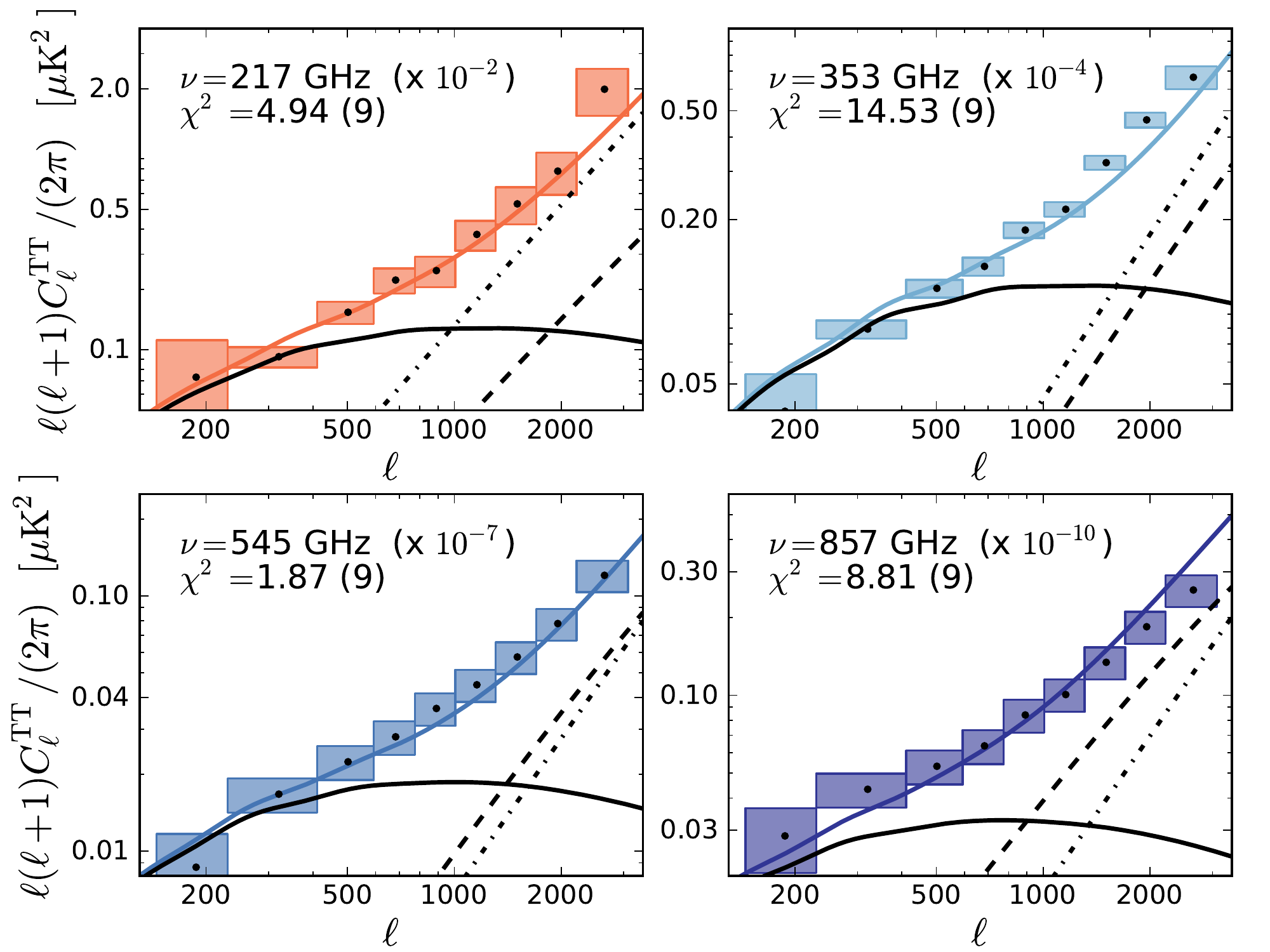}
  \caption[]{PIR auto-spectra with the best-fit mean emissivity $j$ reconstruction
    model fit for the CIB auto and CIB-lensing cross-spectra (solid
    coloured). The $\chi^2$ values are defined as in
    Fig.~\ref{fig:cross_fit}. The one and two-halo contributions are
    shown as the dashed and solid black lines, respectively, while
    shot noise is the dot-dashed black line.}
  \label{fig:autos_fit}
\end{figure}

The linear bias model considers only linear clustering, and so when
fitting the auto-spectra we restrict ourselves to $\ell<500$,
where non-linear contributions are negligible.
Because we do not consider the high-$\ell$ modes, we also neglect
the shot-noise contribution to the auto-spectra.
The best-fit model is shown as the coloured dashed lines
in Fig.~\ref{fig:cross_fit}, with $\chi^2$ values of
13.4, 16.8, 25.2, 21.8, 9.1, and 9.4 if we break up the $\chi^2$
contribution per frequency from 100 to 857\,GHz, leading to an overall
$\chi^2$ of 95.7 for $N_{\rm dof}=59$.  We see that the model captures
some features of the data, but we also have evidence it is
significantly missing some as well. This is perhaps not surprising
given the simplicity of the model. The two-dimensional marginal
distributions of  the $z_c, \sigma_z$ parameters are  shown in
Fig.~\ref{fig:hall_bias}. Although we allowed for these parameters to
be frequency dependent we note that the point $z_c=1$ and
$\sigma_z=2.2$ is in a region of high probability at all frequencies,
and gives a redshift distribution for the emissivity density roughly
consistent with our expectations, rising toward $z=1$ due to the
$\chi^2$ term and then only slowly falling off toward higher redshifts.

Althought it is useful to see the extent to which such a simple model
can explain our data, we now turn to make a stronger connection
between the properties of the infrared light and the distribution of
the underlying dark matter applicable into the non-linear regime. 

\subsubsection{An extended halo model based analysis}
\label{sec:model_independent}

In this section we use a description of the CIB motivated by the halo
model, which has been used successfully to describe the transition between the
linear and non-linear clustering regimes for optical galaxies.
We use the halo model to
attempt to reconstruct the CIB emissivity as a function of
redshift. This is an extension of the approach taken in PER, where the
modelled CIB emissivity at high redshift was treated as a single bin with
the amplitude constrained by the data.
The goal of this approach is to isolate the high-redshift
contribution to the CIB, which is difficult to probe using observations
of individual galaxies, due to their low brightness. The power of such
an approach is further demonstrated in PIR. 

We replace the linear bias used in Sect.~\ref{sec:model_linbias} with
a halo model and an HOD prescription that assigns galaxies to
host dark matter halos (see PER for references and definitions). It allows a
consistent description of the linear and non-linear part of the galaxy
power spectrum and its redshift evolution.
Because it is built on the clustering of dark matter halos, the halo model
allows us to describe the clustering of DSFGs and the gravitational lensing
caused by the halos in a consistent way.  However, it is important to realise
that the HOD prescription was developed to describe stellar mass within dark
matter halos -- an application for which it has been thoroughly tested --
while here we are applying it to star formation within halos. The
accuracy of this approach needs to be further quantified. However,
it provides a good phenomenological description of our data as well as
other CIB measurements, but also of other  astrophysical probes of the
relation between dark matter and light (e.g., \citet{Leauthaud:2011gw,Hikage:2012zk}). 

Unlike the model presented in PER we use a single HOD to
describe our data at all frequencies. This is made possible by
allowing for a deviation from the B11 emissivity model.
Note however that we will still consider the CIB emissivity to
depend only on redshift and not on the galaxy 
host halo mass, a simplification highlighted in \citet{Shang:2011mh}
that will be relaxed in the PIR model. The emissivity of
the CIB is inhomogeneous, due to spatial variations in the number density
of galaxies:
\be
\frac{\delta j_{\nu}}{\bar{j}_{\nu}}(\n,z) =
 \frac{\delta n_{\rm g}}{\bar{n}_{\rm g}}(\n,z) \equiv \delta_{\rm g} (\n,z).
\ee
Here $j(\n,z)$ is the CIB emissivity at redshift $z$ with mean $\bar{j} (z)$,
$n_{\rm g}(\n,z)$ is the number
density of DSFGs with mean $\bar{n}_{\rm g} (z)$,
and $\delta_{\rm g}(\n,z)$ is the DSFG overdensity, with power spectrum
$\left< \delta_{\rm g}({\bf k},z) \, \delta_{\rm g}({\bf k}',z)^{\ast} \right>
 = (2\pi)^3 \, \delta({\bf k-k}') \, P_{\rm gg}(k,z)$.
We calculate this power spectrum, including the constituent 1 and 2-halo terms,
using the procedure described in appendix~C of PER, with the constraint
$\alpha_{\rm sat}=1$, a theoretically favoured value \citep{Tinker:2009mx}.
We remove the relationship between $M_{\rm sat}$, a characteristic satellite
mass scale, and $M_{\rm min}$, the halo mass at which a halo has a
50\,\% probability of containing a central galaxy that was enforced in
PER (i.e., $M_{\rm sat}=3.3 M_{\rm min}$), and allow both $M_{\rm sat}$
and $M_{\rm min}$ to vary independently.

\begin{figure}[!t]
  \centering
  \includegraphics[width=88mm,clip=true,trim=0cm 0 0 0]{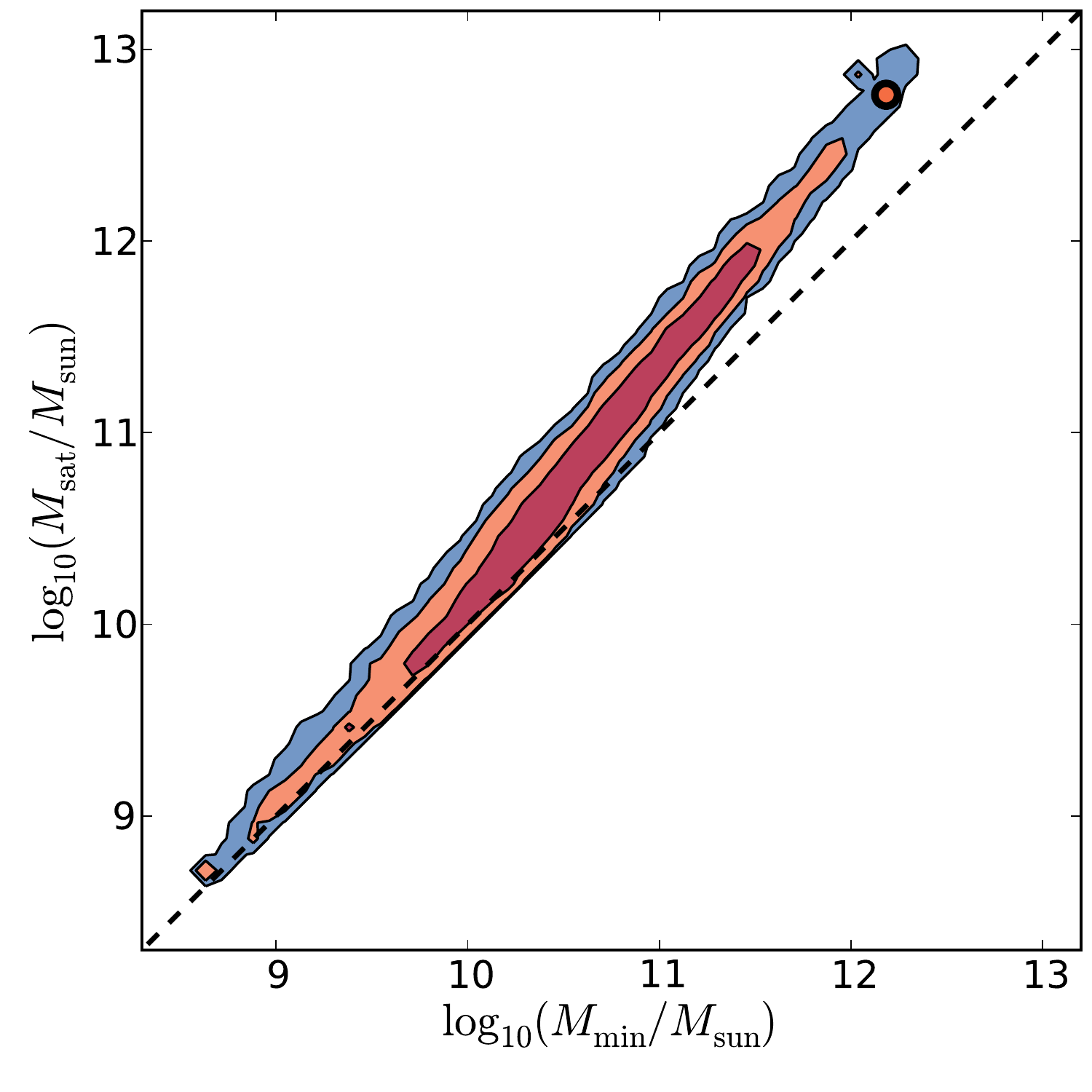}
  \caption[]{Marginalized 2-D distribution of
        $\log_{10}{\left(M_{\rm min}/\Msolar\right)}$
        and $\log_{10}{\left(M_{\rm sat}/\Msolar\right)}$ for our overall
	HOD model when the CIB-lensing cross-spectra are combined with the CIB
	auto-spectra and FIRAS measurements. The orange dot shows the
        best-fit value. The contours correspond to 68\%, 95\% and
        99.7\% confidence intervals.}
\label{fig:mmin}
\end{figure}

At redshift $z < 1$ we fix the emissivity to the B11 value, but
at higher redshift we assume that the emissivity is constant within
$z$-bins and solve for the amplitude of the bins.
Two factors affect the number of bins that we choose.
The auto-spectra  have a $\bar{j}^2$ dependence, and so if the true
value of $\bar{j}$ has a strong $z$ dependence within a bin then the
best-fit emissivity in the bin will be difficult to interpret.
The best-fit bin values could be significantly different from those that
would be obtained by binning the true emissivity.
However, as more bins are used and the number of parameters increases,
it becomes more difficult to determine the best-fit parameters and the
parameters will be highly correlated.  After investigation using simulations,
we found that three bins was a good compromise, given the expected slow
redshift evolution.
The bins are defined by: $1< z\le1.5$; $1.5<z\le3$; and $3<z\le7$.
As in Sect.~\ref{sec:model_linbias} we use the FIRAS results at
217--857\,GHz to add an integral constraint on the emissivity.
The CIB auto and lensing cross-spectra are \citep{Song:2002sg}:
\be
\begin{split}
\label{eqn:cibcib}
C_\ell^{\nu \nu'} &= \int d\chi \; W^{\nu}(\chi) W^{\nu'}(\chi) \;
 P_{\rm gg}(k=\ell/\chi,\chi); \\
C_\ell^{\nu \phi} &= \int d\chi \; W^{\nu}(\chi) W^{\phi}(\chi) \;
 P_{\delta {\rm g}}(k=\ell/\chi,\chi)\ .
\end{split}
\ee
Since we fix $\bar{j}$ at $z < 1$, the model spectra consist of a low
redshift part that is independent of the emissivity parameters, and a
contribution from $z > 1$ that is
proportional to $\bar{j}_{\nu} \bar{j}_{\nu'}$ for the auto-spectra 
and $\bar{j}_{\nu}$ for the lensing cross-spectra.

\begin{figure}[!t]
  \centering
  \includegraphics[width=88mm,clip=true,trim=0.8cm 0 0.3cm 0]{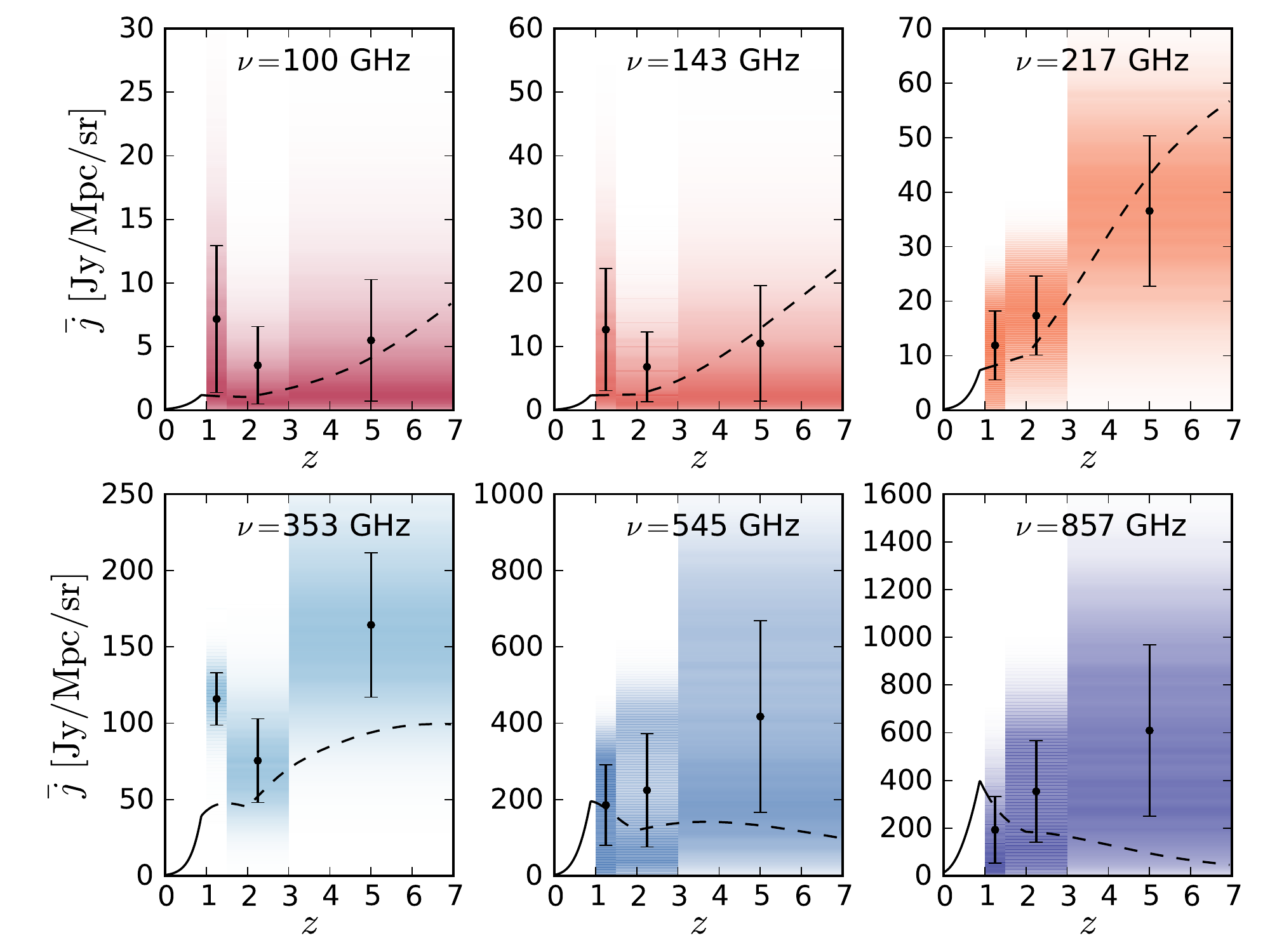}
  \caption[]{Reconstructed mean emissivity, $\bar j$, for each frequency as a
    function of redshift. The solid line at low $z$ and the dashed
    line at higher $z$ correspond to the B11 model. The B11 emissivity
    model at $z>1$ is not used, and is shown only for reference. The
    black error bars correspond to the 68\% C.L. while the color
    shading display the full posterior distribution.}
  \label{fig:jrecon}
\end{figure}

Overall, the halo-based model contains two halo parameters that describe
the galaxy clustering and are independent of frequency,
and three $j$ amplitudes at each frequency, giving a total of 20 parameters
for the six frequencies of interest to us.
The auto and cross-spectra have a total of 120 $\ell$-bins, with four additional FIRAS data points.
Solving for the likelihood described in
Sect.~\ref{sec:modeling}, gives the best-fit models shown
in Figs.~\ref{fig:cross_fit} and \ref{fig:autos_fit} as solid lines.
The $\chi^2$ values in each panel are the contribution to the total $\chi^2$
from the data within the panel.
The combined reduced $\chi^2$ is 102.1 for $N_{\rm dof}=104$,
indicating a good fit. 
The constraints we find on $M_{\rm sat}$ and $M_{\rm min}$ are shown in
Fig.~\ref{fig:mmin}.  We force $M_{\rm sat} \ge M_{\rm min}$ in the MCMC
fitting procedure,
with the dashed line in Fig.~\ref{fig:mmin} showing equality.
The red cross corresponds to the parameter values that give the
minimum $\chi^2$ in the fit, and are
$\log_{10}{\left(M_{\rm min}/\Msolar \right)} = 12.18$ and
$\log_{10}{\left(M_{\rm sat}/\Msolar \right)} = 12.76$,
which gives $M_{\rm sat}/M_{\rm min} = 3.80$.
The mean parameter values are
$\log_{10}{\left(M_{\rm min}/\Msolar \right)} = 10.53 \pm 0.62$ and
$\log_{10}{\left(M_{\rm sat}/\Msolar \right)} = 10.80 \pm 0.74$.
The best-fit value of $M_{\rm min}$ is consistent with those derived in PER
at multiple frequencies, even though we now set $\alpha_{\rm sat}=1$
and reconstruct the mean emissivity as a function of redshift.
The associated mean emissivity parameters
are given in Table~\ref{tab:jval} and displayed in Fig.~\ref{fig:jrecon},
where we also plot the B11 model for reference.
As can be seen in Fig.~\ref{fig:jrecon},
we remain consistent with the B11 model in most redshift bins.

\subsection{Interpreting the reconstructed emissivities}

We now illustrate one interesting consequence of this measurement and
show how using the constrained emissivities, $j_\nu(z)$, we can
estimate the star formation rate (SFR) density at different
redshifts. Following \citet{penin2011b}, we begin by writing the
emissivity as an integral over the galaxy flux densities:  
\begin{equation}
j_\nu(z) = \left(a {d\chi\over dz}\right)^{-1}
  \int S_\nu \frac{d^2N}{dS_\nu \, dz} dS_\nu\ .
\end{equation}
Here $S_\nu$ is the flux density, and $d^2N/dS_\nu \, dz$ is the
number of galaxies per flux element and redshift interval. The galaxies
contributing to the CIB  can be divided into various populations (labelled
as $p$) based on the galaxy SED, e.g., according to galaxy
type or dust temperature: 
\begin{equation}
j_\nu(z) = \left(a {d\chi\over dz}\right)^{-1} \sum_{p}
 \int S_\nu \frac{d^2N_{p}}{dS_\nu \, dz} dS_\nu\ .
\label{eqn:jtypes}
\end{equation}
If we define $s_{\nu}$ as the flux density of an $L_{\rm IR}=L_\odot$
source with the SED of a given population, i.e.,
$S_{\nu} = s_{\nu}L_{IR}$ (with $L_{\rm IR}$ in units of ${\rm L}_\odot$),
then we can write Eq.~\ref{eqn:jtypes}
as \citep{penin2011b}: 
\begin{equation}
\label{lab2}
j_\nu(z) = \left(a {d\chi\over dz}\right)^{-1}{dV\over dz}\, \sum_{p} s_{\nu}
 \int L_{\rm IR} \frac{d^2N_{p}}{dL_{\rm IR} \, dV} dL_{\rm IR}.
\end{equation}
The contribution to the infrared luminosity density from a given population is
\begin{equation}
\rho_{{\rm IR},p} = \int L_{\rm IR}
 \frac{d^2N_{p}}{dL_{\rm IR} \, dV} dL_{\rm IR}.
\end{equation}
We assume a simple conversion between $L_{\rm IR}$ and the star formation
rate density, $\rho_{\rm SFR}$,
using the Kennicutt constant $K$ \citep{kennicutt1998}. Since by
definition $\rho_{\rm SFR} = K\sum_{p}\rho_{{\rm IR},p}$, we can
rewrite Eq.~\ref{lab2} as:  
\begin{equation}
j_\nu(z) = (1+z) \, \chi^2 \, \frac{\rho_{\rm SFR}}{K}
 \left ( \frac{\sum_{p} s_\nu \, \rho_{{\rm IR},p}}
  {\sum_{p}\rho_{{\rm IR},p}}\right ),
\end{equation}
where the final term in brackets is the effective SED of infrared
galaxies, which we write as $s_{\nu,\rm eff}$. We derive these SEDs
following the evolution model of \citet{Bethermin:2012a} using 
  \citet{Magdis:2012ys} templates. The construction of these effective
  SEDs will be explained in detail in future work. Finally, we obtain
  the conversion factor between mean emissivity and SFR density, 
\begin{equation}
\rho_{\rm SFR}(z) = \frac{K}{(1+z)\, \chi^2(z) \, S_{\nu,\rm eff}(z)}
\, j_\nu(z)\ .
\label{eqn:sfr_eqn}
\end{equation}
Using Eq.~\ref{eqn:sfr_eqn} we find the coefficients for each of the redshift
bins and frequencies used in Table~\ref{tab:jval}.

\begin{figure}[!t]
  \centering
  \includegraphics[width=88mm]{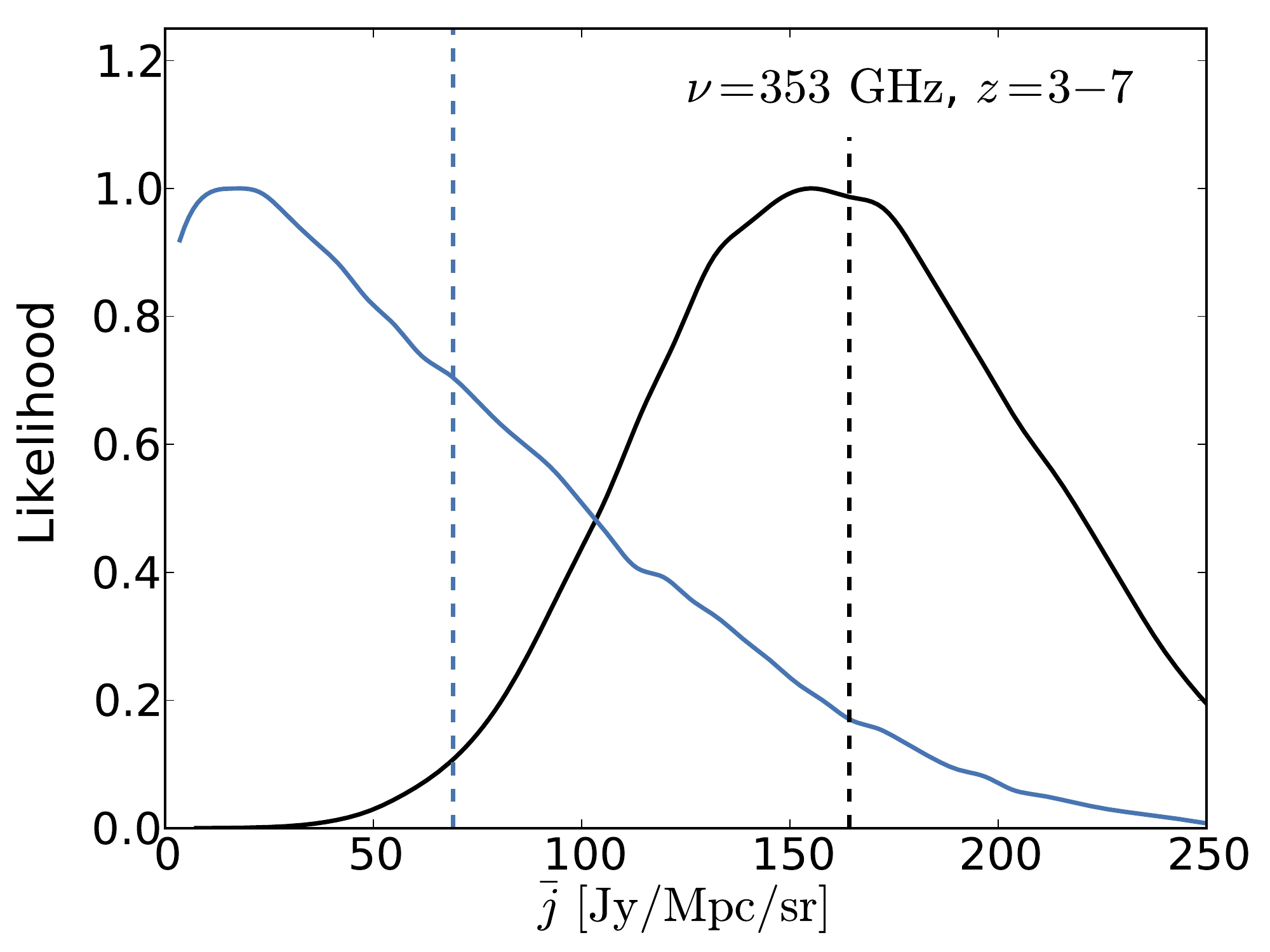}
  \caption[]{Marginalized 1-D distribution of the emissivity in the high redshift bin at
    353\,GHz with (black line) or without (blue line)
    including the CIB-lensing correlation. Its inclusion helps to
    constrain the emissivity at high redshift, transforming an upper limit
    into a detection.}
  \label{fig:jcomp}
\end{figure}

\begin{table*}[!t]                 
\begingroup
\newdimen\tblskip \tblskip=3pt
\caption{Reconstructed emissivity as a function of redshift and
  associated star formation rate.  At each frequency and for each of the
  three redshift bins the first quantity
  corresponds to the mean emissivity in the corresponding redshift bin,
  $\bar j(z)$, in ${\rm Jy}\,{\rm Mpc}^{-1}\,{\rm sr}^{-1}$,
  while the second corresponds to the SFR density,
  $\rho_{\rm SFR}$, in ${\rm M}_\odot\,{\rm Mpc}^{-3}\,{\rm yr}^{-1}$.}
\label{tab:jval}                            
\nointerlineskip
\vskip -3mm
\footnotesize
\setbox\tablebox=\vbox{
   \newdimen\digitwidth 
   \setbox0=\hbox{\rm 0} 
   \digitwidth=\wd0 
   \catcode`*=\active 
   \def*{\kern\digitwidth}
   \newdimen\signwidth 
   \setbox0=\hbox{+} 
   \signwidth=\wd0 
   \catcode`!=\active 
   \def!{\kern\signwidth}
\halign{\hbox to 0.80in{#\leaderfil}\tabskip 1.5em&
 \hfil#\hfil& \hfil#\hfil& \hfil#\hfil&
 \hfil#\hfil& \hfil#\hfil& \hfil#\hfil\tabskip=0pt\cr
\noalign{\doubleline}
\omit& \multispan2 \hfil$1 < z \leq1.5$\hfil&
 \multispan2 \hfil$1.5<z\leq3$\hfil&
 \multispan2 \hfil$3<z\leq7$\hfil\cr
\noalign{\vskip -5pt}
\omit& \multispan2 \hrulefill&
 \multispan2 \hrulefill&
 \multispan2 \hrulefill\cr
\omit& *$\bar j(z)$& $\rho_{\rm SFR}$& *$\bar j(z)$& $\rho_{\rm SFR}$&
 *$\bar j(z)$& $\rho_{\rm SFR}$\cr
\noalign{\vskip 3pt\hrule\vskip 3pt}
100\,GHz&7.16$\pm$5.77&1.96$\pm$1.58&3.53$\pm$3.05 &0.655$\pm$0.564&5.49$\pm$4.78&0.271$\pm$0.236\cr
143\,GHz&12.7$\pm$9.60&*1.37$\pm$0.964&6.82$\pm$5.46&0.438$\pm$0.351&10.5$\pm$9.05&0.178$\pm$0.153\cr
217\,GHz&11.9$\pm$6.33&0.310$\pm$0.165&17.3$\pm$7.23&0.282$\pm$0.118&36.6$\pm$13.8&0.182$\pm$0.068\cr
353\,GHz&116$\pm$17.1&0.671$\pm$0.099&75.5$\pm$27.5&0.286$\pm$0.104&164$\pm$47.3&0.320$\pm$0.092\cr
545\,GHz&185$\pm$106&0.320$\pm$0.183&224$\pm$148&0.317$\pm$0.210&417$\pm$251&0.659$\pm$0.396\cr
857\,GHz&193$\pm$139&0.144$\pm$0.104&354$\pm$212&0.317$\pm$0.190&609$\pm$359&*1.37$\pm$0.809\cr
\noalign{\vskip 3pt\hrule\vskip 3pt}}}
\endPlancktablewide                 
\endgroup
\end{table*}                        

 \begin{figure}[!t]
   \centering
   \includegraphics[width=88mm, clip=true, trim=1.3cm 0 0 0]{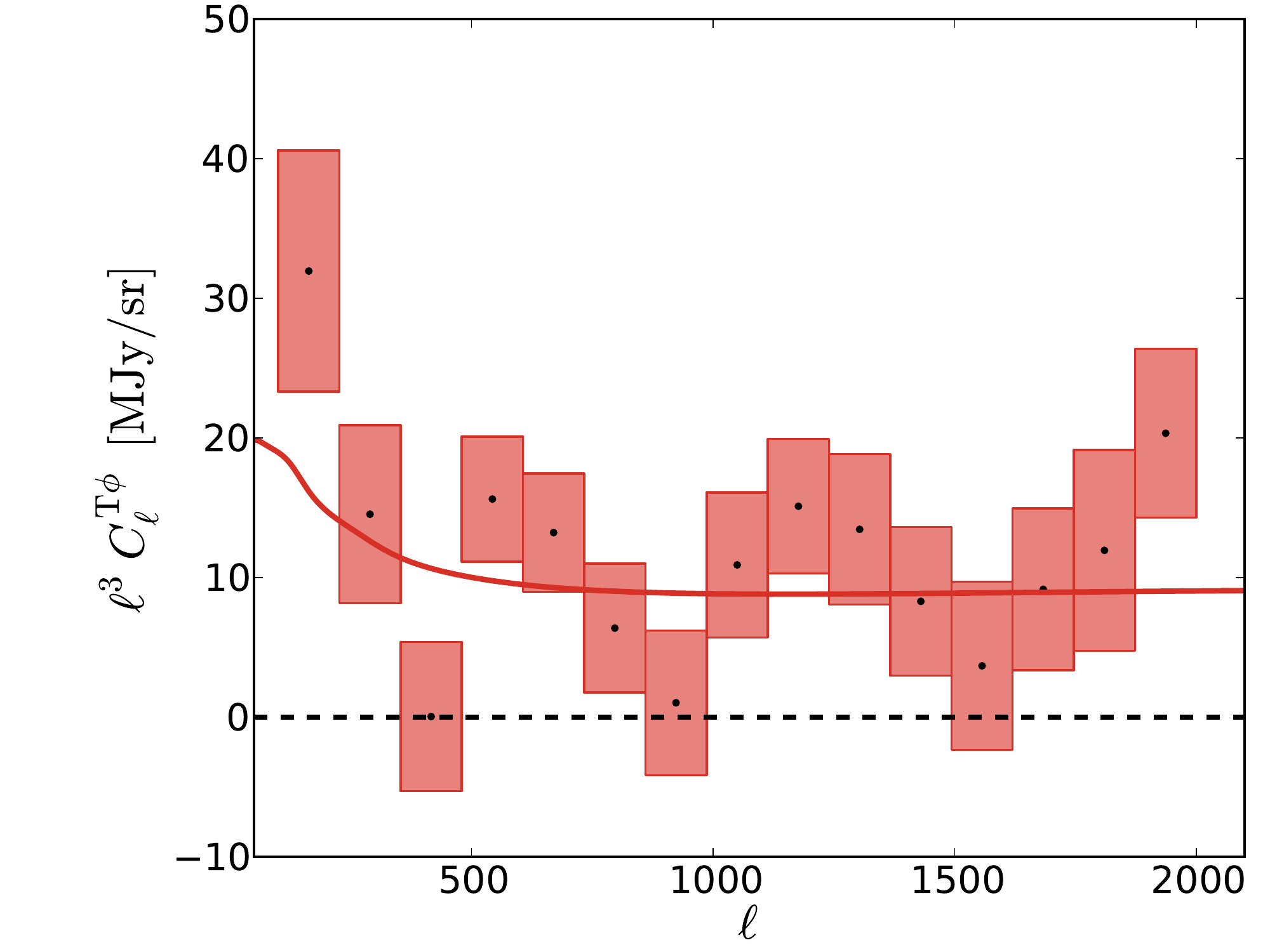}
   \caption[]{Correlation between the lensing potential and the
     IRIS map at 100$\,\mu$m using our  nominal lens reconstruction. We clearly see a correlation and
     estimate the significance to be 9\,$\sigma$, ignoring possible
     systematic effects. The solid line represents a simple
     reasonnable prediction for this signal.}
   \label{fig:iris_results}
 \end{figure}

\subsection{Discussion and outlook}
\label{sec:disc_summary}

In the previous section we described a model that simultaneously fits
the CIB auto-spectra and the CIB-lensing
cross-spectra, at all frequencies and with a single HOD
prescription. Given that we use an emissivity function that is close to
the B11 emissivities (to within our uncertainties),
we expect predictions of the galaxy number counts derived from our best-fit
emissivity to agree with current estimates \citep{Bethermin:2012b}.
The fact that our measurement is consistent with previous models
of the CIB lends support to our current understanding of its origin.
For example, the characteristic mass scale at which halos host galaxies,
$M_{\rm min}$, is consistent with values derived in PER, and is consistent
with, but slightly higher than, the value derived more
recently in \cite{Viero:2012he},
$\log_{10}{\left(M_{\rm min}/\Msolar\right)}=9.9 \pm 0.5$ (although a direct
comparison could be misleading given the different model assumptions).
In particular, it is clear that our model has
limitations, some of which have been partially addressed in recent work
\citep{Shang:2011mh,Bethermin:2012c,DeBernardis:2012bf,Bethermin:2012a,Viero:2012he,Addison2012}
and are points of focus in PIR, amongst them the mass independence
of the emissivity. Another question worth further investigation is the
dependence of our results on the binning scheme chosen for the
emissivity, which will be addressed in a future paper.

Given the consistency of our model with the PER results, the information
added by our
cross-spectrum measurement is worth quantifying. As an example, we show in
Fig.~\ref{fig:jcomp} the highest redshift emissivity bin in the 353\,GHz band.
Adding the CIB-lensing cross-spectrum information tightens the
constraint on the high-redshift part of the emissivity.
This statement also holds for the other frequencies and
stems from the fact that the CMB lensing kernel peaks at high redshift,
making the cross-correlation more sensitive to the high-redshift CIB signal
than the CIB auto-spectrum, as is illustrated in
Fig.~\ref{fig:clcc_clpp_redshifts.pdf}.  Although
this gain does not translate into a substantial improvement in
$M_{\rm min}$, it leads to interesting constraints on the SFR
density, as can be seen in Table~\ref{tab:jval}.

The results at frequencies above 217\,GHz
each lead to around $2\,\sigma$ evidence for 
a non-zero SFR density for $1.5<z<3$ and for $3<z<7$. The values
inferred are consistent with other probes of the SFR in these redshift ranges,
as compiled for example in Fig.~1 of
\citet{Hopkins:2006bw}. Assuming that each frequency is independent, we
obtain SFR densities for the three redshift bins of 0.423 $\pm$ 0.123, 0.292 $\pm$ 0.138 and 0.226 $\pm$ 0.100
${\rm M}_\odot\,{\rm Mpc}^{-3}\,{\rm yr}^{-1}$, respectively where the
errors are 68\% C.L.. We note that the $j$ distributions are rather
non-Gaussian so that the 95\% C.L. become 0.228, 0.246 and 0.191 
respectively. This roughly $2\,\sigma$ detection per bin compares very
favourably with other published measurements. These constraints clearly illustrate how this particular correlation can be used to better isolate the high redshift component of
the CIB and improve our constraints on the star formation rate at high
redshift. Such constrains will improve with future measurements, in particular
if we can increase the signal-to-noise ratio in our lower frequency
channels, where the high redshift contribution is the greatest. This will likely
require an accurate removal of the CMB, our dominant source of noise at
low frequencies. A more thorough discussion of this possibility will be
given in PIR.

To fully utilize the richness of the correlation will require more studies.
Future work could involve using more sophisticated halo models specifically
designed to model star formation within halos, as well as relaxing some
of the assumptions made here, such as the mass independent luminosity
function. In addition the use of map-based methods that enable estimates of the
galaxy host halo mass by stacking the lensing potential maps is worth pursuing,
as is the extension to other data-sets.
For illustration purposes, we show in Fig.~\ref{fig:iris_results} raw
measurements of the correlation between our lensing potential map and the IRIS
map at 100\,$\mu$m that was introduced in Sect.~\ref{sec:iris}. We use
our nominal mask and lens reconstruction, with no \hi\ cleaning performed
on the IRIS map. We
clearly see a strong correlation, whose significance we estimate to be
9\,$\sigma$, ignoring any possible systematic effects. To guide the eye
we have added a prediction (not a fit) based on the HOD model
presented in \cite{penin2011b}. The full analysis of this signal is
beyond the scope of this paper, but it illustrates possible future uses
of the lensing potential map. In this case the IRIS wavelength range will help us
to isolate the low-redshift contribution to the CIB. 

To conclude, we have presented the first measurement of the correlation
between lensing of the CMB and the CIB. \Planck's unprecedented
full-sky, multi-frequency, deep survey enables us to make an internal
measurement of this correlation. Measurements with high
statistical significance are obtained, even after accounting for possible
systematic errors. The high degree of correlation that is measured (around
80\,\%) allows for unprecedented visualization of lensing of the
CMB and holds great promise for new CIB and CMB focused science.
CMB lensing appears promising as a probe of the origin of the CIB,
while the CIB is now established as an ideal tracer of CMB lensing. 

\begin{acknowledgements}

Based on observations obtained with \Planck\ (\url{http://www.esa.int/Planck}), an ESA science mission with
instruments and contributions directly funded by ESA Member States,
NASA, and Canada. The development of \Planck\ has been supported by:
ESA; CNES and CNRS/INSU-IN2P3-INP (France); ASI, CNR, and INAF (Italy); NASA and DoE
(USA); STFC and UKSA (UK); CSIC, MICINN and JA (Spain); Tekes, AoF and
CSC (Finland); DLR and MPG (Germany); CSA (Canada); DTU Space
(Denmark); SER/SSO (Switzerland); RCN (Norway); SFI (Ireland);
FCT/MCTES (Portugal); and PRACE (EU).  A description of the Planck
Collaboration and a list of its members, including the technical or
scientific activities in which they have been involved, can be found at
\url{http://www.sciops.esa.int/index.php?project=planck&page=Planck_Collaboration}. We
acknowledge the use of the {\tt HEALPix} package, and the LAMBDA
archive (\url{http://lambda.gsfc.nasa.gov}).

\end{acknowledgements}



\bibliographystyle{aat} 

\bibliography{main_bib_pep,main_bib,planck_cib_lensing,Planck_bib}

\appendix

\section{Statistical Errors}
\label{app:stat_errors}

\begin{figure*}[!t]
  \centering
  \includegraphics[width=120mm, clip=true, trim=0.5cm 0 0 0]{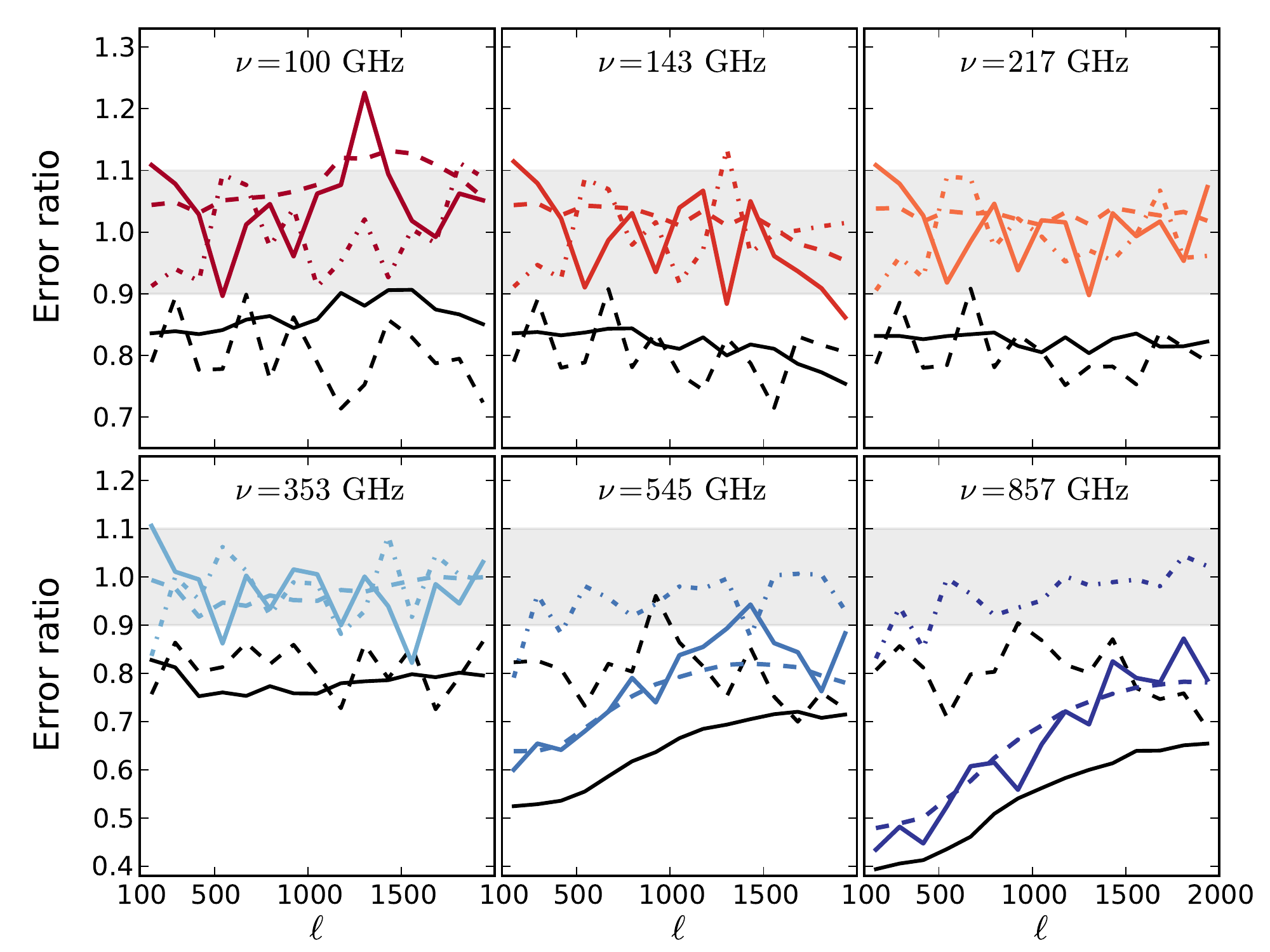}
  \caption[]{Ratio of various error estimation procedures to the errors
    obtained with the data-based analytical estimate.
    At each frequency the numerator is given by:
    (i) the scatter within an $\ell$-bin in simulations (solid black line);
    (ii) the scatter within an $\ell$-bin in the data (solid dashed black line);
    (iii) the scatter of bins across simulated realizations (solid coloured
    line);
    (iv) the analytical errors calculated from the simulations (dashed
    coloured line);
    (v) the scatter across realizations for the cross-correlation between the
    simulated temperature map and the lensing potential reconstructed from
    the data
    (coloured dot-dashed line).
    The grey envelope is the precision of the simulated errors expected
    from 100 simulations (shown as a spread around unity).} 
  \label{fig:error_ratio}
\end{figure*}

In this section we compare six different methods to estimate our
statistical errors.
This comparison is used to validate our claim
(presented in the main text) that we can obtain our errors from the
naive analytical errors calculated from the data, i.e., method 1 below.
The six different methods we compare are:
\begin{enumerate}
  \item	The naive, Gaussian, analytical errors estimated
	from the data through the measured total power of the $\rm T$ and
        $\phi$ fields, respectively
	$\hat{C}_\ell^{\rm TT}$ and $\hat{C}_\ell^{\phi \phi}$ and
	the model cross-spectrum $C_\ell^{{\rm T} \phi}$.
  \item As above but instead of the data maps we use one of our
        simulations of the CIB and CMB temperature maps 
        described in Sect.~\ref{sec:simulations}. The lens
        reconstruction is obtained from the simulated maps using the same
        procedure that we use for the data. 
  \item The scatter directly {\it within} individual $\ell$-bins in
        the data-determined cross-spectrum. 
  \item As above but the scatter is measured in each $\ell$-bin for
        each simulation realization, and the errors are averaged over 100
        realizations.
  \item The scatter {\it of the bins\/} is calculated using the cross-spectra
        measured from our simulated maps. This is a direct measurement of the
        statistical error we require (to the extent that our simulations are
        realistic), and will differ from the scatter {\it within\/} the
        $\ell$-bins, for example due to noise correlations between
	different multipoles of the cross-spectrum.
  \item The error in the cross-spectrum of the reconstructed lensing potential
        in simulated maps with the measured temperature maps. This will only
        give part of the contribution to the error since the temperature maps
        are fixed, but it is still a useful cross-check.
\end{enumerate}

In Fig.~\ref{fig:error_ratio} we show a comparison of the errors found
from our six measurement methods. The precision achievable with 100
simulations is indicated by the grey envelope. We show the errors in each
$\ell$ bin from the different methods divided by the data-based
analytical estimate. To discuss the implications of these results we
shall focus on the 100~GHz panel first. The scatter measured within
an $\ell$ bin is fairly consistent in the simulations
(method 4, black solid line) and in the data (method 3, black dashed line)
giving us confidence in our simple simulation procedure. This rules out
important systematic contributions
and shows that our signal is mostly Gaussian, as expected.
Note that the consistency with the simulations is not surprising, since at
low frequencies we are dominated by CMB and instrumental noise, which are
well understood.  In addition, the fact that the analytical errors
calculated on the simulations (method 2, coloured dashed line) are mostly
within the grey shaded region, and are therefore close to the analytical
errors calculated from the data (method 1), gives us further
confidence in the simulations.
To the extent that the simulations are accurate, the scatter of the
$\ell$-bins in simulations (solid coloured line) is the error that we require.
The fact that it is essentially all within the shaded envelope means that
this method gives errors that are consistent with the analytical errors
measured using the data, justifying
our nominal choice for calculating the errors at low frequencies.

However, comparing the black lines with the coloured lines clearly indicates
that in the data and simulations obtaining the error bars by measuring
the scatter within the $\ell$-bins leads to an underestimation of the
errors by approximately 20\,\%. Given the fact that this difference is
observed in both the data and the simulations, we exclude any instrumental
systematic effect as its cause
and explain it as being due to noise correlations within the $\ell$-bins.
Such a correlation is expected, since
most of the lens reconstruction signal in the $\ell$-range of interest to us
comes from modes in the CMB map within
a relatively narrow range at $\ell\simeq 1500$, and
so multipoles in the lens reconstruction are correlated.
We have also checked that the mask induced $\ell$-bin coupling is negligible,
given the bin width we have chosen, and is always smaller than 2\,\%.

All of these conclusions remain valid up to 353\,GHz.
However, at 545 and 857\,GHz, we see by looking at the errors measured
using simulations (solid black for method 4, solid coloured for method 5,
and dashed coloured for method 2) that the errors deduced from the
analytical estimates measured from the data are
substantially higher than those we measure in the simulations.
This is easily explained through the fact that we are omitting any
foreground emission in our simulations. The relative contribution
of Galactic foreground emission is higher at low $\ell$, which is expected
because the Galactic cirrus emission has a steep power spectrum.
Overall, the amplitude of this contribution is also consistent with what is seen in
Fig.~\ref{fig:error_by_l}.

Since the scatter within the $\ell$-bins measured in simulations
(black dashed line)
is about 20\,\% lower than the data-based analytical estimates at all
frequencies, we use the data-based analytical estimates as the basis for
our statistical errors.  We could alternatively scale the scatter-determined
errors by 20\,\% and obtain consistent results.
This approach accounts for the foreground emission seen at
545\,GHz and 857\,GHz, but will in practice neglect the contribution
to the errors from
the non-Gaussian part of the foregrounds. However, we show in Sect.~\ref{sec:cib_bispectrum}
that this contribution is small enough that we can ignore it.

The remaining method to discuss is obtained from the cross-spectrum of the 
reconstructed lensing potential in simulated maps with the data-measured
temperature maps
(method 6, coloured dashed-dotted line in Fig.~\ref{fig:error_ratio}).
At 545 and 857\,GHz the CIB signal is dominant over a large $\ell$ range,
and so the error obtained from this method is equal to the ``signal''
terms in Eq.~\ref{eqn:error_breakdown},
which are the orange and green lines in Fig.~\ref{fig:error_by_l}.
These two lines make up a significant fraction of the total error and
provide a reasonable approximation to the true error at high-$\ell$.
However, at low-$\ell$ where
Galactic emission is important, and at 100--353\,GHz where the CMB and
instrumental noise are the largest components, the orange and green
curves do not accurately describe the total error. We can see from
Fig.~\ref{fig:error_ratio} that the errors obtained using this method are
close to the errors measured using the other techniques.  However, this
method will underestimate the true errors, since the scatter in the CMB
and noise components is neglected.

Note that the results presented in Fig.~\ref{fig:error_ratio} are all
computed using the 40\,\% Galaxy mask, but we have checked that they hold
when using the 20\,\% and 60\,\% Galaxy masks (which are discussed in
detail in Sect.~\ref{sec:inst_and_syste}) and that the results show
the appropriate $f_{\rm sky}$ scaling. 

\raggedright
\end{document}